\documentclass[fleqn,usenatbib,useAMS]{mnras}
\usepackage[T1]{fontenc}
\usepackage{ae,aecompl}

\usepackage{newtxtext,newtxmath}
\usepackage{graphicx, url, times}
\usepackage{booktabs, threeparttable}
\usepackage{multirow}
\usepackage{amsmath}

\graphicspath{{pics/}}

\newcommand{\tabincell}[2]{\begin{tabular}{@{}#1@{}}#2\end{tabular}}

\newcommand{\hkpc}{{\ifmmode{h^{-1}{\rm kpc}}\else{$h^{-1}$kpc}\fi}}

\newcommand{\Fig}[1]{Fig.~\ref{#1}}
\newcommand{\Eqn}[1]{Eq.~\ref{#1}}
\newcommand{\Tbl}[1]{Table~\ref{#1}}
\newcommand{\Sec}[1]{Sec.~\ref{#1}}

\def\Mpc{\rm Mpc}
\def\Gpc{\rm Gpc}
\def\Msun{M_{\rm \odot}}
\def\ttt{\textsc{TheThreeHundred}}
\def\ill{Illustris-1}
\def\tng{TNG100-1}

\newcommand{\papertitle}{A stochastic model to reproduce the star formation history of individual galaxies in hydrodynamic simulations}

\newcommand{\affa}{School of Physics and Astronomy, Sun Yat-sen University, Zhuhai 519082, China}
\newcommand{\affj}{CSST Science Center for Guangdong-Hong Kong-Macau Great Bay Area, Zhuhai 519082, China}
\newcommand{\affh}{Zhejiang University-Purple Mountain Observatory Joint Research Center for Astronomy, Zhejiang University, Hangzhou 310027, China}
\newcommand{\affi}{Purple Mountain Observatory, No 8 Yuanhua Road, Nanjing 210034, China}
\newcommand{\affb}{School of Physics \& Astronomy, University of Nottingham, Nottingham NG7 2RD, UK}    
\newcommand{\affc}{Departamento de F\'isica Te\'{o}rica, M\'{o}dulo 15, Facultad de Ciencias, Universidad Aut\'{o}noma de Madrid, 28049 Madrid, Spain}
\newcommand{\affd}{Centro de Investigaci\'{o}n Avanzada en F\'{\i}sica Fundamental (CIAFF), Universidad Aut\'{o}noma de Madrid, 28049 Madrid, Spain }
\newcommand{\affe}{International Centre for Radio Astronomy Research, The University of Western Australia, 35 Stirling Highway, Crawley, Western Australia 6009, Australia}

\newcommand{\afff}{Institute for Astronomy, University of Edinburgh, Royal Observatory, Edinburgh EH9 3HJ, United Kingdom}

\newcommand{\affg}{Department of Mathematics and Theories, Peng Cheng Laboratory, No.2 Xingke 1st Street, Nanshan District, Shenzhen 518000, China}

\newcommand{\affk}{School of Physics and Astronomy, China West Normal University, No. 1 Shida Road, Nanchong 637002, China}

\newlength{\figwidth}
\newlength{\resplot}
\hypersetup{ pdfauthor={Yang Wang}, pdftitle={\papertitle},
pdfsubject={pdfsub}, urlcolor=blue, }

\title[SFH in simulations ]{\papertitle}
\author[Wang et al.]{Yang Wang$^{1,2,3}$
	\thanks{Contact Email: \href{mailto:wangy18@pcl.ac.cn}{wangy18@pcl.ac.cn}},
		Nicola R. Napolitano$^{2,3}$	\thanks{Contact Email: \href{mailto:napolitano@mail.sysu.edu.cn}{napolitano@mail.sysu.edu.cn}},
    Weiguang Cui$^{4,5}$,
	Xiao-Dong Li$^{2,3}$	\thanks{Contact Email: \href{mailto:lixiaod25@mail.sysu.edu.cn}{lixiaod25@mail.sysu.edu.cn}},
	Alexander Knebe $^{5,6,7}$,\newauthor
	Chris Power $^{7}$,
	Frazer Pearce $^{8}$,
	Lin Tang$^{9,2,3}$
	Gustavo Yepes$^{5,6}$,
	Xi Kang$^{10,11}$,
	\\
	$^{1}$\affg\\
	$^{2}$\affa\\
	$^{3}$\affj\\
	$^{4}$\afff\\
	$^{5}$\affc \\
	$^{6}$\affd \\
	$^{7}$\affe \\
	$^{8}$\affb\\
	$^{9}$\affk\\
	$^{10}$\affh\\
	$^{11}$\affi\\
}
\begin{document}
% \author[0000-0002-1512-5653]{Yang Wang}
% \affil{\affa}

% \author{Xi Kang}
% \affil{\affh}

% \author{Xiaodong Li}
% \affil{\affa}

% \author{Alexander Knebe}
% \affil{\affc}
% \affil{\affd}
% \affil{\affe}

% \author{Chris Power}
% \affil{\affe}

% \author{Frazer Pearce}
% \affil{\affb}

% \author{Gustavo Yepes}
% \affil{\affc}
% \affil{\affd}

% \author{Lin Tang}
% \affil{\affa}

% \author{Weiguang Cui}
% \affil{\affc}

\label{firstpage}
\pagerange{\pageref{firstpage}--\pageref{lastpage}}
\maketitle
\begin{abstract}
	The star formation history (SFH) of galaxies is critical for understanding galaxy evolution.
	Hydrodynamical simulations enable us to precisely reconstruct the SFH of galaxies and establish a link to the underlying physical processes.
	In this work, we present a model
	to describe individual galaxies' SFHs from three simulations: \ttt, \ill, and \tng.
	This model divides the galaxy SFH into two distinct components: the "main sequence" and the "variation". The "main sequence" part is generated by tracing the history of the $SFR-M_*$ main sequence of galaxies across time.
	The "variation" part consists of the scatter around the main sequence, which is reproduced by fractional Brownian motions.
	We find that:
	(1) The evolution of the main sequence varies between simulations;
	(2) fractional Brownian motions can reproduce many features of SFHs;
	however,
	discrepancies still exist;
	and (3) The variations and mass-loss rate are crucial for reconstructing the SFHs of the simulations.
	This model provides
	a fair description of the SFHs in simulations.
	On the other hand, by correlating the fractional Brownian motion model to simulation data, we provide a 'standard' against which
	to compare simulations.

\end{abstract}

\begin{keywords}
	methods: numerical  -- galaxies: evolution
\end{keywords}

\section{Introduction}
\label{sec:intro}
Observations over the last few decades have yielded a wealth of data about galaxies ranging from the local universe to high redshift.
Numerous studies of galaxy statistics and scaling relations at various epochs have been conducted using these data   \citep[e.g., ][]{Faber1976,Kormendy1977,Djorgovski1987,Baldry2004b,Wel2014,Barro2017}.
These have provided strong evidence for the long-term evolution of the scaling relations between various properties of galaxies.
Nevertheless, the relationship between the evolution of individual galaxies and the evolution of scaling relations as a whole is not yet well-understood.
Unlike stellar evolution, where the evolution track of a star with a given mass can be clearly defined, we do not yet have a credible model for tracing the evolution of individual galaxies.

Among all galaxy properties, the star formation rate (SFR) plays a critical role in the evolution of galaxies.
The SFR is shown to be tightly connected to the galaxy stellar mass, $ M_*$,
as $SFR\propto M_*^\beta$, with $\beta\sim1$ up to at least $z=3$ \citep{Brinchmann2004,Daddi2007,Elbaz2007,Noeske2007}.
The linear relationship between the logarithm of stellar mass and the logarithm of SFR is also referred to as the galaxy main sequence (MS).
Observations indicate that the intercept of MS
evolves with redshift as $\sim(1+z)^{2.2}$
\citep{Pannella2009,Stark2013,Schreiber2015,Boogaard2018}.
The increase
of intercept with time indicates a higher specific star formation rate,
which is most likely caused by the higher accretion rate of cold gas at high redshifts \citep{Lilly2013,Tacchella2013,Tacchella2018,Genzel2015,Cui2021}.
The scatter of the MS is small,
about $0.2$-$0.4 $ dex,
and steady across time, indicating that there is no strong evidence of evolution with redshift \citep{Whitaker2012,Speagle2014,Schreiber2015}.
This stability of the MS is interpreted as the outcome of quasi-steady state of gas inflow, outflow, and consumption during galaxies' evolution\citep{Bouche2010,Daddi2010,Dave2012,Dekel2014,Rodriguez-Puebla2016a}.

The above-mentioned MS features describe a statistical mean behavior of the galaxies
while  the physics involved in the history of each individual galaxy suffers from various perturbations, causing deviations from the mean relations for the quantities of galaxies.
These variations appear to be dispersed randomly distributed (although it is possible that they are not random).
Additionally, SFH encodes the information pertaining to these physics.
Hence, we reasonably expect that we can
gain additional knowledge about this physics if we would be able to model
relatively precise SFHs of individual galaxies.

So far, there have been numerous approaches to modelling individual galaxies' SFHs.
One popular method is utilizing the stellar population synthesis
and the spectral energy distribution (SED) fitting.
A galaxy's SED contains information about its stellar populations within a galaxy.
Using specific models parametrized with the stellar population parameters like age, metallicity and the timing of star formation episodes \citep[see e.g., ][]{Bruzual2003},
SED fitting provides a metric to optimize the parameters and consequently build up the SFH.
This approach, however, is very challenging because
the fitting process contains numerous degeneracies \citep{Papovich2001,Shapley2001,Muzzin2009,Conroy2013,Carnall2019,Leja2019}.
For instance, \cite{Ge2018} found that the SED fitting may be largely biased
for dust-rich galaxies.
Furthermore, the SED modelling can only reproduce the overall shapes of SFH but ignoring short-time variations\citep{Gallazzi2009,Ocvirk2006,Zibetti2009,Leja2019}.
These latter leave their imprint
on the destruction of giant molecular clouds (GMCs), SN feedback,  cosmic rays and photoionization feedbacks
\citep{Iyer2020,Tacchella2020}.
To compensate for this deficiency, dedicated investigations have been conducted to characterize these short-time-scale variations in SFH
\citep{Sullivan2000,Boselli2009,Wuyts2011,Guo2016a,Broussard2019,Emami2018,Faisst2019,Wang2020a,Wang2020,ChavesMontero2021}.

Another way to study SFH is by using hydrodynamic simulations, which can
follow the formation and evolution of individual galaxies in a self-consistent manner.
In recent years, there have been many extensive studies about SFH predictions, utilizing a large number of hydrodynamic cosmology and zoom-in simulations.
For example, \cite{Tacchella2016} investigated the SFH in the zoom-in simulations VELA \citep{Ceverino2014,Zolotov2015} and discovered a tight correlation between the SFH variation and the gas processes such as gas compaction, depletion, and replenishment.
Based on the EAGLE simulation, \cite{Schaye2015} and \cite{Matthee2019} have found that the scatter of MS comes from a combination of short- and long-time-scale fluctuations in SFHs.
The short-time-scale fluctuations are related to self-regulation from cooling, star formation and outflows, while the long-time-scale fluctuations are due to the dark matter halo growth.
Similar
conclusions have been made and discussed in Illustris \citep{Sparre2015}, IllustrisTNG \citep{Torrey2018},  FIRE simulation \citep{Sparre2017}, and the NIHAO simulation \citep{Blank2021}.
However,
different sub-grid physics implementations in hydrodynamic simulation
can result in significantly different SFH predictions across hydrodynamic simulations.
\cite{Iyer2020}, for example, has analyzed the power spectrum density of individual SFH based on six cosmological simulations, two zoom-in simulations, and additional semi-analytic and empirical models.
They discovered that there are obvious discrepancies between the SFHs produced by different simulations/models, even though the stellar mass function of galaxies in these simulations and models all accord well with observations.
Therefore, further improvement of the recipes for baryon models in hydrodynamic simulations is still required, and a deeper understanding of the physics behind the SFHs can help us on this endeavor.

In this paper, we present a mathematical model that can be used to mimic the evolution tracks of galaxies' SFHs in simulations.
When SFHs are described in a universal form, comparisons between SFHs from different simulations become easier.
In our model, we make the basic assumption that
an individual galaxy SFH follows a simple pattern:
1) it grows in lockstep with the trend of the main sequence (hence abbreviated as "MS part", denoted as $\Psi_{\rm MS}$);
2) it evolves randomly from there, producing a path that
deviates from the MS part and follows a Brownian motion (hence abbreviated as "variation part",  denoted as $\Delta$).
Motivated by \citet{Kelson2014}, we simulate the variation part using fractional Brownian motion.
A similar attempt was made by \cite{Caplar2019}, who proposed a stochastic process model for the variation of SFH characterized by a broken power law.
We validate this model by applying it to galaxy cluster re-simulation The Three Hundred (hereafter \ttt, also abbreviated as ``The300''), the simulation Illustris-1 and the simulation TNG100-1.

This paper is structured as follows:
We describe the data that we utilize in \Sec{sec:data}.
We construct the model in \Sec{sec:model}, which is divided into two parts:
\Sec{sec:ms} generates the MS part of the SFH model based on the evolution of $SFR-M_*$ scale relations from simulations;
\Sec{sec:var} generates the stochastic model for the variation part of the SFH.
We merge two parts in \Sec{sec:fullsfh} to build up a complete SFH model and assess its performance.
Finally, \Sec{sec:con} summarizes our model's findings and discusses their validity and future direction.

\section{Simulation Data}
\label{sec:data}
The simulation data from \ttt, \ill\ and \tng, as well as the
methods
to measure the star formation history of simulated galaxies, are briefly introduced in this section.

\subsection{\textsc{TheThreeHundred}}

{\sc The  Three Hundred} project\footnote{\url{https://the300-project.org}} consists of 324 re-simulated clusters and 4 field regions
extracted from the MultiDark Planck simulation, MDPL2
\citep{Klypin2016}. The MDPL2 simulation has cosmological parameters
of $\Omega_M=0.307, \Omega_B=0.048, \Omega_{\rm \Lambda}=0.693, h=0.678, and \sigma_8=0.823$.
All the clusters and fields have been simulated using the
full-physics hydrodynamic codes {\sc Gadget-X} \citep{Rasia2015} and {\sc Gadget-MUSIC} \citep{Sembolini2013},
which are improved versions of {\sc Gadget2} \citep{Springel2005a}.
In the re-simulation region, the mass of a dark matter particle is $12.7\times10^8h^{-1}\Msun$ and the mass of a gas particle is $2.36\times10^8h^{-1}\Msun$.
The mass of star particles varies from $3.60\times10^7h^{-1}\Msun$ to $1.65\times10^8h^{-1}\Msun$ with $99\%$ of them
being less massive than $4.60\times10^7h^{-1}\Msun$.
The softening length is $6.5\ h^{-1}\rm{kpc}$.
Each cluster re-simulation consists of a spherical region of radius $15h^{-1}\Mpc$ at $z=0$ centred on one of the 324 largest objects within the host MDPL2 simulation box, which is $1h^{-1}\Gpc$ on a side. The host halos of galaxies range in mass from
$2.54\times10^{10}h^{-1}\Msun$ to $2.63\times{10}^{15}h^{-1}\Msun$.
The largest halos within each of the 324 cluster re-simulations vary from $8.15\times10^{14}h^{-1}\Msun$ to $2.63\times{10}^{15}h^{-1}\Msun$.
A more detailed description of the 324 clusters and the simulation codes can be found in \cite{Cui2018}.

\subsection{Illustris-1}
The \ill\ simulation is a cosmological hydrodynamic simulation with a comoving volume of $(106.5{\rm Mpc})^3$.
It employs the moving mesh code AREPO \citep{Springel2010}.
Its cosmological parameters are consistent with WMAP9 data release \citep{Hinshaw2013},
%assuming that 
i.e., $\Omega_{\rm \Lambda}=0.7274$, $\Omega_{\rm m}=0.2726$, $\Omega_b=0.0456$, $\sigma_8=0.809$, $n_s=0.963$ and $h=0.704$.

The simulation contains $1820^3$ dark matter particles and $1820^3$ initial hydrodynamic cells.
The mass resolution of dark matter particles is $6.26\times10^6M_{\rm \odot}$ and the initial mass resolution of baryons is $1.26\times10^6M_{\rm \odot}$.
The simulation evolves the initial condition from redshift $127$ to $0$ with $136$ output snapshots.
Besides gravitation, it accounts for hydrodynamics and baryon processes such as gas cooling and photo-ionization, star formation, ISM model, stellar evolution, stellar feedback and AGN feedback.
The star formation histories are provided by the SubLink merger trees \citep{Rodriguez-Gomez2015}.
Readers can refer to \cite{Nelson2015} for the data release of \ill\ simulation.
More details of the simulation
can be found in \cite{Vogelsberger2014a,Vogelsberger2014,Genel2014,Sijacki2015}.

\subsection{TNG100-1}
The \tng\ simulation is a cosmological, large-scale gravity + magnetohydrodynamical simulation with the moving mesh code AREPO \citep{Springel2010}.
Its cosmological parameters are consistent with Planck2015 \citep{Collaboration2016},
i.e.,  $\Omega_{\rm \Lambda}=0.6911$, $\Omega_{\rm m}=0.3089$, $\Omega_b=0.0486$, $\sigma_8=0.8159$, $n_s=0.9667$ and $h=0.6774$.
Its box size is $110.7^3Mpc^3$.
The simulation contains $1820^3$ dark matter particles and $1820^3$ initial hydrodynamic cells.
The mass resolution of dark matter particles is $7.5\times10^6M_{\rm \odot}$ and the initial mass resolution of baryons is $1.4\times10^6M_{\rm \odot}$.
%For 
The simulation evolves the initial condition from redshift $127$ to $0$ with $100$ output snapshots.
Compared with \ill, the \tng\ simulation includes an updated physical model to simulate the formation and evolution of galaxies \citep[see][]{Pillepich2018a}
The TNG model updates the recipes for star formation and evolution, chemical enrichment, cooling and feedbacks \citep{ Weinberger2017,Pillepich2018,Nelson2018}.
It also presents a revised AGN feedback model to control the massive galaxies \citep{Weinberger2017} and galactic winds model to shape the low mass galaxies \citep{Pillepich2018}.
Readers can refer to \cite{Nelson2019} for the data release of \tng\ simulation.
More details can be found in the introductory paper series of \tng. \citep{Pillepich2018,Springel2018,Nelson2018,Naiman2018,Marinacci2018}.

\subsection{Galaxy Samples}

The SUBFIND \citep{Springel2001a} algorithm is used to locate the substructures.
Within a subhalo, all gases, stars, and dark matter particles (or cells) are associated with a single galaxy.
The galaxy's stellar mass ($M_*$) is defined as the sum of the masses of all stellar particles contained inside a single substructure.

The SFH of $243810$ galaxies with $M_*(z=0)>10^9M_{\rm \odot}h^{-1}$
is extracted from the galaxy catalogue in \ttt.
The largest galaxy has a stellar mass of $10^{13.4}M_{\rm \odot}h^{-1}$.
There are a few super large galaxies in this sample, which are in fact central brightest cluster galaxies(BCG) plus intra-cluster light(ICL).
Their number is quite small, so they do not have too much influence on building up the models.
Therefore, we do not try to separate the ICL for them.
There are few super large galaxies in these sample which are in fact central Brightest Cluster Galaxies (BCG) plus Intra-Cluster Light (ICL).
Their number is quite small thus do not have  too much influence on building up the models.
Therefore we do not try to separate the ICL for them.
Readers can refer to \cite{Cui2022} for more details.
To avoid contamination at the periphery of re-simulations, the selected galaxies are confined to a distance of $15 Mpc$ from the host cluster's center at $z=0$.

At redshift $0$, the final \ill\ catalog has $4366546$ substructures.
Among these, we select the SFH of $21795$ galaxies with $M_*(z=0)>10^8M_{\rm \odot}/h$ for our analysis.
The largest galaxy has a stellar mass of $10^{12.0}M_{\rm \odot}/h$.

At redshift $0$, the final \tng\ catalog has $4371211$ substructures.
Among these, we select the SFH of $28388$ galaxies with $M_*(z=0)>10^8M_{\rm \odot}/h$ for our analysis.
The largest galaxy has a stellar mass of $10^{12.1}M_{\rm \odot}/h$.

\subsection{The SFR of Simulated Galaxies}

In simulations, each gas cell/particle has its own star formation rate (SFR) to guide its star formation process in the subsequent phase.
A galaxy SFR is computed by adding the SFR of all its gas cells/particles.
From the an observational point of view, this is an instantaneous star formation rate, which is not measurable in practice.
Therefore numerous studies derive the SFR estimates from the stellar mass formed over the last $N$ Myr  \citep{Donnari2019,Hahn2019}.
In these approaches the SFR measurement is strongly dependent on the choice of the timescale, $N$.
Since in our work we focus exclusively on the SFH from the three reference simulations, in the following we will use the instantaneous SFR for %the three simulations, 
to minimize the uncertainties introduced by the timescale of SFR estimates.
This is also more appropriate for our analysis, where we want to model the impact of stochastic events that can produce sudden variations of the SFR.

Due to the time resolution limitation of snapshots, we must relinquish fluctuation information with a timescale shorter than the interval between snapshots
(about  $\delta t=0.135 Gyr$).
Principally, our work is
using the instantaneous SFR to represent the average SFR in the following $\delta t$.
Because the time scale of instantaneous SFR is shorter than $\delta t$, this will introduce bias to fluctuation at this time scale.
However, other methods to evaluate the SFR within a time scale can not avoid contamination from mergers and the death of stars
\cite[e.g., ]{Matthee2019}.
There is not a perfect method to probe into the SFR variance down to a very short time scale.
Readers should keep in mind that the variations around a time scale of $0.135 Gyr$ or shorter are not accurate in this work.

Notably, many simulated galaxies might have their instantaneous SFR of $0$ at some time.
For brevity, we will refer to this as the "0SFR" stage.
Galaxies at "0SFR" stage typically lack of cold gas, or contain just hot gas cells/particles.
These "0SFR" galaxies are more likely the result of a resolution effect.
At that time, galaxies may have a small volume of gas can not be resolved, or a relatively low SFR value yet are numerically recognized as having zero SFR.
In the SFH of a simulated galaxy, a "0SFR" stage will show up as
%may occur on occasion, 
sudden 0-peak, followed by a jump to a non-zero SFR. This is a condition that is unlikely to occur in real galaxies.
Additionally, "0SFR" stage can appear in a prolonged quenched phase that sometimes continues until redshift $z=0$.
To avoid biased estimates from spurious events and $-\infty$ values in $\log SFR$, we will either reset the value of "0SFR" data points or omit them from our inferences, depending on our objectives.
We will specify which of these options we take it in the following, when necessary.

Finally, for the sake of brevity, we will adopt the following definitions throughout the rest of the paper:
\begin{gather}
	m\equiv\log(M_*/(h^{-1}M_\odot)) \label{Equy}\\
	\Psi\equiv\log(SFR/(M_\odot yr^{-1})). \label{Equx}
\end{gather}

\begin{figure}
	\includegraphics[width=1\linewidth]{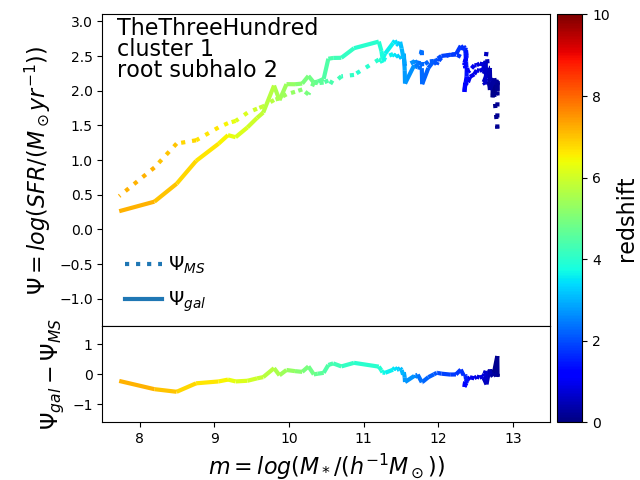}
	\caption{The evolution of star formation history of one single galaxy (solid lines).
		The x axis represents the logarithm of stellar mass, while the y axis represents the logarithm of SFR.
		The colors of the lines indicate the redshifts.
		The SFR of the main sequence, $\Psi_{\rm MS}$, with the same stellar mass and redshift as that galaxy is represented with a dotted line as a reference.
		The lower panel depicts the SFR deviation from the main sequence.
	}
	\label{FigSFH}
\end{figure}

\section{The SFH model}
\label{sec:model}

\begin{figure*}
	\includegraphics[width=0.33\linewidth]{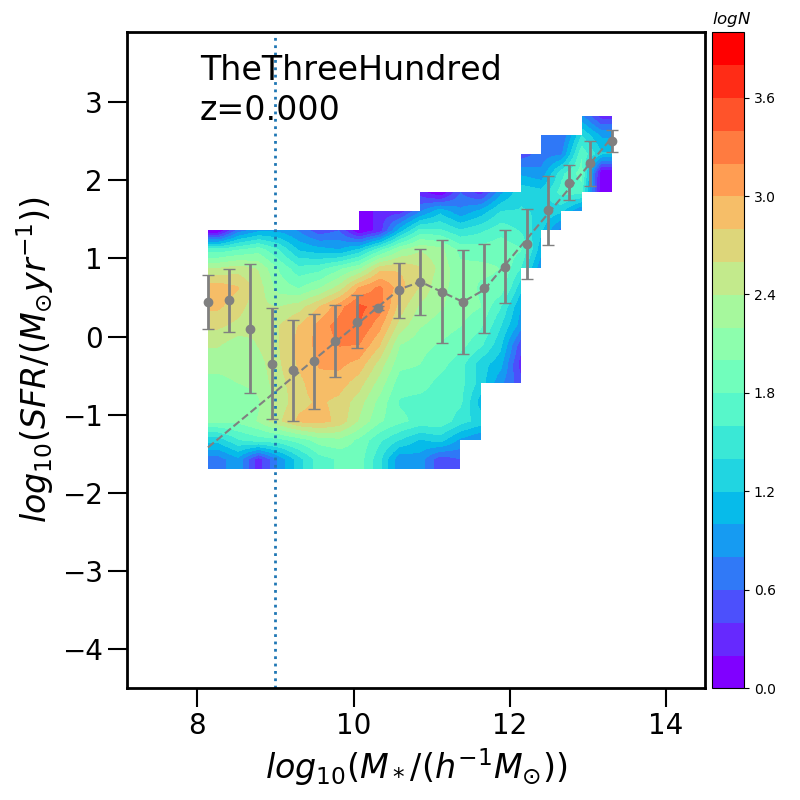}
	\includegraphics[width=0.33\linewidth]{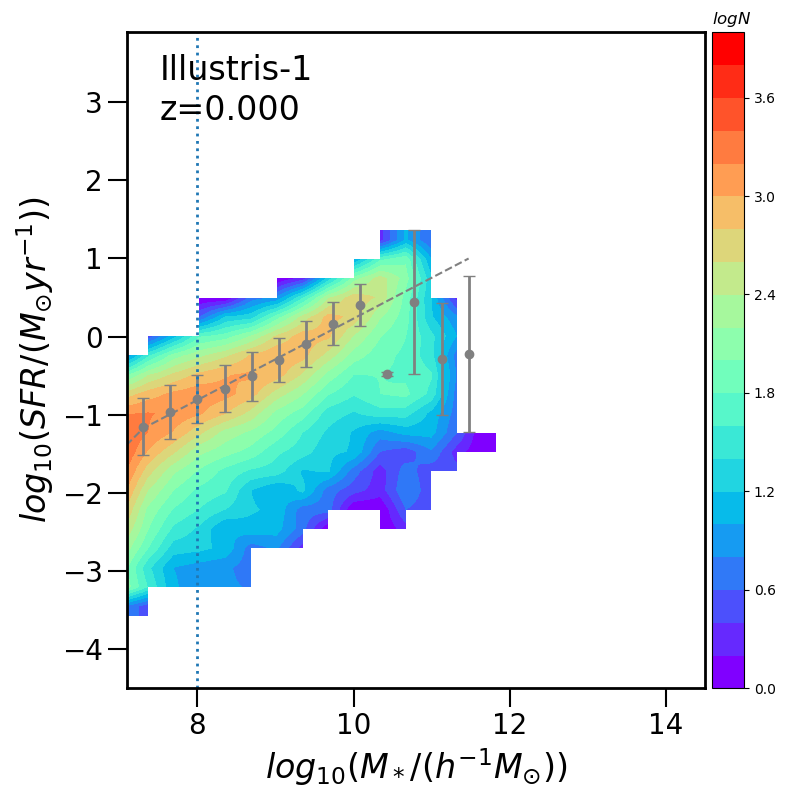}
	\includegraphics[width=0.33\linewidth]{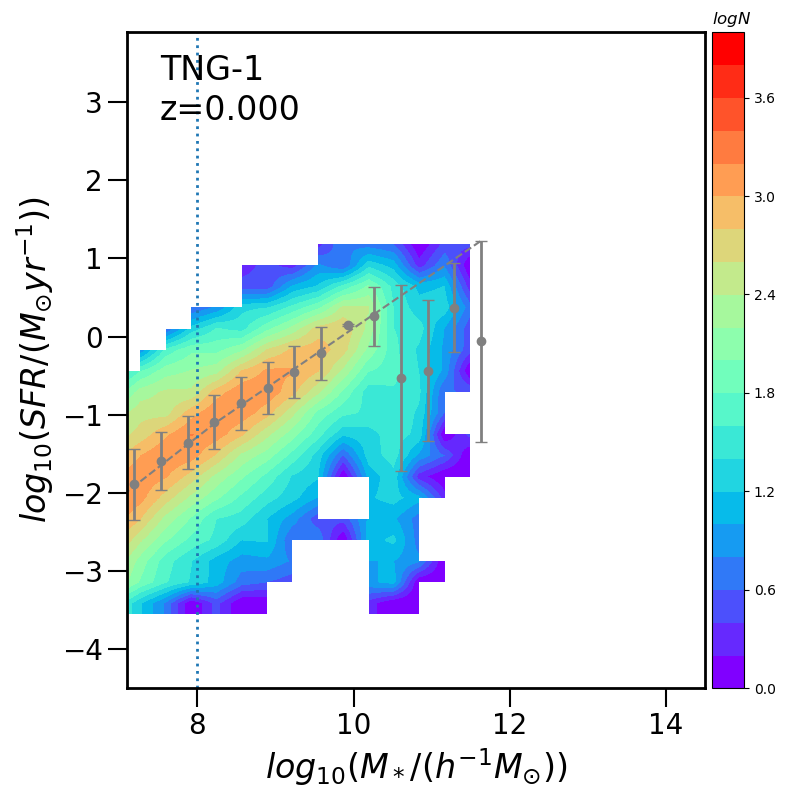}
	\caption{The contour of $SFR-M_*$ distribution of galaxies at $z=0$ in  simulations \ttt\ (left), \ill\ (center) and \tng\ (right).
		The grey dots indicate the peaks in the SFR density distribution for each stellar mass bin, while the error bars indicate the area of  $1\sigma$.
		The dots without error mean that the SFR distribution could not be fitted well by a log-normal distribution.
		The grey dashed lines indicate linear regression to the dots of SFR peaks, which is the main sequence SFR $\Psi_{\rm MS}(m,z=0)$.
		The vertical dotted lines depict the samples' lower stellar mass limit.
	}
	\label{FigMSz0}
\end{figure*}

In this section, we present our SFH models for the three simulations.
%In simulations, 
We have found out that the SFH of a single galaxy tends to follow the evolution of the main sequence star formation rate, $SFR_{\rm MS}$,
%The definition of $SFR_{\rm MS}$ will be given 
which will be detailedly defined in section \ref{sec:ms1}.
In \Fig{FigSFH}, we illustrate the $\Psi(m)$ trajectory of one galaxy from the \ttt\ simulation as a solid line color coded by the actual redshift.
%with a solid line in \Fig{FigSFH}.
%The colors of lines indicate the corresponding redshift for a specific stellar mass.
As a reference, the curve of $\Psi_{\rm MS}(m,z)\equiv\log(SFR_{\rm MS}(m,z)/(M_\odot yr^{-1}))$ is also indicated by a dotted line.
This  $\Psi_{\rm MS}$ curve depicts the main sequence star formation rate when the stellar mass and redshift are the same as the investigated galaxy.
As it can be seen, the simulated galaxy SFR does not depart significantly from the MS SFR,
%The deviation from MS is within $1$ dex, 
remaining within $1$ dex during its evolution, as shown in the bottom panel.
Such small deviation
%can be detected in 
is typical of the majority of the star formation histories, except
%when they are 
for quenched galaxies.

Based on this
%scenario
picture, we
%considered to build up 
propose a SFH model with the following form:
\begin{equation}
	\Psi_{\rm gal} = \Psi_{\rm MS}+\Delta.
\end{equation}
%This model consists of two components: a  
where $\Psi_{\rm MS}$ is the MS component, which represents the evolution of the MS, and $\Delta$ is the variation part, which represents the deviation from the MS.
We assume that these two components are unrelated in order to model them independently, as we detail in the following two subsections.

\subsection{modelling the Main Sequence Part of SFH}
\label{sec:ms}

\subsubsection{Mesuring the main sequence SFR}
\label{sec:ms1}

The first part of our model can be derived from the analytical function for the main sequence star formation rate $\Psi(m,z)\equiv \log SFR(m,z)$.
Many observations indicate a strong correlation between the stellar mass and SFR of star-forming galaxies, dubbed the ``main sequence''\citep[e.g][]{Noeske2007,Wuyts2011,Whitaker2012,Schreiber2015}.
Such correlation is also recovered in simulations, as illustrated in \Fig{FigMSz0}.
However, the precise formula describing this relation depends on the sample selection and measuring method \citep[see.][for reference]{Pillepich2018,Donnari2019,Bisigello2018}.
%In this case, we have to explain how  the main sequence is quantified.
Here below, we describe the procedure adopted to quantify the main sequence in this paper.

We found that galaxies within a stellar mass bin $[m-\Delta m/2, m+\Delta m/2]$ have their $\ln SFR$ generally following a Normal distribution.
\begin{equation}
	P(\ln SFR)\sim N(\mu,\sigma^2) =\frac{1}{\sigma\sqrt{2\pi}}e^{-\frac{(\ln SFR-\mu)^2}{2\sigma^2}}
\end{equation}
The mass bin $\Delta m$ is chosen to have a width of $0.3$ dex in \ttt\ and $0.4$ dex in \ill\ and \tng.
We can locate the peak of the SFR distribution at $SFR_{\rm peak}(m)=e^{\mu(m)}$ in each mass bin about $m$ by fitting the Normal distribution to the $ln SFR$ histogram.
Only $0SFR$ galaxies are excluded from the samples in this procedure.
The mean and scatter of SFR distribution, $\mu$ and $\sigma$, vary with the stellar mass and the redshift.
The $\mu$ has a linear relationship with $m\equiv\log M_*$, while the $\sigma$ is found to be approximately between $0.2$ to $0.5$ dex.
The grey dots in \Fig{FigMSz0} indicate the location of the $SFR_{\rm peak}$.
The error bars indicate the standard deviation of Normal distribution fits, which represents the scatters of SFR in each mass bins.

Then we use the following fitting equation to fit a collection of $\log SFR_{\rm peak}(m)$ data at each redshift $z$:
\begin{gather}
	\Psi_{\rm MS}=k(z)m+\Psi_0(z) \label{EquFit}
\end{gather}
$k$ is the slope and $\Psi_0$ is the intercept.
After fitting with \Eqn{EquFit}, we obtained the main sequence SFR function $\Psi_{\rm MS}(m,z)$, which is a function of stellar mass and redshifts.
In practice, as illustrated in \Fig{FigMSz0}, our fitting to the $\Psi_{\rm MS}$ omits the data points with large scatters at low mass or high mass end.

The resulted main sequence exhibits distinct slopes at low mass and high mass regions.
This is evident in the \ttt\ simulation.
MS appears to have a consistent at $z=0$
in \ill\ and \tng.
However, at higher redshifts, their MSs also have distinct slopes for low and high masses.
\Fig{FigMSzH} shows the $SFR-M_*$ distribution at higher redshifts.
From this figure, we can deduce that MS is bending.
To extract a universal function of main sequence SFR, we adopt piecewise linear function.
For the MS in the \ttt\ simulation,
we adopt a three-fold linear function:
\begin{gather}
	\Psi_{\rm MS} =\begin{cases}
		k_1(m-m_1)+\Psi_1                           & m\leq m_1     \\
		\frac{\Psi_1-\Psi_2}{m_1-m_2}(m-m_2)+\Psi_2 & m_1\leq m<m_2 \\
		k_2(m-m_2)+\Psi_2                           & m\geq m_2     \\
	\end{cases}
	\label{EquMSttt}
\end{gather}
For the MS in the \ill\ and \tng\ simulations, we adopt a two-fold linear function:
\begin{equation}
	\Psi_{\rm MS} =\begin{cases}
		k_1(m-m_1)+\Psi_1 & m< m_1     \\
		k_2(m-m_1)+\Psi_1 & m\geq m_1. \\
	\end{cases}
	\label{EquMSill}
\end{equation}

\begin{figure}
	\includegraphics[width=1\linewidth]{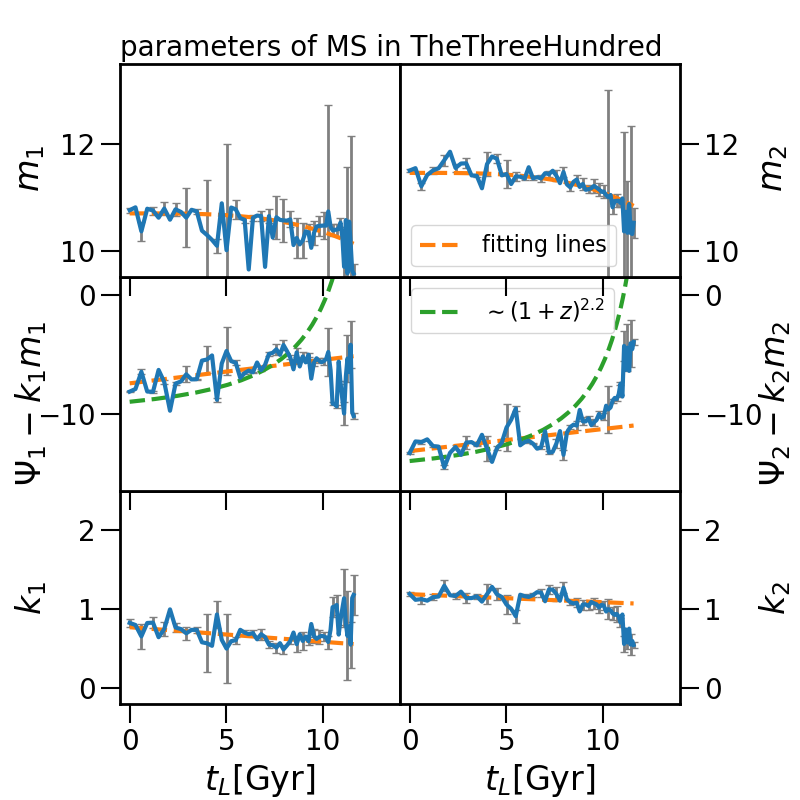}
	\caption{The fitting parameters for the MS,  $k_1$ (slope of lower mass MS), $k_2$ (slope of higher mass MS),  $\Psi_1-k_1m_1$ (intercept of lower mass MS),  $\Psi_2-k2m_2$ (intercept of higher mass MS), $m_1$ (turning point 1) and $m_2$ (turning point 2), in the \ttt\ simulation as functions of the lookback time $t_L$. The orange lines represent the best fits to the parameter histories. The green dashed lines depict observational trends $(1+z)^{2.2}$ of the changes of intercept of MS.}
	\label{FigParattt}
\end{figure}

\begin{figure}
	\includegraphics[width=1\linewidth]{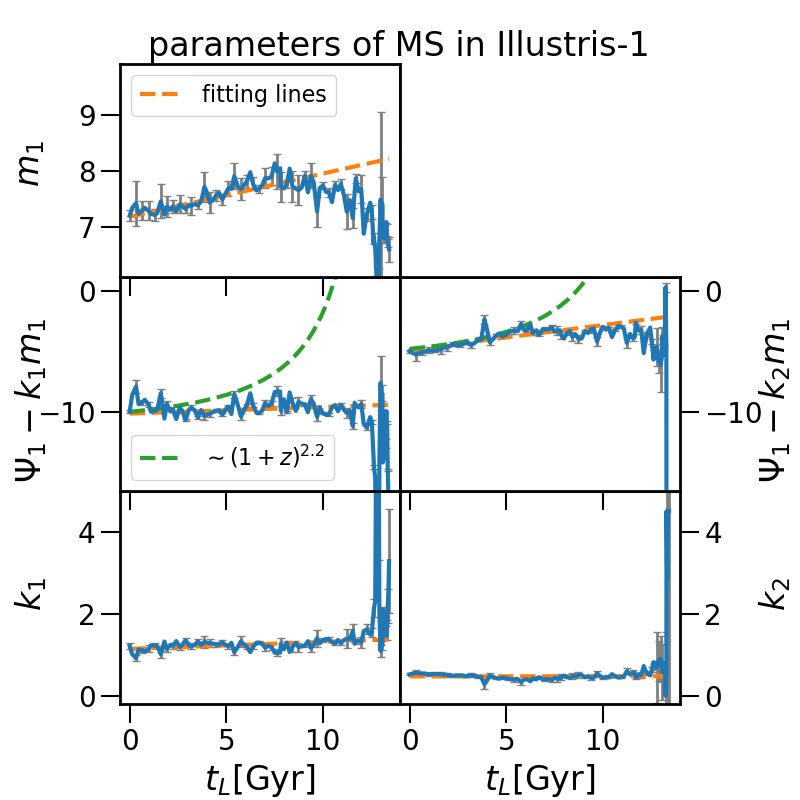}
	\caption{The fitting parameters, $k_1$ (slope of lower mass MS), $k_2$ (slope of higher mass MS),  $\Psi_1-k_1m_1$ (intercept of lower mass MS), $\Psi_1-k_2m_1$ (intercept of higher mass MS) and $m_1$ (turning point), in the \ill\ simulation as functions of the lookback time $t_L$.
		Note that in \ill\ the MS has only one turning point.
		Description of lines styles is similar to Fig.\ref{FigParattt}}
	\label{FigParaill}
\end{figure}

\begin{figure}
	\includegraphics[width=1\linewidth]{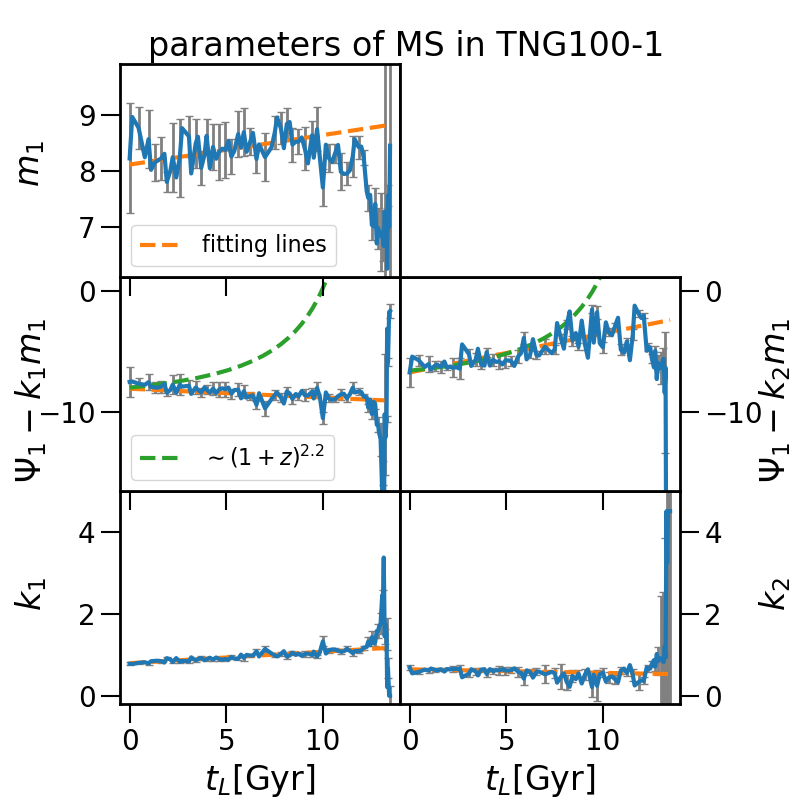}
	\caption{Similar to Fig.~\ref{FigParaill} but for the \tng\ simulation.}
	\label{FigParatng}
\end{figure}

Both \Eqn{EquMSttt} and \Eqn{EquMSill}have time-dependent parameters $k_1$, $k_2$, $m_1$, $\Psi_1$, $m_2$ and $\Psi_2$.
Their evolution as a function of the universal lookback time $t_L$ are shown in \Fig{FigParattt}, \Fig{FigParaill} and \Fig{FigParatng}.
To facilitate modelling, we substitute the universal lookback time $t_L$ for the redshift $z$.
And in all following context we will use $t_L$ by default.
We plot the $\Psi_1-k_1m_1$ as a function of $t_L$ in these figures rather than the $\Psi_1$, because the former one represent the intercept in
\Eqn{EquMSttt} and \ref{EquMSill}.

The best-fit values for these parameters are indicated by orange dashed lines in the same figures,
and for completeness,
the final forms of fitting functions are shown below.

For the \ttt\ simulation :
\begin{equation}
	\begin{aligned}
		k_1           & = -0.018\pm0.004t_L+0.77\pm0.02                              \\
		k_2           & = -0.010\pm0.003t_L+1.19\pm0.02                              \\
		m_1           & = (-1.50\pm4.16)\times10^{-4} t_L^{3.36\pm1.20}+10.70\pm0.02 \\
		m_2           & = (-1.28\pm2.63)\times10^{-4} t_L^{3.46\pm0.90}+11.45\pm0.02 \\
		% \Psi_1-k_1m_1 & =1.25\pm0.55t_L^{0.11\pm0.12}-8.16\pm0.50                    \\
		% \Psi_2-k_2m_2 & =(5.22\pm12.5)\times10^{-7}t_L^{6.73\pm1.00}-12.57\pm0.11
		\Psi_1-k_1m_1 & =0.20\pm0.04t_L-7.44\pm0.20                                  \\
		\Psi_2-k_2m_2 & =0.19\pm0.04t_L-13.16\pm0.21
	\end{aligned}
	\label{EquParattt}
\end{equation}

For the \ill\ simulation:
\begin{equation}
	\begin{aligned}
		k_1           & = 0.017\pm0.004t_L+1.14\pm0.03   \\
		k_2           & = 0.0004\pm0.0020t_L+0.47\pm0.02 \\
		m_1           & = 0.077\pm0.006 t_L+7.17\pm0.03  \\
		\Psi_1-k_1m_1 & =0.054\pm0.023t_L-10.13\pm0.12   \\
		\Psi_1-k_2m_1 & =0.20\pm0.02t_L-4.88\pm0.09
	\end{aligned}
	\label{EquParaill}
\end{equation}

For the \tng\ simulation:
\begin{equation}
	\begin{aligned}
		k_1           & = 0.028\pm0.001t_L+0.79\pm0.009   \\
		k_2           & = -0.0086\pm0.0023t_L+0.65\pm0.01 \\
		m_1           & = 0.053\pm0.013 t_L +8.11\pm0.08  \\
		\Psi_1-k_1m_1 & =-0.073\pm0.014t_L-8.09\pm0.12    \\
		\Psi_1-k_2m_1 & =0.32\pm0.03t_L-6.74\pm0.29
	\end{aligned}
	\label{EquParatng}
\end{equation}

The parameters of all MSs from simulations are summarized in \Tbl{TabMSPara}.
In \Tbl{TabMSPara}, we also give two MSs from previous observational works.
Readers who are interested in the comparison with observations can refer to appendix \ref{sec:mscompare}.

\subsubsection{Evolution trend of MS}
\label{sec:ms2}

\begin{table*}
	\centering
	\caption{A summary of the MS's parameters. All MSs are in the form of $\Psi(m,t_L)=k(t_L)m+\Psi_0(t_L)$,
		where $t_L$ is the lookback time in Gyr. The errors are not displayed in this table for brevity's sake. To facilitate comparison of parameters from different simulations within the same mass range, the second and fourth columns indicate the bound of the range of $\log M_*$.}
	\label{TabMSPara}
	\begin{tabular}{|c|cc|cc|c}
		\toprule
		 & \multicolumn{2}{c}{<-- Lower Mass}
		 &
		 & \multicolumn{2}{c}{Higher Mass -->}                \\
		\toprule
		 & $k$
		 & $m_1$
		 & $k$
		 & $m_1$ (and $m_2$)
		 & $k$                                                \\
		\hline
		The300
		 & --
		 & $\sim 8$
		 & $-0.018t_L+0.77$
		 & \tabincell{c}{$-1.50\times10^{-4}t_L^{3.36}+10.70$ \\ $-1.28\times10^{-4}t_L^{3.46}+11.45$}
		 & $-0.010t_L+1.19$
		\\
		Illustris-1
		 & $0.017t_L+1.14$
		 & $0.077t_L+7.17$
		 & $0.0004t_L+0.47$
		 & $\sim 12$
		 & --
		\\
		TNG100
		 & $0.028t_L+0.79$
		 & $0.053t_L+8.11$
		 & $-0.0086t_L+0.65$
		 & $\sim 12$
		 & --
		\\
		Speagle et al. (2014)
		 &
		 & $9.7$
		 & $0.026t_L+0.49$
		 & $11.1$
		 & -
		\\
		Iyer et al. (2018)
		 &
		 & $\sim 7$
		 & $0.017t_L+0.57$
		 & $\sim 11$
		 & -
		\\
		\toprule
		 & $\Psi_0$
		 & $m_1$
		 & $\Psi_0$
		 & $m_1$ (and $m_2$)
		 & $\Psi_0$
		\\
		\hline
		The300
		 & --
		 & $\sim 8$
		 & $0.20t_L-7.44$
		 & \tabincell{c}{$-1.50\times10^{-4}t_L^{3.36}+10.70$ \\ $-1.28\times10^{-4}t_L^{3.46}+11.45$}
		 & $ 0.19t_L-13.16$
		\\
		Illustris-1
		 & $0.054t_L-10.13$
		 & $0.077t_L+7.17$
		 & $0.20t_L-4.88$
		 & $\sim 12$
		 & --
		\\
		TNG100
		 & $-0.073t_L-8.09$
		 & $0.053t_L+8.11$
		 & $ 0.32t_L-6.74$
		 & $\sim 12$
		 & --
		\\
		Speagle et al. (2014)
		 &
		 & $9.7$
		 & $-0.11t_L-5.07$
		 & $11.1$
		 & -
		\\
		Iyer et al. (2018)
		 &
		 & $\sim 7$
		 & $-0.042t_L-6.04$
		 & $\sim 11$
		 & -
		\\
		\bottomrule
	\end{tabular}
\end{table*}

\begin{figure}
	\includegraphics[width=1\linewidth]{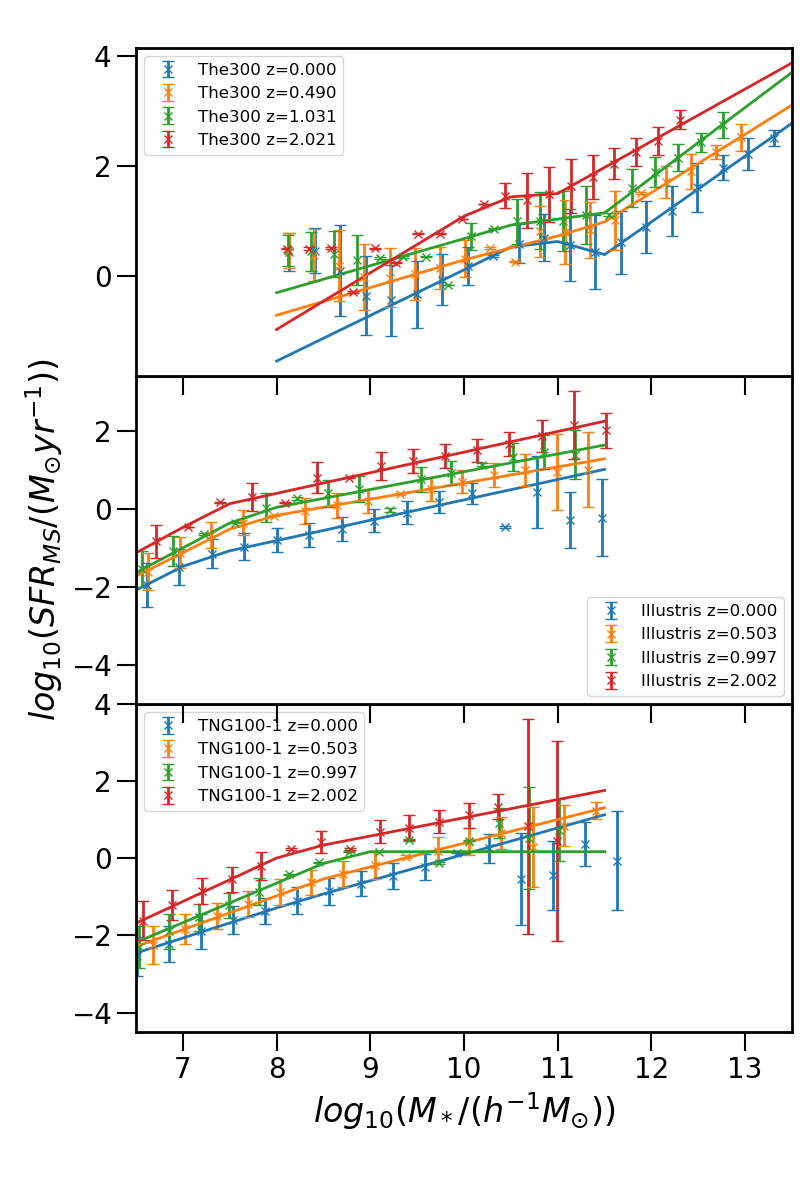}
	\caption{$\Psi_{\rm MS}$ comparison between simulations \ttt\ (top panel), \ill\ (middle panel) and \tng\ (bottom panel) at different redshifts.
		The colors of lines indicate the redshifts as legends show.}
	\label{FigMScompare}
\end{figure}

The MS show an evolving pattern in all simulations.
To illustrate how the MSs varies, in \Fig{FigMScompare} we compare the MS at redshifts $0$, $0.5$, $1$ and $2$ for the three simulations.
In general, the MSs descend throughout time, while their slopes are merely changed.
The time dependent functions of MS's slopes and intercepts  are also summarized in \Tbl{TabMSPara},
%In \Tbl{TabMSPara} 
where the parameters within a close mass range are listed in one column.
According to \Fig{FigParattt},\Fig{FigParaill},\Fig{FigParatng} and \Tbl{TabMSPara}, the MS in each simulation has distinct slopes and intercepts.

The slope of MS ($k_1$ or $k_2$)
denotes how many more stars form when the galaxy stellar mass is increased.
It can be related to the specific star formation rate ($SFR/M_*$).
A slope corresponding to $k=1$ indicates that the specific star formation rate is
independent of the stellar mass.
For $k>1$, the specific star formation rate increases with stellar mass, implying that massive galaxies form stars more efficiently.
On the contrary, for $k<1$
the specific star formation rate decreases with stellar mass.
In \ttt, the slope of MS is smaller than $1$ in the low mass range and greater than $1$ in the high mass range.
In comparison to \ttt,
the trend in \ill\ is completely reversed.
In \tng, the slopes of two mass ends are both less than $1$.
Unlike \ttt\ and \ill, the slopes at low and high mass bins are more similar in the \tng\ simulation.

Additionally, the slopes' evolving tendencies vary between simulations.
In \ttt, both the slopes of the high mass MS ($k_2$) and the low mass MS ($k_1$) decline slightly as $t_L$ increases.
In \ill\ and \tng, the high mass slopes decrease as $t_L$ increases, whereas
the low mass slopes increase as $t_L$ increases.
In \ttt, the time dependency of slopes is considerably more obvious.
The high mass slopes in \ill\ and \tng\ appear to be nearly constant.

The intercept of MS ($\Psi_1-k_1m_1$, $\Psi_1-k_2m_1$ or $\Psi_1-k_2m_2$) differs in three simulations.
In \ttt, the intercepts of both low and high mass MS increase as $t_L$ increases in a similar slope, albeit the latter one is approximately $5.7$ dex smaller.
In \ill\ and \tng, the intercepts of higher mass MS increase as $t_L$ increases, whereas the intercepts of  low mass MS slightly decrease with  $t_L$.
At an early epoch (about $t_L>8.5 Gyr$), all simulations exhibit abrupt decreases or increases in intercepts.
We use a linear fitting to describe the evolution of intercepts that omits this part.
Such linearly time dependent intercept was also adopted in \cite{Speagle2014} and \cite{Iyer2018}.
This form seems to contradict previous literature claiming that the MS intercept grows with redshifts as $(1+z)^{2.2}$ \citep{Brinchmann2004,Daddi2007,Elbaz2007,Noeske2007,Whitaker2012,Schreiber2015}.
In \Fig{FigParattt}, \Fig{FigParaill} and \Fig{FigParatng},
we represent this trend of $(1+z)^{2.2}$ as green dashed lines (converting $z$ to $t_L$).
Only the massive end of the MS in \ttt\ has a comparable pattern.
However,
be aware that the intercepts in \cite{Speagle2014}, \cite{Iyer2018} and this work are the intercepts located at $logM_*=0$, while in some other works they tend to use the intercept at $logM_*=9\sim10$.
The difference between intercepts at different $logM_*$ depends on the slope of MS, which is time-dependent.
It's not fair to compare intercepts at different stellar masses.

The  MS of \ttt\ appears considerably different with those two Illustris runs.
It should be mentioned, however, that the low mass MS in \ttt\ is roughly in the same mass range with the high mass MS in \ill\ and \tng\ (see \Fig{FigMScompare}).
When the low mass MS in \ttt\ is compared to the high mass MS in \ill\ and \tng, it appears to be more consistent.
While the \ttt\ galaxies are mainly in clusters,  the \ill\ and \tng\ galaxies lives in various kinds of environments.
One may argue that this is an explanation for the MS disparity between simulations.
However, some prior studies asserted that the underlying physics governing MS evolution are relatively insensitive to the environments\citep{Peng2010,Koyama2013,Lin2014}.
Therefore, we considered the difference between \ttt\ and Illustris runs to be mostly due to  their  sub-grid physics recipes, rather than the effect from environments.

The difference of MS across different mass range
should mainly be attributed to the
the varied sub-grid physics at different masses.
Many semi-analytical models have demonstrated that the intercept of MS is
significantly connected with the gas inflow and outflow rates with halo mass
\citep[e.g., ][]{Dave2011,Dave2012,Guo2008}.
\cite{Sparre2015} asserts that this holds true for simulations as well.
Although the gas flow in simulations can not be directly controlled,
the strength of feedbacks and cooling rate can have an effect.
The explicit division of MS into two mass halves
reflects the fact that these simulations adopts quite different feedback
models in different mass range in order to match the stellar mass function in all mass range.
Although the resolution effect is another  possible reason,
the values of MS turning point  minimize this probability.
\ill\ and \tng\ have a very close resolution.
However, the knee point in \tng\ is about $1$ dex larger than that in \ill.
They should be the comparable if resolution effect dominates the difference in MS slopes.
On the other hand, no simulation maintains constant values for turning points.
If the difference in MS between mass ranges is purely a resolution effect, we would anticipate turning points to be independent of time, as  resolution does not change over time in a single simulation.
The temporal dependence of turning points, $m_1$ and $m_2$,
is best fitted by a power function in \ttt.
In \ill\ and \tng,  their $m_1$ initially increases and subsequently decreases.
Concerning the uncertainty at the early epoch, we  fit  $m_1$ at $t_L<10Gyr$ using a linear function (left top panel in \Fig{FigParaill} and \Fig{FigParatng}).

\subsubsection{Uncertainties in MS measurement}
\label{sec:msuncert}
\cite{Donnari2019} demonstrates that some factors will affect the MS measured.
When comparing theoretical models with observations, as well as  between observational data themselves, the uncertainties brought by measurements
must be carefully examined.

When researchers compare the MS from observations to that from simulations, they typically use an average SFR over a certain time scale, such as 10, 50, 100 or 1000 Myr\citep{Iyer2020,Donnari2019}.
Although the simulations provide the galaxy's instantaneous SFR, it can not be observed in observations.
However, the varying timescales can result in different MS\cite[see appendix A in ][]{Donnari2019}.
Because the goal of our work is to find the pattern of SFHs in simulations rather than to compare them to observations,
we employ the instantaneous SFR to focus the more intrinsic variables.
Similarly, we determine the SFR and stellar mass of all particles bound to a subhalo rather than taking a specific aperture, e.g., 30 kpc.

The MS may also be affected by sample selection due to its definition.
Usually, the median or mean $\log SFR$ of a certain range of star forming galaxies is defined as the  $\Psi_{\rm MS}(M_*)$.
The cut for sample selection results in inconsistency in the star-forming main sequence across various works \citep{Somerville2015}.
To avoid the uncertainty from selection effect, some works employ more complex approaches to define the MS.
\cite{Renzini2015} presented an objective definition for the MS, defining it as the ridge line that connects the peaks on the $SFR-M_*$ contour.
\cite{Bisigello2018} used multiple-Gaussian function to decompose the $SFR-M_*$
distribution into three sequence: star burst, main sequence, and quenched galaxies.
\cite{Hahn2019} identified the MS using a flexible data-driven approach termed Gaussian mixture modelling.
All of these methods consider all galaxies while determining the MS, without making any selection on galaxy samples.
Our method is fairly similar to theirs.
We begin by identifying the ridge line using a single Gaussian function and then fitting it with a linear function obtain the MS.
As with previous works, this method is less affected by the sample's range.
Therefore, we use all galaxies, except those with $SFR=0$, to define the MS.

\subsubsection{A summary of the MS part }
\label{sec:ms3}

The MS part of the SFH can be
modelled using the fitting formula for the main sequence of the $\Psi-m$ distribution.
For instance,
we can build the evolution of main sequence
by combining \Eqn{EquMSttt},\ref{EquMSill} with \Eqn{EquParattt}, \ref{EquParaill}, \ref{EquParatng},
to obtain:
\begin{equation}
	\Psi_{\rm MS}(t_L,m)=k(t_L)m(t_L)+\Psi_0(t_L)
	\label{EquMSpart}
\end{equation}
$k(t_L)$ and $\Psi_0(t_L)$ can be obtained from \Tbl{TabMSPara}.

One might construct a stellar mass growth history for galaxies as well as the SFR history, assuming that galaxies
evolve exactly following the modelled MS.
In principle, if the growth of a galaxy's stellar mass can be represented as a function of time, the SFR history $\Psi_{\rm MS}(t_L)$ can be modelled via \Eqn{EquMSpart}.
For an {\it{in situ}} growth of stellar mass, the $m(t_L)$ is an integration to $\Psi(t_L)$. By applying the time dependent regression of slope $k$ and intercept $\Psi_0$ to the function for {\it in situ} mass growth, we can obtain the following results:
\begin{equation}
	\frac{dM_*(t_L)}{dt_L}= -10^{\Psi_0}M_*^k
	\label{EquMt}
\end{equation}
\Eqn{EquMt} describes a SFH model with mass growth following the MS without any perturbation.
We refer to it as the "MS model" here after.
Solving \Eqn{EquMt} is difficult.
We give a short description of the analytical solution to it in appendix \ref{ap1}.
In practice, we try to solve it in a numerical way.
We will discuss its performance in \Sec{sec:fullsfh}.

\subsection{modelling the SFH variation}
\label{sec:var}
\subsubsection{modelling method and result}

\begin{table*}
	\centering
	\caption{A summary of the parameters of the modelled variation history  $\Delta(t_L)=\alpha t_L+\beta+\mathscr{A}\times B_H(t_L), \mathscr{A}\sim\mathscr{N}(\mu_A,\sigma_A)$ }
	\label{TabModPara2}
	\begin{tabular}{c|ccccc}
		\toprule
		                &
		$\alpha$        &
		$\beta$         &
		$\mu_A$         &
		$\sigma_A$      &
		$H$               \\
		\hline
		The300          &
		$0.034\pm0.012$ &
		$ -0.23\pm0.08$ &
		$0.54\pm0.10$   &
		$0.22\pm0.08$   &
		$0.20\pm0.05$     \\
		Illustris-1     &
		$0.061\pm0.012$ &
		$ -0.21\pm0.10$ &
		$0.44\pm0.09$   &
		$0.27\pm0.05$   &
		$0.052\pm0.040$   \\
		TNG100          &
		$0.059\pm0.014$ &
		$ -0.22\pm0.11$ &
		$0.41\pm0.10$   &
		$0.24\pm0.04$   &
		$0.070\pm0.065$   \\

		\bottomrule
	\end{tabular}
\end{table*}

Apart from the evolution along the main sequence,
our model has to reproduce the observed scatter in the $M_*$ - SFR diagram, by accounting for the variation in the SFH of individual galaxies.
We quantify the
variation as an offset of a galaxy's position in the $M_*$ - SFR diagram relative to the main sequence:
\begin{equation}
	\Delta(t_L) = \Psi_{\rm gal}(t_L)-\Psi_{\rm MS}(t_L,m(t_L))
\end{equation}

These offsets are quite likely to occur in a stochastic process.
According to previous studies \citep{Kelson2014,Caplar2019},
fractional Brownian motion (fBm) can describe the pattern followed by individual galaxies.
For a standard Brownian motion $B(t)$, the increments $B(t)-B(s)$ are  stationary and independent and follow the normal distribution $\mathscr{N}(0,\sigma^2|t-s|)$.
Fractional Brownian motion is a Brownian motion with increments  weighted
by the kernel $(t-s)^{H-1/2}$\citep{Mandelbrot1968}.
The parameter $H$, satisfying $0<H<1$, shows the self-similarity property of a stochastic process.
When $H=0.5$, the fBm becomes a standard Brownian motion.
When $H<0.5$, a given step is more likely to be followed by a reversed step; that is, if $B_H(t+1)-B_H(t)$ is deviates from the mean, the subsequent step $B_H(t+2)-B_H(t+1) $ will attempt to revert to the mean.
When $H>0.5$, the stochastic process exhibits a long-term trend.

We proposed a model based on a stationary fBm with a small inclination.
This model is applied to the simulation data to find out whether it is true. The model is described by the following equation:
\begin{equation}
	\begin{aligned}
		\Delta(t_L) = \alpha t_L+\beta+ & \mathscr{A}\times {\rm B}_{\rm H}(t_L),       \\
		                                & \mathscr{A} \sim \mathscr{N}(\mu_A, \sigma_A)
		\label{EquModel}
	\end{aligned}
\end{equation}

The formal part $\alpha t_L+\beta$ part describes the overall trend of $\Delta$. $\alpha$ is the average slope and $\beta$ is the average intercept.
With this item, our model can match the $\Delta(t_L)$ regardless of whether this process is stationary  ($\alpha=0$) or non-stationary ($\alpha \neq 0$)

The latter part of the equation, $\mathscr{A}\times {\rm B}_{\rm H}(t_L)$, is scaled fractional Brownian motion.
The fBm $B_H(t_L)$ is generated in Python using the {\it fbm}module\footnote{\url{https://pypi.org/project/fbm/}}.
To begin, we build a fBm series with $Np=400$ points for each realization.
It will produce a series $B_H$ subject to the constraint $B_H(m)-B_H(n) \sim \mathscr{N}(0, (\frac{|m-n|}{400})^{2H})$, where $m$ and $n$ are integers between $0$ to $400$, respectively, and $H$ is the Hurst parameter.
We
then select the points from index $200$ to $200+[T/\delta t]$ as the series we want.
The start point is arbitrarily chosen to $200$.
In this case, the initial fluctuation in modelled SFH is a normal distribution rather than $0$.
$T$ denotes the overall duration of a galaxy's SFH in unit of Gyr.
$\delta t$ is the time interval.
We set $\delta t$ to be $0.135 Gyr$, so that the number of data points of
our modelled variation history is close to that of SFHs with the same age from simulations.
Be aware that the time interval does not affect the majority of the properties of fBm series.
For example, when the H parameter is the same, the fBm series with $Np$ points and $\delta t$ time interval is equivalent to the fBm series with $2Np$ and $ 0.5\delta t$ at a time scales of $\tau > \delta t$,
while the latter offers additional information at time scales of $0.5\delta t<\tau<\delta t$.

Given that the variation history $\Delta(t)$ of each individual galaxy may have a different amplitude,
we rescale the fractional Brownian motion for each individual galaxy history using a random number $\mathscr{A}$.
$\mathscr{A}$ obeys a normal distribution with a mean of $\mu_A$ and a variance of $\sigma_A$.
$\mu_A$ and $\sigma_A$ are free parameters.

In summary, this model comprises five free parameters: $\alpha$, $\beta$, $\mu_A$, $\sigma_A$ and $H$.
We generate the best fitting models for each simulation by tweaking these parameters.
In practice, we create modelled time series $\Delta_{\rm model}(t_L)$, i.e., the variation of SFH, for each individual galaxies using a set of $\alpha$, $\beta$, $\mu_A$, $\sigma_A$ and $H$.
The numbers of $\Delta_{\rm model}(t_L)$ series are the same as the number of sampled SFHs from their corresponding simulations.
The length (age) of $\Delta_{\rm model}(t_L)$ also follows the same distribution of length of corresponding simulated SFHs.
We first apply an initial estimation of free parameters to the model, and then derive some statistics of the modelled variation histories.
The same statistics are also applied to variation histories from simulations.
By comparing those statistics,
we tweak the free parameters.
We make use of the statistics of the following features:

\begin{itemize}
	%[i)]
	\item[i)] the distribution of average variation $\overline{\Delta}$;
	\item[ii)] the distribution of root square mean variation $\overline{\Delta^2}$;
	\item[iii)] the distribution of star burst time $t_b$;
	\item[iv)] the distribution of star burst duration $\tau_b$;
	\item[v)] the distribution of quenching time $t_q$;
	\item[vi)] the distribution of quenched duration $\tau_b$.
\end{itemize}

The former two features, $\overline{\Delta}$ and $\overline{\Delta^2}$, qualify the amplitudes of $\Delta_(t_L)$.
The "0SFR" points introduce a large bias in averaging the amplitudes, and are thus excluded from the SFHs when calculating $\overline{\Delta}$ and $\overline{\Delta^2}$.

The latter four features, $\tau_b$, $\tau_q$, $t_b$ and $t_q$, are related to the star burst events and quenching processes of the SFH.
In this work, we define the galaxies located above
$2\sigma$\footnote{$2\sigma$ is about $0.65 dex$ in all three simulations.} from the main sequence as being in a ``star burst stage'', and galaxies located below $2\sigma$ from the main sequence as being in a ``quenched stage''.
$t_b$ (or $t_q$) is the time point when a galaxy enters the star burst (or quenched) stage.
Multiple star burst or quenching times can exist within a single SFH.
$\tau_b$ (or $\tau_q$) is the cumulative amount of time a galaxy spends in the star burst (or quenched) stage.
These four features pertain solely to the timing in SFHs.
"0SFR" data points are not removed when calculating the $t_q$ and $\tau_q$.
Because, while their SFR values are imprecise, their timing values are regarded to be correct and physically meaningful in characterizing the variation histories.

\begin{figure}
	\centering
	\includegraphics[width=0.9\linewidth]{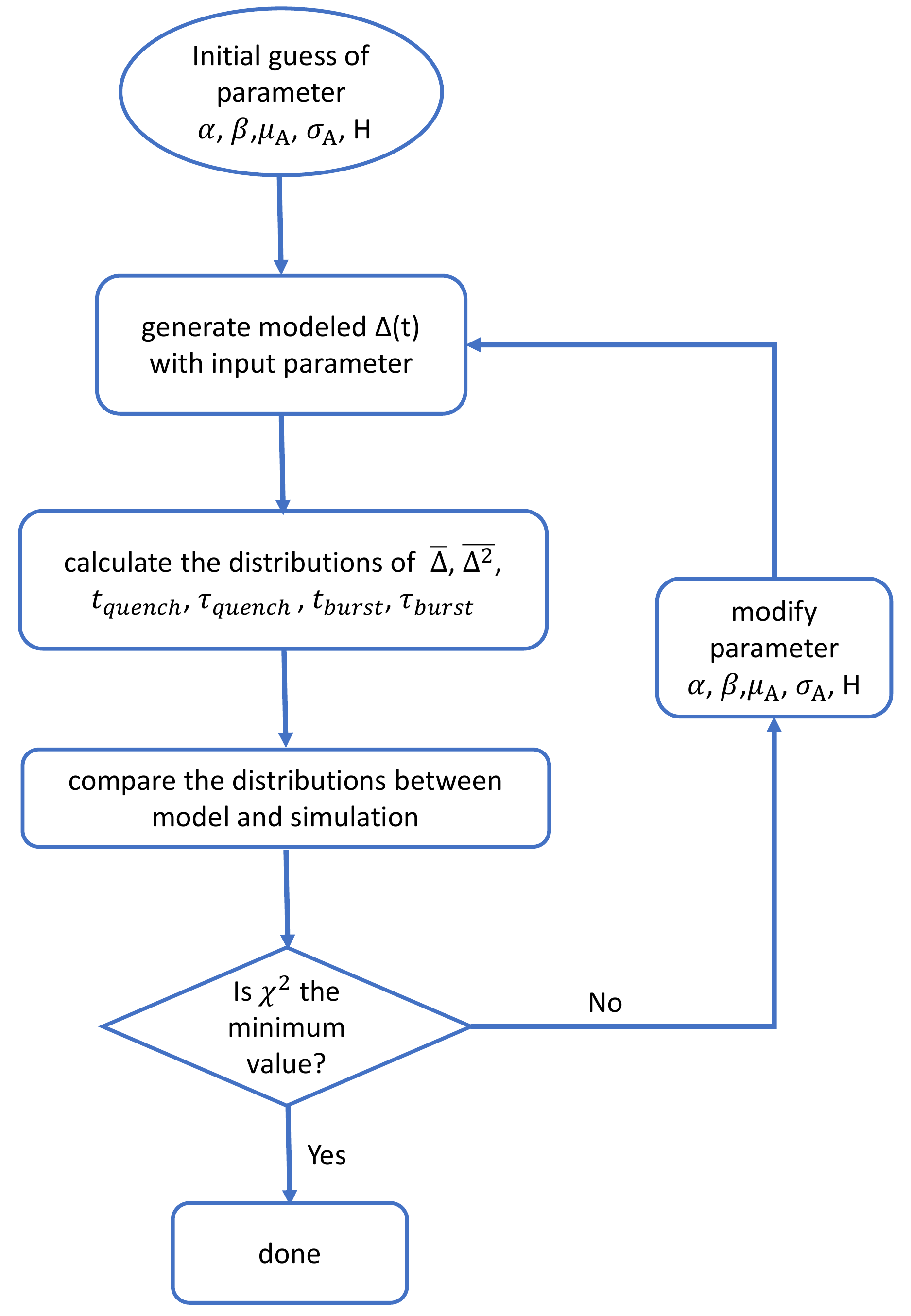}
	\caption{The flow chart of how to build up the modelled variation histories.}
	\label{FigFlow}
\end{figure}

We generate and compare the distributions of each feature using both models and simulations.
To calibrate the comparison,
we utilize the sum of mean squared differences:
\begin{equation}
	\begin{aligned}
		\chi^2=\overline{(P(\overline{\Delta_{\rm model}})-P(\overline{\Delta_{\rm sim}}))^2}+\overline{(P(\overline{\Delta_{\rm model}^2})-P(\overline{\Delta_{\rm sim}^2}))^2} \\
		+\overline{(P({t_{\rm b,model}})-P({t_{\rm b,sim}}))^2}
		+\overline{(P({t_{\rm q,model}})-P({t_{\rm q,sim}}))^2}                                                                                                                  \\
		+\overline{(P({\tau_{\rm b,model}})-P({\tau_{\rm b,sim}}))^2}
		+\overline{(P({\tau_{\rm q,model}})-P({\tau_{\rm q,sim}}))^2}
	\end{aligned}
\end{equation}
We can
finely tune the input parameters, by adjusting them iteratively and recomputing the $\chi^2$ until it reaches a minimal value.
\Fig{FigFlow} illustrates the process mentioned above graphically.
Due to the random nature of the process used to generate histories, the best fitting parameters are not exactly the same in each time of fitting.
Therefore, we perform fitting for $50$ times and get the mean and variance of the fitting parameters.
The best fitting parameters are listed in \Tbl{TabModPara2}.

\begin{figure*}
	\includegraphics[width=0.9\linewidth]{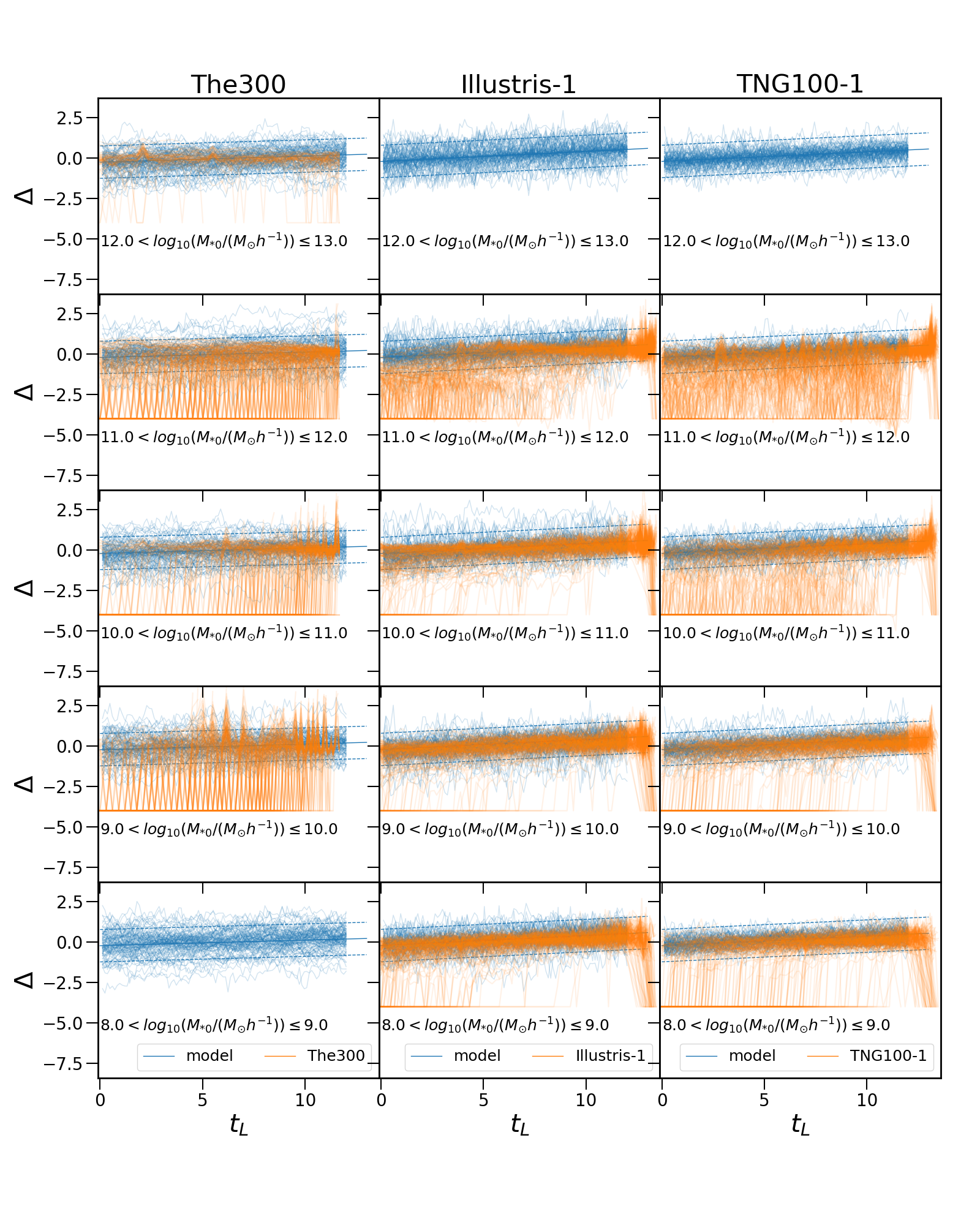}
	\caption{
		The variation history $\Delta$ of single galaxies  in different stellar mass bins as function of lookback time $t_L$. Each subplot depicts the $\Delta(t_L)$ of $100$ simulated galaxies (orange lines) and $100$ modelled galaxies (blue lines).
		The blue dashed  lines show the variances in $1 dex$ from the median $\Delta(t_L)$ value in simulations.
		The galaxies are divided into 5 bins according to their stellar mass at redshift $0$, as indicated in the lower left corner of each plot.
		Each row represents the results in one stellar mass bin.
		Each column shows the results from one simulation as described by the title.
		The bottom left-hand, top middle, and top right-hand panels do not contain galaxies samples from simulations due to the mass limit.
		To illustrate the whole curves of each SFH, the "0SFR" points are moved to the positions of $\Delta=-4$.
	}
	\label{FigTra2}
\end{figure*}

Each of the three models has positive average slope $\alpha$ and negative average intercept $\beta$.
This suggests that, on average, the trajectories of SFHs in these simulations tend to travel
from above the main sequence to below the main sequence, which is in agreements with earlier findings \citep[see][]{Iyer2020,Matthee2019}.
The average slopes are
of \ill\ and \tng\ are greater than those of \ttt.
This implies that, on average, the SFHs in \ttt\ are more likely to be parallel to the main sequence.

The parameters concerning the amplitudes of variations, $\mu_A$ and $\sigma_A$,
are relatively similar in three simulations.
Only in \ttt\ is the $\mu_A$ slightly larger than in \ill\ and \tng, reflecting a more varied SFH there.

The $H$ parameter in three simulations are much smaller than $0.5$.
This implies that the variations tends to converge around $0$.
In other words, the SFHs in simulations tend to follow the main sequence.
The $H$ parameter in \ttt\ ($0.20$) is significantly larger than those in \ill\ ($0.052$) and \tng\ ($0.070$).
This indicates that the trends toward returning to the main sequence are significantly stronger in \ill\ and \tng.
Readers may note that \cite{Kelson2014} proposed a Hurst parameter of $0.9$ for his SFH model, which looks quite different from our models.
We emphasize that the small value of $H$ here is solely for the variation history.
In \cite{Kelson2014}, he chose the value of  $H$ for SFHs, i.e., the MS part + variation part in this work.
The entire SFH exhibits very significant long-term trends, which results in a larger $H$.
We can also obtain a value of $H\sim 0.7$
by measuring the Hurst parameters of SFHs in three simulations.
It is difficult to tell which value is closer to the truth,
since both simulations and theories in \cite{Kelson2014} are capable of reproducing realistic galaxy populations.
The discussion of this distinction requires additional investigation, which is beyond the purpose of this work.

\Fig{FigTra2} gives an overview on how well the variation histories in the models converge with those in simulations.
The samples are separated into different bins according to their stellar mass at $z=0$.
In each stellar mass bin, we randomly select $100$ variation histories from simulations and $100$ from corresponding models.
These variation histories are plotted together in each sub panels of \Fig{FigTra2} for comparison.
Note that our current models are independent of the stellar mass.
Without considering the "0SFR" points, the modelled variations histories are well in agreement with those from simulations.

\begin{figure*}
	\includegraphics[width=1\linewidth]{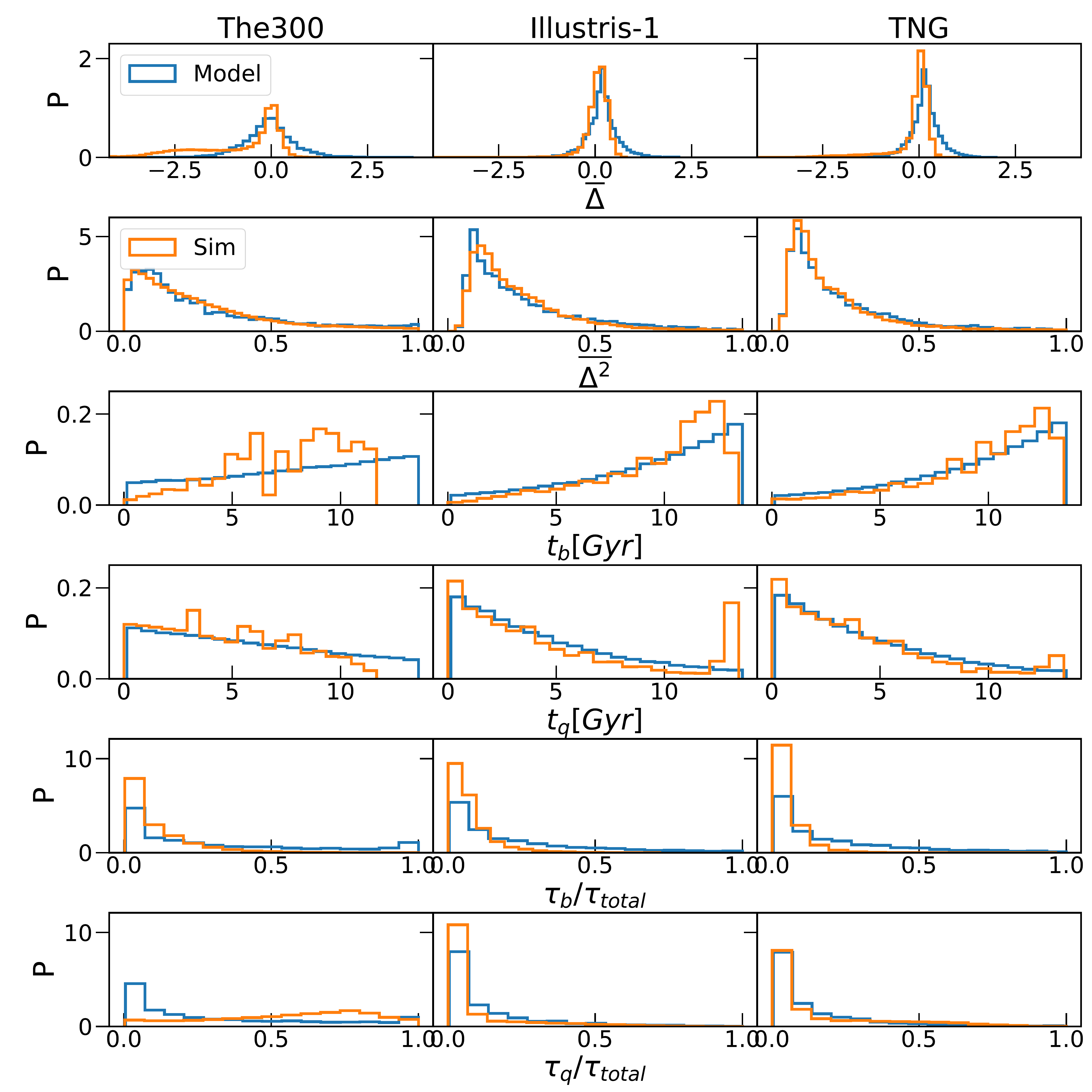}
	\caption{
		Histograms of mean $\Delta$ (1st row), mean $\Delta^2$(2nd row), time of entering star burst (3rd row), time of quenching (4th row),  duration of galaxy star burst stage (5th row) and duration of quenched stage (last row).
		Each column represents the statistics from one simulation,
		\ttt, \ill\ and \tng\ from left to right respectively.
		The orange lines show the distribution of simulated galaxies, while the blue lines represent their corresponding models.
	}
	\label{FigBQ}
\end{figure*}

However, the variation histories in  simulations show some appreciable dependence on the stellar mass.
For galaxies with a stellar mass more than
$10^{12}M_{\rm \odot}h^{-1}$ and between  $10^{10}M_{\rm \odot}h^{-1}$ and  $10^{11}M_{\rm \odot}h^{-1}$ in \ttt, their variation histories are more concentrated to $0$, which makes them less similar to the model.
In \ill\ and \tng, the simulated and modelled variation histories show a higher degree of agreement
and merely no mass dependency.
Variation histories of galaxies above $10^{11}M_{\rm \odot}h^{-1}$ in \ill\ have shallower slopes of overall trends,
which is different from the model.
As illustrated in \Fig{FigTra2}, the models capture the variations within $1 dex$.
When a galaxy's SFR falls below $1 dex$ below MS, e.g., when it enters a quenching stage, the random walk model can no longer predict its
trajectory.

\begin{figure*}
	\includegraphics[width=0.33\linewidth]{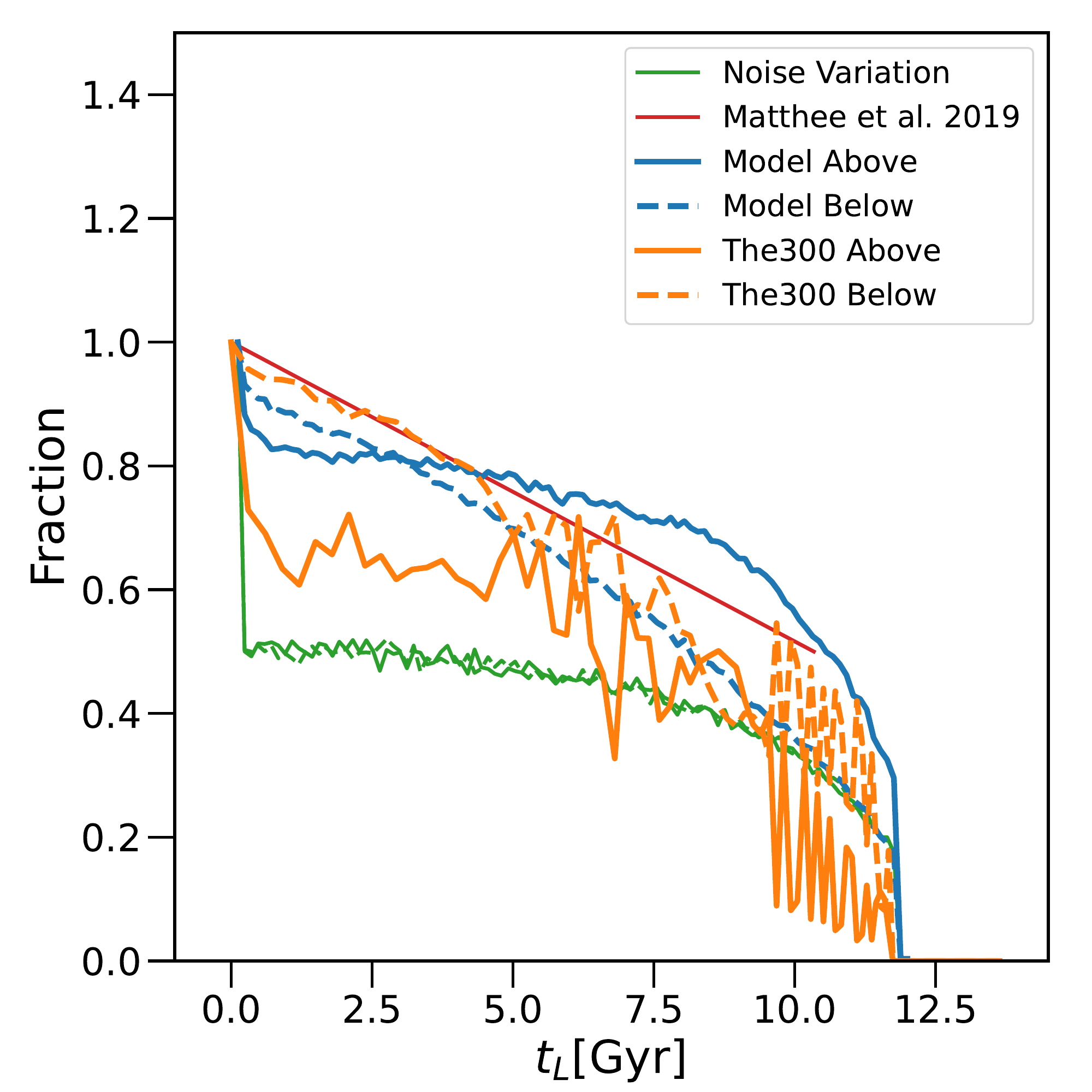}
	\includegraphics[width=0.33\linewidth]{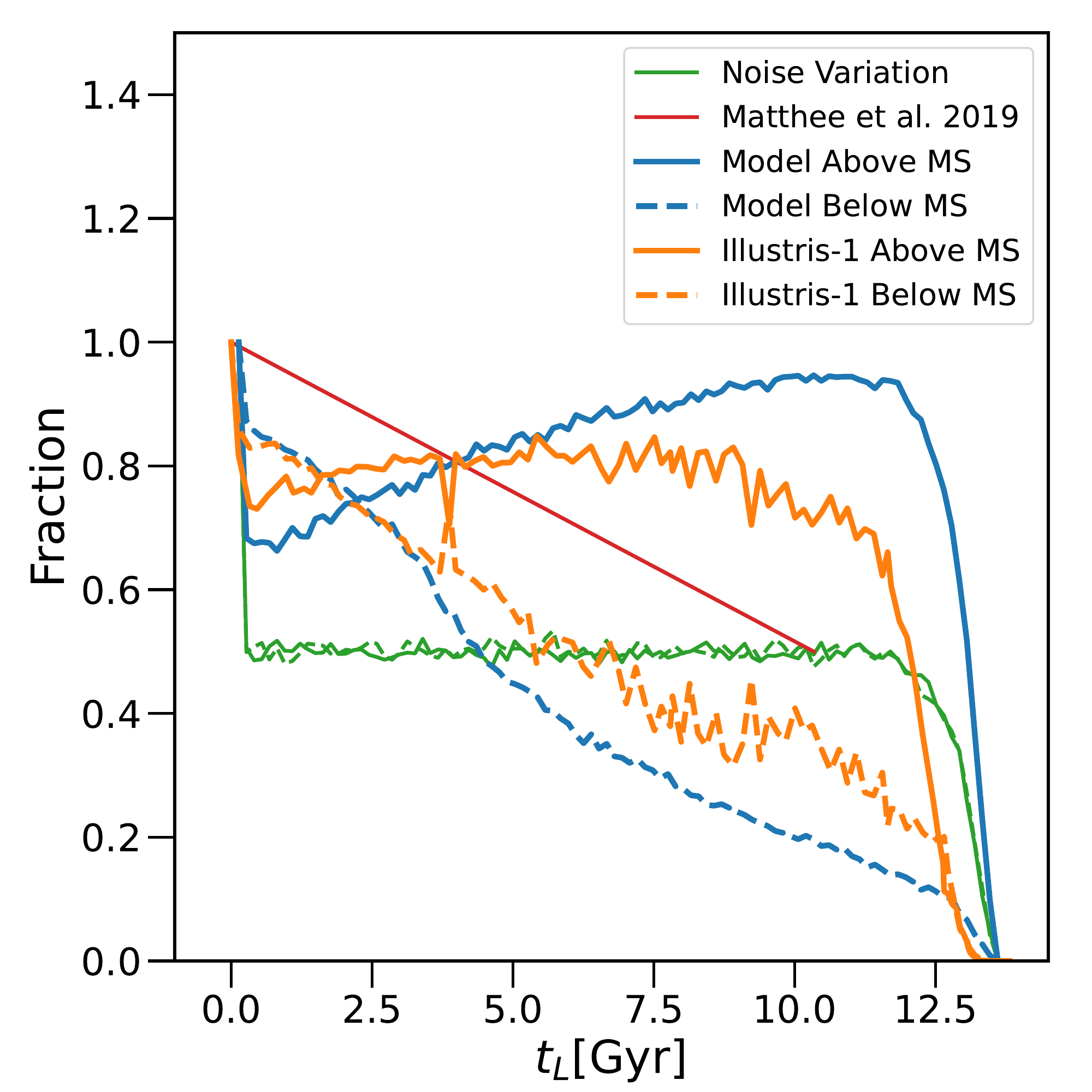}
	\includegraphics[width=0.33\linewidth]{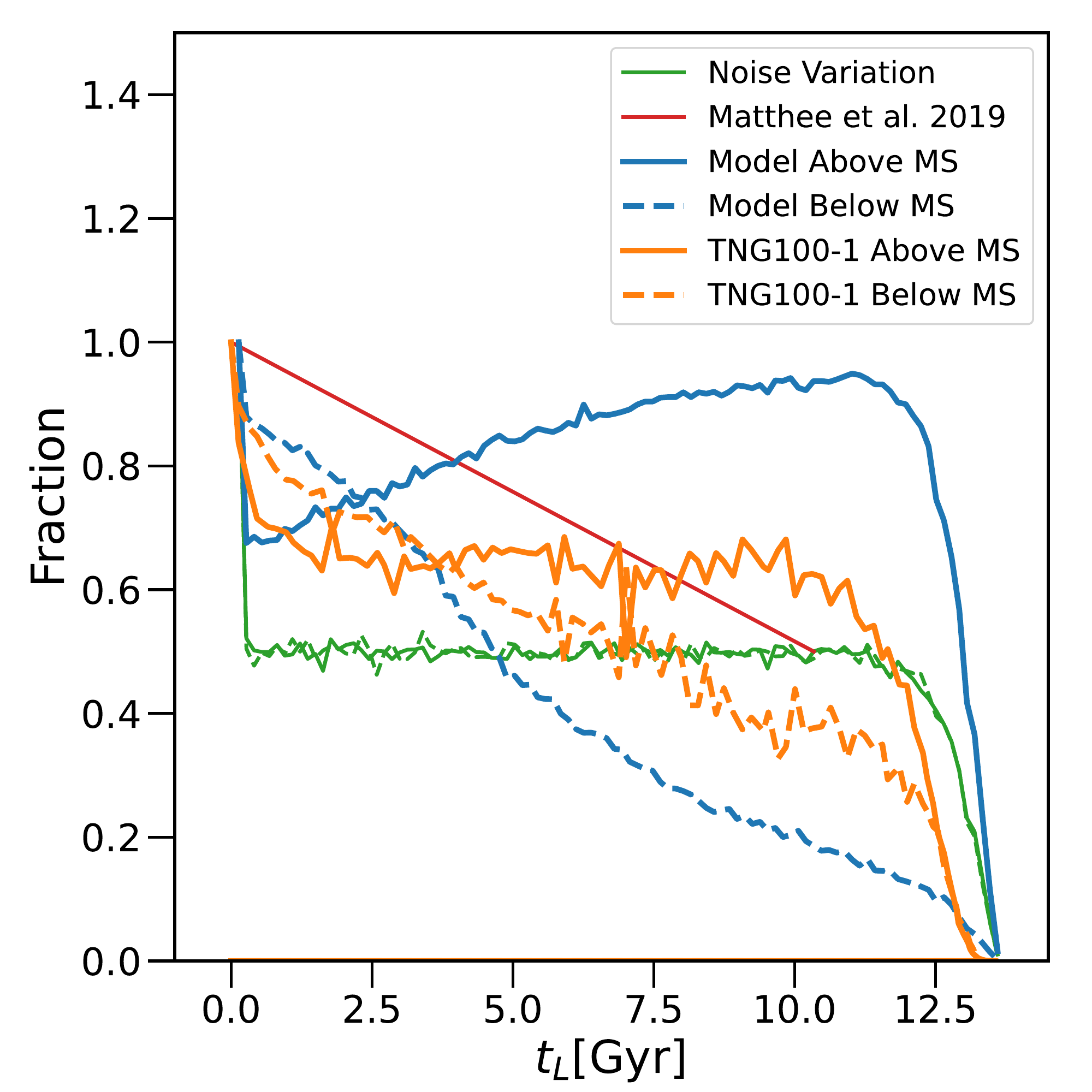}
	\caption{
		The fraction of galaxies located above and below the main sequence in reference to the galaxy sample at $z=0$.
		Solid lines represent the change of the fractions of galaxies above the main sequence, whereas dashed lines represent those below the main sequence.
		Orange lines show the simulated history curves, whereas blue lines show the results from corresponding models.
		Green lines show the evolution of galaxy fractions for variation histories made of white noise (random fluctuation).
		Red line represents the outcome of \protect\cite{Matthee2019}.
		For reference, \protect\cite{Matthee2019} used galaxies with $z=0.1$. We shift their  result to start at $z=0$ for a fair comparison with this work.
		From left to right, the panels show the results from \ttt, \ill\ and \tng.
	}
	\label{FigFracHist}
\end{figure*}

\subsubsection{The goodness of modelling}

As mentioned above, we use the distributions of six parameters to constrain our variation history models.
Prior evaluating our model's performance,
we have to demonstrate how well these parameters are matched.
\Fig{FigBQ} show the distributions of six parameters corresponding to our best-fit models for three simulations.

In particular, top two rows of \Fig{FigBQ} show that the distributions of $\overline{\Delta}$ and $\overline{\Delta^2}$ in our models are closely matched with simulations.
The primary divergence
is the presence of tails at negative end in the distribution of $\overline{\Delta}$ from simulations(especially \ttt\ and \tng), which can not be reproduced by our models.
Indeed, the Brownian motion patterns
yield a Normal distribution of $\overline{\Delta}$ by definition.
The existence of quenched stage in simulations that can not be replicated using Brownian motion should take responsibility to this tail.
On the other hand, the distributions of $\overline{\Delta^2}$ of modelled histories nicely
reproduce the distributions from simulations  with rather good accuracy.

The histograms of $t_b$, $t_q$, $\tau_b$, and $\tau_ q$ are shown in the third to sixth rows of \Fig{FigBQ}.
The $t_b$ distributions from all simulations (orange lines) peak at larger lookback times, indicating that star burst occurs at early epochs, whereas the $t_q$ distributions peak at lower lookback times, indicating that quenching occurs at late epochs.
In the majority of cases,
our model accurately captures these characteristics,
with some larger deviations for \ttt.
In the left two columns, we plot the duration of the star
burst and quenched stages with respect to the total life time of galaxies.
As we see, all distributions peak at quite low ratios ($\sim0.1$), meaning that both star burst and quenched phases last for less than $10\%$  of the galaxy's life time.
Unlike the $t_b$ and $t_q$ distribution, the $\tau_b$ and $\tau_q$ distributions from our model show some deviations. In particular, they  predict slightly longer  star burst
phases.
Once again, the largest discrepancies from our model predictions are found for the \ttt\ simulation, which may warrant further discussions.

The \ttt\ simulation shows that there is an excess of longer quenched phases ($\tau_q/\tau_{\rm total}>0.5$ Gyr, see second column).
We remark that this could be the cause of the discrepancies found in the MS scatter discussed above in histograms of $\overline{\Delta}$ and $\overline{\Delta^2}$.
The SFHs with long durations of quenched stages populate the very negative part of the $\overline{\Delta}$ distribution.
One possibility to recover this behavior would be patch extra quenching process into our Brownian motion model.
However, this would likely require some fine tuning that is beyond the scope of this paper and will be addressed in forthcoming works.
Hence, we did not attempt to reproduce the  $\tau_q$ in \ttt, as our models were still capable of accurately representing the distribution of $t_b$, $t_q$, $\tau_b$, $\tau_q$ in all simulations.

\subsubsection{The position relative to the MS}
\label{sec:postoMS}

The first test on our model follows the approach presented in \cite{Matthee2019}.
They examine the trend of median fluctuations of SFR
by selecting sub-sets of galaxies that are above the main sequence at $z=0.1$ and measuring the fraction of these galaxies that remain above the main sequence at other cosmic times.
They show (in their Figure 4) that the fraction of galaxies located above the main sequence drops linearly from $100\%$ at $t_{univ}=12Gyr$ to $50\%$ at $t_{univ}=3Gyr$.
Based on this evidence, they assert that
current galaxy SFRs retain memory of the
past star formation history.

We select subgroups of galaxies above (or below) the MS at $z=0$ and trace their progenitors to find out what fraction of them remains above (or below) the MS line.
Our results are shown in \Fig{FigFracHist}.
We calculate both the evolution of the fraction of galaxies above (solid lines) and below (dashed lines) the main sequence.
The data from \cite{Matthee2019} are added  to \Fig{FigFracHist} as red lines for reference.
Additionally, the results from a variation history model constructed using white noise (i.e., random fluctuation) are shown in this figure as green line.
White noise means that the fluctuation has no memories of its former existence.
So their fractions of galaxies above or below the MS remain constant of $50\%$ at all other times except the time as reference.
The fractions drop to $0$ at earlier time.
Because the life time of SFHs is not infinity.
Tracing on their progenitors will come to a stop at some time, resulting in the demise of fraction.
Because the SFHs in \ttt\ have shorter life time than other twos,
we observe that their curves begin to decline at lower $t_L$.

From $t_L=0$ to earlier epochs, the fraction of galaxies below the MS decreases rapidly,
while the fraction of galaxies above the MS has a sharp decline followed by a mild bend.
Our model can well reproduce the curves for the fraction of galaxies below the MS, but there are obvious inconsistencies for the fraction above the MS.
In our models, a galaxy located above the MS is more likely to maintain its position compare than in simulations.
There are two possible explanations.
If some galaxies above the MS have already experienced quenched stages at earlier time, the fraction of galaxies above MS will drop more quickly.
On the other hand, if the variation histories contain large proportion of noisy-like fluctuations, as indicated by the green lines, the fractions will drop to $0.5$ quickly.
We hypothesize that
while there are more galaxies in the quenched stages (including "0SFR" points) in \ttt,
the noisy-like fluctuation appears to be more prominent in \ill\ and \tng.
This assumption will be proved by other outcomes in following sections.

The curve in \cite{Matthee2019} is different from the curves in the simulations we examine.
We confirm that this distinction is due to the incline of stochastic process.
Our variation history is specified as non-stationary stochastic process with an inclination $\alpha t_L+\beta$ (see \Eqn{EquModel}).
When $\alpha=0$ and $\beta=0$, the fractions above and below the MS
behave identically to the curve described in \cite{Matthee2019}.

\subsubsection{PSD of variation history}

\begin{figure*}
	\includegraphics[width=0.33\linewidth]{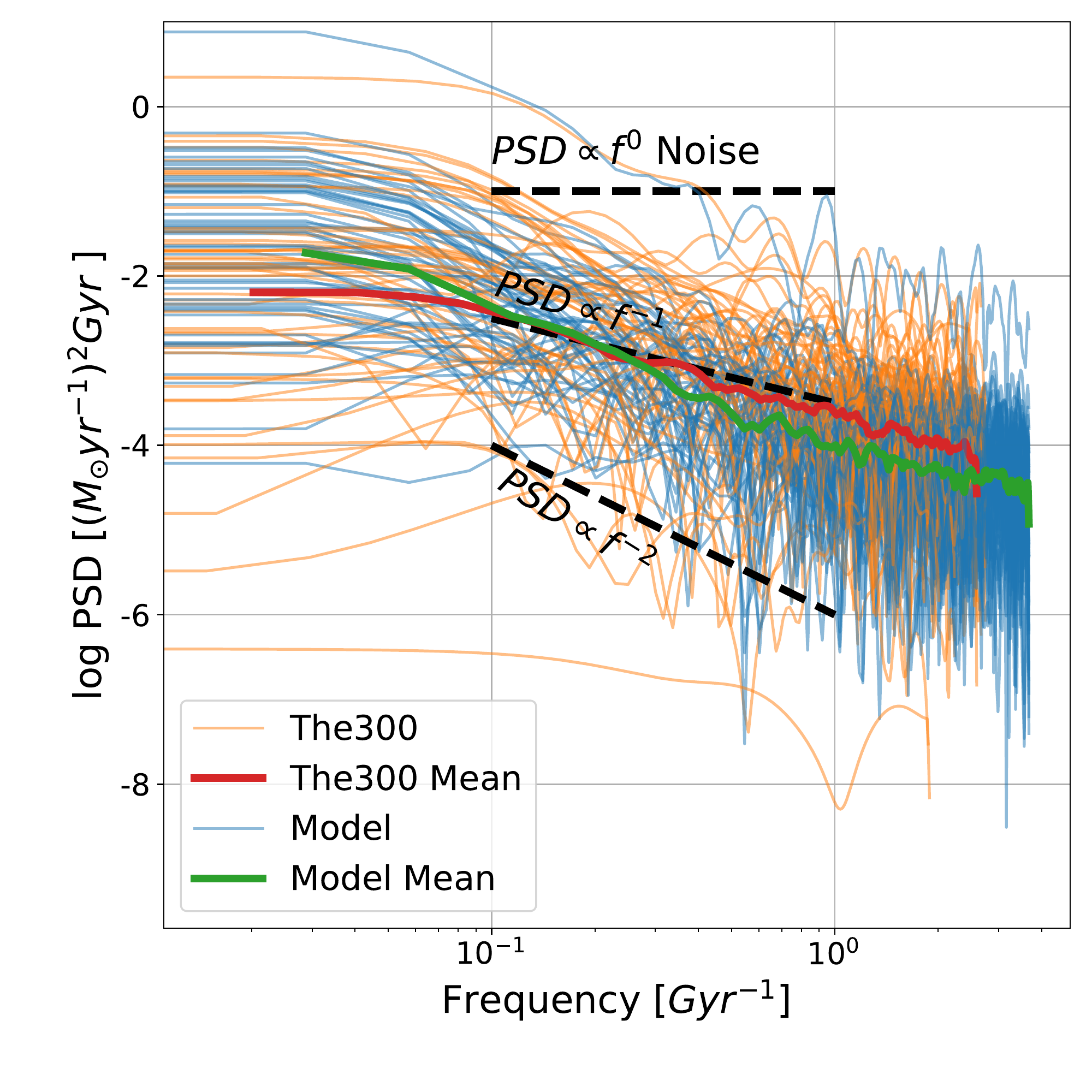}
	\includegraphics[width=0.33\linewidth]{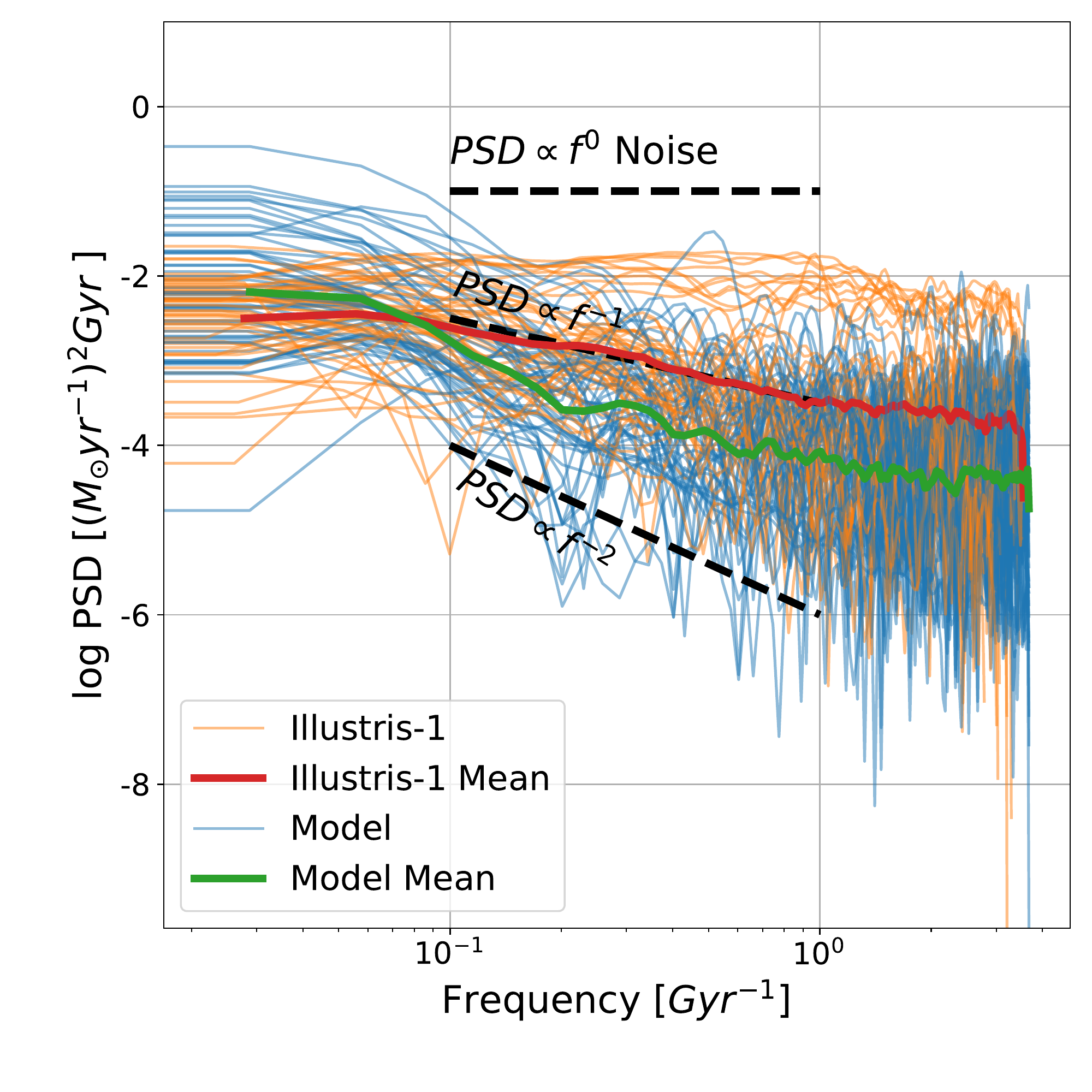}
	\includegraphics[width=0.33\linewidth]{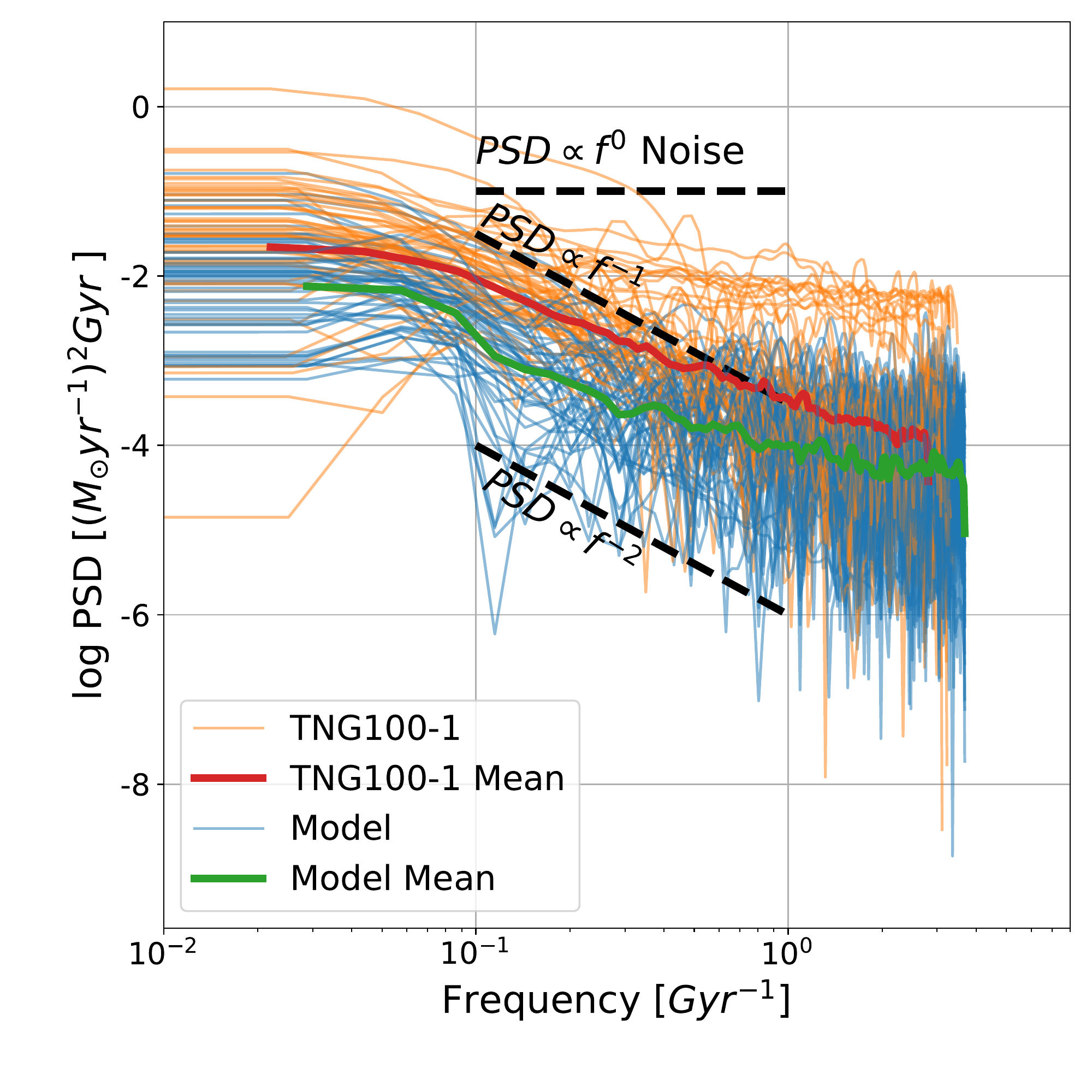}
	\caption{
		The power spectrum density of $\Delta(t_L)$.
		PSDs of variation histories of $50$  randomly chosen galaxies from simulations are shown in orange lines.
		$50$ PSDs of corresponding modelled variation histories are plotted in blue lines.
		From left to right panels, the results are for data from the \ttt, \ill\ and \tng\ simulations, respectively.
		The mean of $50$ PSDs of models or simulations
		are plotted with green or red thick lines.
		Slopes of $-2$, $-1$ and $0$ are also plotted with black dashed lines as reference.
	}
	\label{FigPSD}
\end{figure*}

\begin{figure*}
	\includegraphics[width=0.33\linewidth]{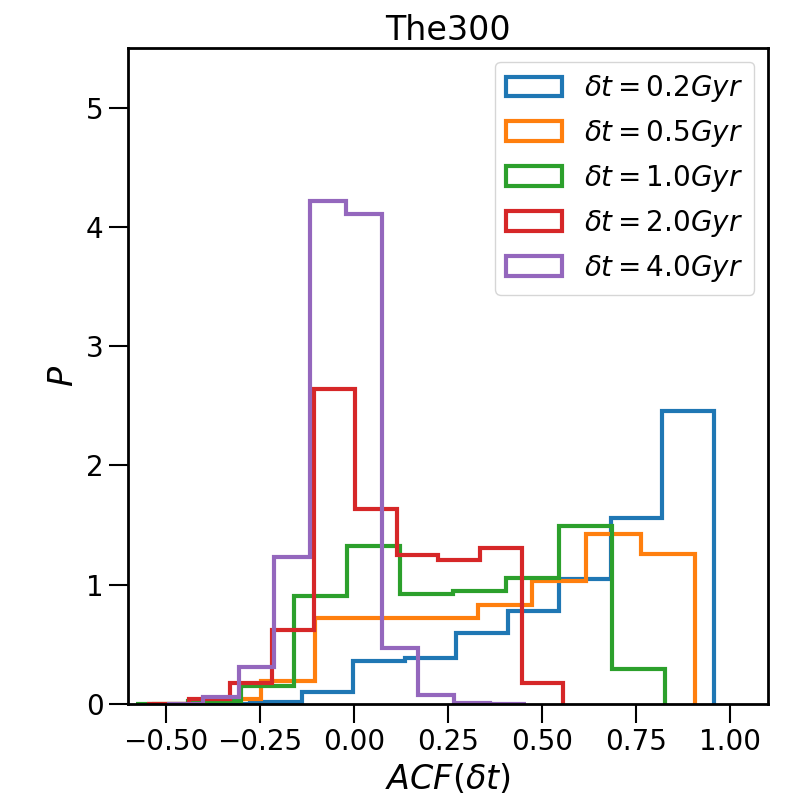}
	\includegraphics[width=0.33\linewidth]{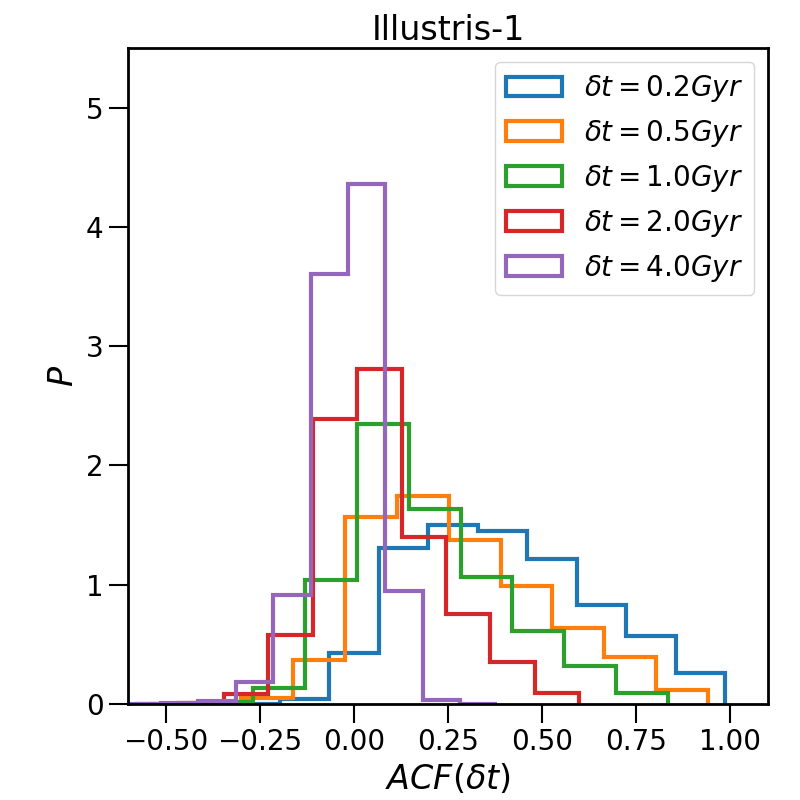}
	\includegraphics[width=0.33\linewidth]{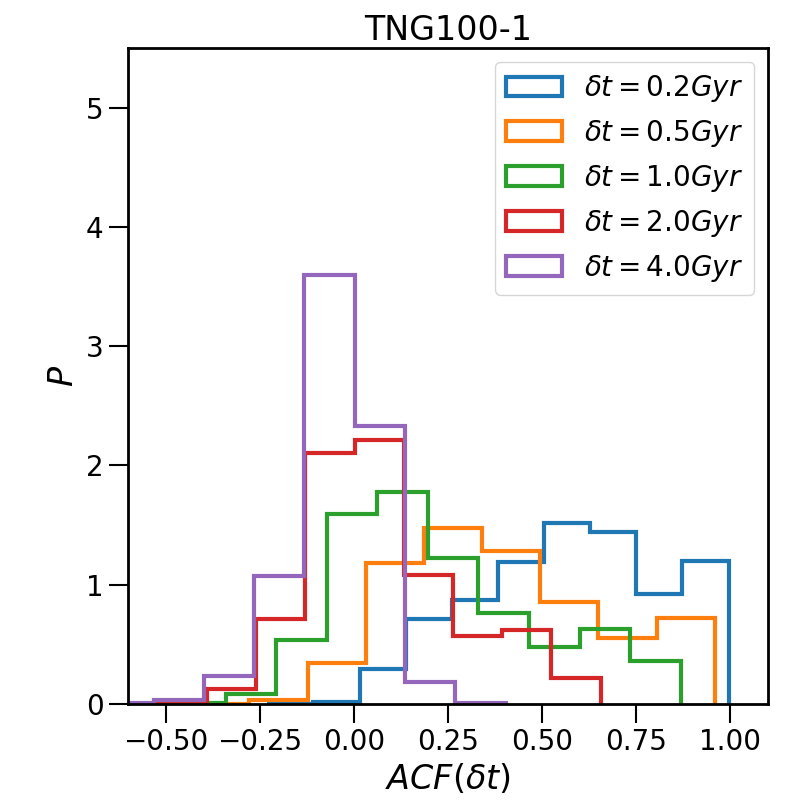}
	\includegraphics[width=0.33\linewidth]{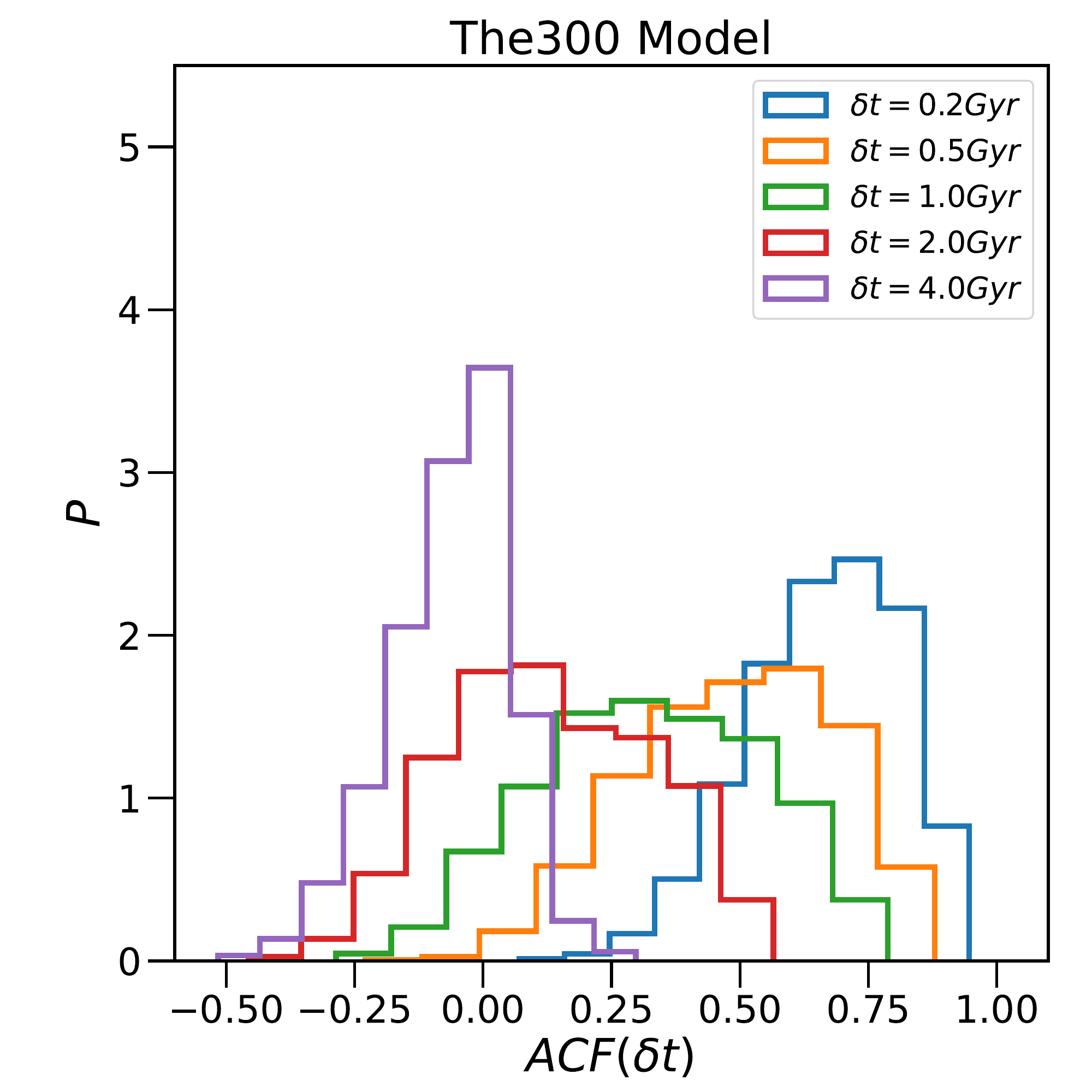}
	\includegraphics[width=0.33\linewidth]{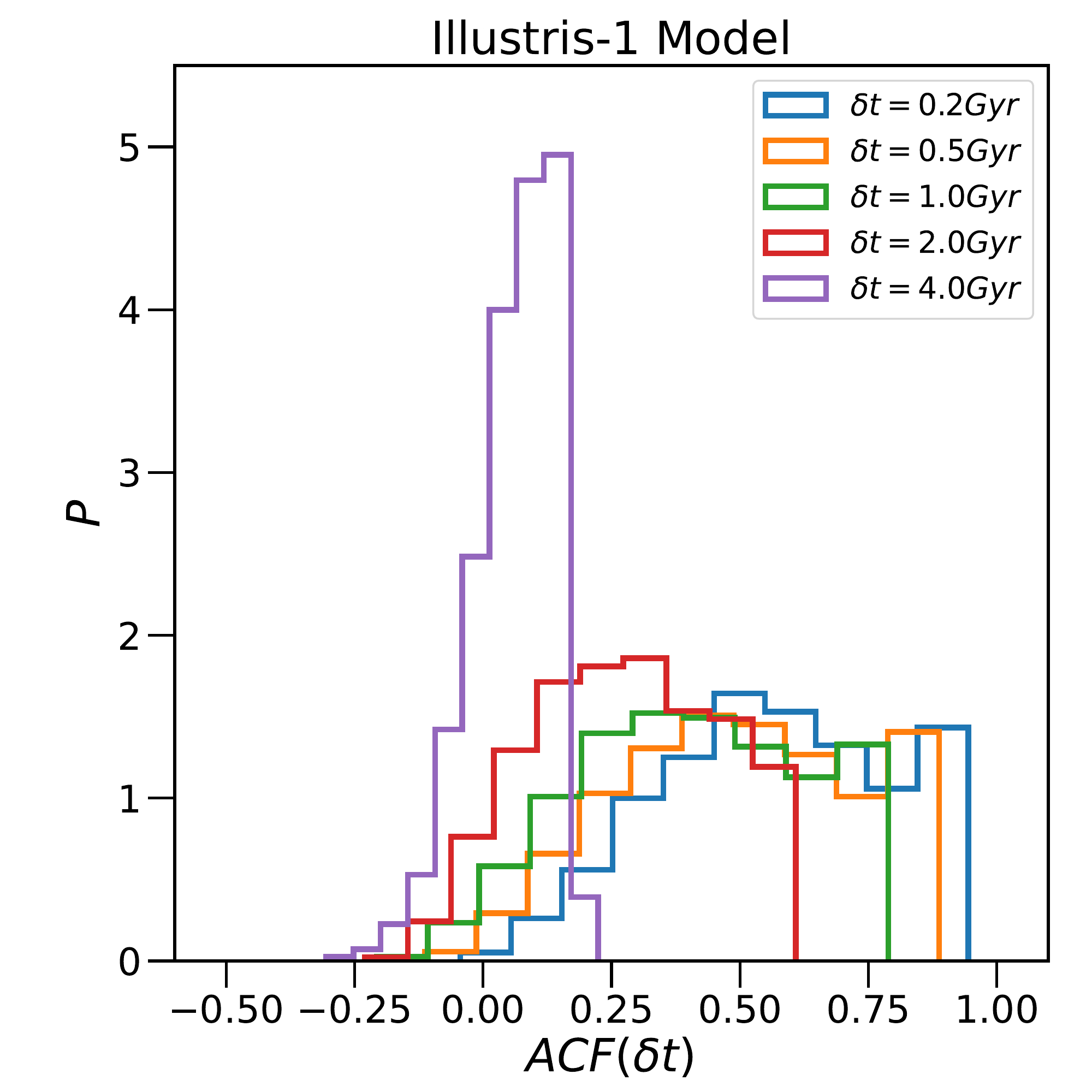}
	\includegraphics[width=0.33\linewidth]{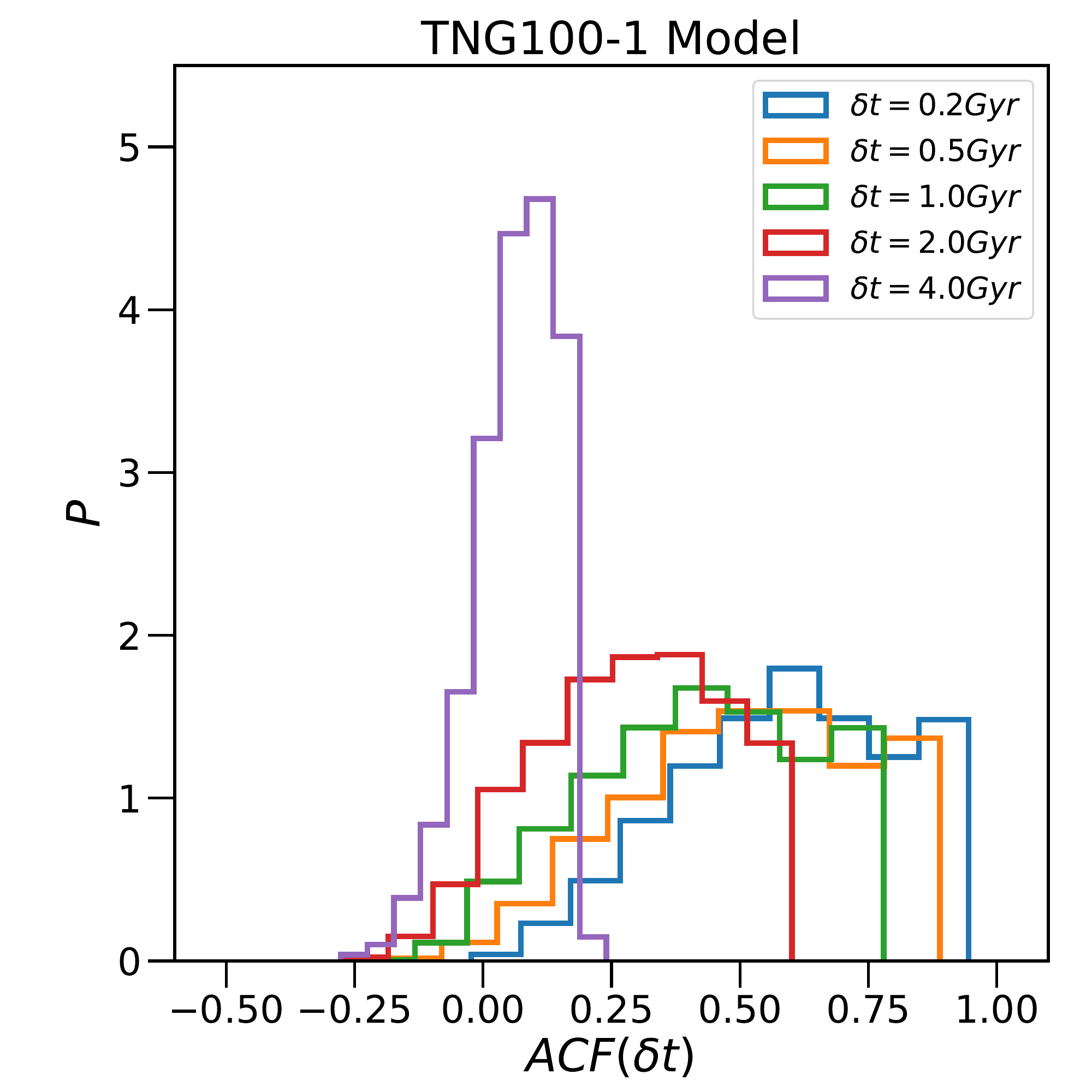}
	\caption{
		Distributions of auto-correlation coefficient of $\Delta(t_L)$ in simulations (top panels) and their corresponding models (bottom panels).
		Histograms of different colors show the ACF with different time delay, as the legends indicate.
		From left to right, each row shows the results from one simulation,  \ttt, \ill\ and \tng, respectively.}
	\label{FigACF}
\end{figure*}

The power spectrum density (PSD) enables us to quantify the importance of different frequency in any time series.

\Fig{FigPSD} compares the PSDs of $\Delta(t_L)$ obtained from simulations to those derived from our models.
We randomly select $50$ variation histories from simulations and $50$ modelled variation histories, and plot their PSDs together in each panel.
The figure shows that the PSDs of our modelled variation histories closely match variation histories from simulations.

Principally, The PSD of a fractional Brownian motion with Hurst parameter $H$ is
\begin{equation}
	PSD =\begin{cases}
		f^{-(2H+1)} & 0<H \le 0.5 \\
		f^{-2}      & 0.5<H<1     \\
	\end{cases}
	\label{EquHslope}
\end{equation}
\citep[see][]{Mandelbrot1968,Majumdar2018}.
Apart from fractional Brownian motion, our variation history model includes a linear change term $\alpha t_L+\beta$,
which strengthens the long-term component of PSD and hence makes the PSD steeper.
Thus, our models have PSDs with slopes of $< -1.4$ for \ttt, $< -1.10$ for \ill, and $< -1.14$ for \tng, respectively.
The average PSDs showed in \Fig{FigPSD} are consistent with theory.

It is worth noting that  the PSDs of variation histories in \ill\ and \tng\ also exhibit a secondary branch, in which the PSDs are nearly constant over all time scales, i.e., $PSD\propto f^0$.
Because white noise typically has the $PSD\propto f^0$, this part is referred to as {\it white noise mode}.
The white noise mode in \ill\ and \tng\ is the reason for the shallower average PSDs than those in our models.

Most theoretical works on the PSDs of SFHs favor a slope of $-2$ \citep{Tacchella2020,Iyer2020}, which is steeper than
the slopes observed in simulations and in our models.
We need to clarify that the PSDs of variations history discussed in this subsection
are shallower than PSDs of SFHs, because the latter has extra long-term evolutions than the former, adding more power to the low frequency end of the PSDs.
The average slopes of PSDs of SFHs are approximately $-2$ for \ttt,
but are still shallower for \ill\ and \tng\, at $-1.5$.
The existence of variation histories with white noise mode in \ill\ and \tng\ may explain why the slopes are shallower, as we will discuss in more detail
in subsequent sections.

The white noise mode can also explain the over prediction to the fractions of progenitors above MS which is discussed in \ref{sec:postoMS}.

\cite{Tacchella2020} suggests analytical SFH model with break power-law PSD:
\begin{equation}
	PSD(f)=\frac{C}{(\tau_x f)^{\beta_l}+(\tau_x f)^{\beta_h}}
\end{equation}
According to their model, the total PSD is contributed mainly by the inflow process, the regulation of gas flow and the star formation process related to the GMCs.
Different physical processes have PSDs with different break time scale $\tau_x$ and slope $\beta_h$ and $\beta_l$.
This should establish a connection between the slopes and break time scales of PSDs and the physics of SFHs.
Thus, the white noise mode in \ill\ and \tng\ may imply a distinct sub-physics process which is different from that in \ttt.

One thing for sure is that those variation histories in white noise mode are unaffected by the long-term perturbations caused by host halos or mergers.
However, due to the temporal resolution of SFHs, we are unable to identify  fluctuations caused by processes with time scales smaller than $\sim 1 Gyr$
This suggests that these galaxies may be driven by the inner baryonic mechanisms such as stellar feedback, galactic wind, photoionization feedback or SNe \citep[see][]{Iyer2020}.

\subsubsection{ACF of variation history}

\Fig{FigACF} shows the test for convergence of the auto-correlation function between models and simulations.
For one time series $\Delta(t)$, the auto-correlation function (ACF), defined as:
\begin{equation}
	ACF(\delta t)=\frac{Cov(\Delta(t),\Delta(t+\delta t))}{\sigma_{\Delta(t)}\sigma_{\Delta(t+\delta t)}}
\end{equation}.
It illustrates the relationship between data and their preceding points at $\delta t$ time intervals.
It is the Fourier transform of PSD.
The ACF enables us to
quantify the self-similarity of the signal over different time scales.
A highly self-correlated series leads to ACF$\sim 1$, whereas an uncorrelated time series leads to ACF$\sim 0$ and anti-correlation with ACF$\sim -1$.
Hence, we can assess whether a galaxy's SFR is correlated to its precursor's SFR at certain time scales.

\Fig{FigACF} shows the ACF of variation histories $\Delta(t_L)$  obtained in three simulations (top row) and their counterpart in our models (bottom row).
In the \ttt\ simulation, the ACF is close to $1$ when the time delay $\delta t$ is about $0.2 Gyr$.
This suggests that the SFHs retain their former state for a period $0.2 Gyr$.
In \tng\ and \ill\ the ACF at this time scale is much weaker than that of \ttt.
Especially in \ill, the $ACF(0.2 Gyr)$ concentrates on the value $<0.5$, implying that the variation histories in \ill\ are most likely uncorrelated at this time scale.
This is in consistent with the claim made by \cite{Caplar2019} that $\tau_{break}$ is around $200 Myr$.
As suggested by \cite{Caplar2019}, the time scale of $200 Myr$
is more likely to be associated with the baryonic effect.
The ACF finally drops down to $0$ when $\delta t$ increases to $2 Gyr$,
which is close to the dynamical time of dark matter halos.
Therefore, the baryonic effects play a more important role in shaping the SFH.
Moreover, it is possible that some baryonic effects in \ttt
lead to a stronger self-similarity of  variation histories at shorter time scales.

Looking at our model predictions (see bottom row in \Fig{FigACF}), they generally reflect the ACF distributions obtained from simulations,
with a little difference.
For \ttt, the modelled variation histories are less self-correlated than those from simulations when $\delta t<1 Gyr$.
On the contrary, the modelled variation histories are more strongly self-correlated than those from the \ill\ and \tng\ simulations when $\delta t<2 Gyr$.
In \ttt, many SFHs experienced many short time quenches, in which their SFR drops to 0.
The strong self-similarity within a short period in \ttt\ could be explained by those continued "0SFR" points in SFHs.
In the \ill\ and \tng\ simulations, there  are white noise components in their variation histories $\Delta(t_L)$,
as shown in \Fig{FigPSD}.
The white noise components reduce the self correlation of a time series, resulting in ACF values closer to $0$.

Both models and simulations
exhibit an uncorrelation (ACF$\sim0$) at a time scale of $\delta t=4$ Gyr, indicating that the
variations have totally forgotten their previous state prior to $4 Gyr$.

\subsection{The complete form of SFH model}
\label{sec:fullsfh}

In previous sections, we build up the SFHs along MS and their variations separately.
We combine these two parts together to achieve a complete SFH model:
\begin{equation}
	\label{equ:model3}
	\frac{dM_*}{dt_L}=-(1-\mu) 10^{(c+\alpha)t_L+(d+\beta)+\mathscr{A}B_H(t_L)}M_*^{at_L+b}
\end{equation}

We generate the modelled SFHs in the following procedure:
First, we copy the initial galaxies of each SFH from one simulation to form the initial stage of modelled SFHs.
This means that the number of samples, the initial mass, and the initial time in a model are exactly the same as in its corresponding simulation.
The mass growth histories of these modelled galaxies are then generated using the formula \Eqn{equ:model3} until redshift $z=0$ is reached.

In \Eqn{equ:model3} the movement of  the MS part's intercept($\Psi_0=ct_L+d$)
can merge with the inclination of the variation part ($\alpha t_L+\beta$).
We emphasize, however, that exact value of these two items have to be measured and determined in two approaches.

The free parameter $\mu$, also known as the mass-loss rate, is introduced here to represent the less-sufficient stellar mass growth\cite{Speagle2014}.
Except for the true mass loss caused by physical processes, $\mu$ is also affected by the variations of star forming and mergers on time scales shorter than the time step of snapshots.
It is preferable for this mass-loss rate to be time or mass-dependent \citep{Speagle2014,Leitner2011}.
\cite{Jungwiert2001} proposed a recipe of cumulative mass-loss rate $f_{\rm ml}(t)=C_0 ln(t/\tau+1)$, in which $C_0$ and $\lambda$ are free parameters.
The mass-loss rate $\mu(t)$ in \cite{Speagle2014} is  $\sim 0.45$ in the zeroth order and follows $d\mu / dt \sim 2/3 \times$ galaxy age in first order.
However, the situations are more complicated in simulations.
We show the mass and redshift dependence of the mass-loss rate in three simulations in appendix \ref{sec:mu} for readers who are interested in it.
But we will not study it further in this work.

In this work, we simply test the performance of the mathematical model described by \Eqn{equ:model3} with arbitrary constant mass-loss rates of
$0.5$, $0.8$, and $0.75$ for simulations \ttt, \ill\ and \tng\ respectively.
These values are determined by matching the stellar mass functions of models and simulations at $z=0$, which will be shown in next subsection.
Keep in mind that the parameter $\mu$ is the only free parameter to be fitted in this step.
The parameters in the variation part and the MS part keep their values in \Tbl{TabModPara2} hereafter.

\subsubsection{Evolution of stellar mass function}

\begin{figure*}
	\includegraphics[width=0.33\linewidth]{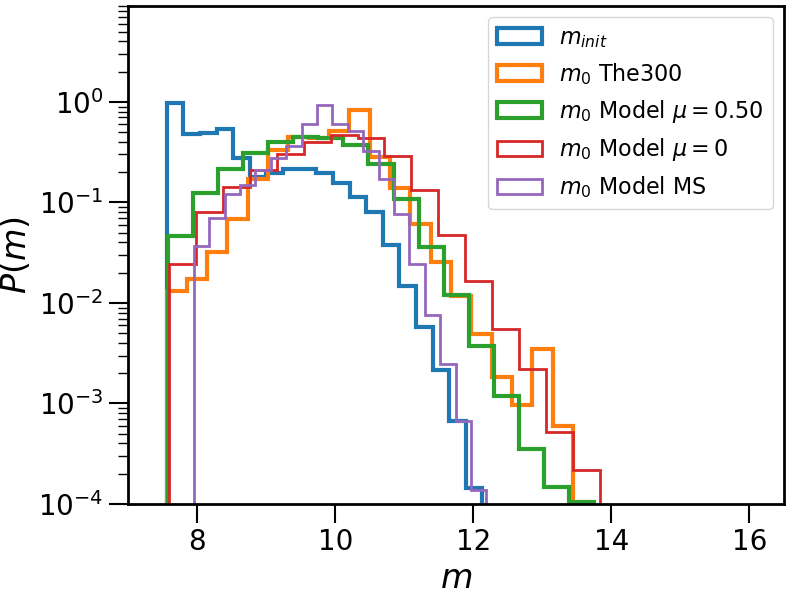}
	\includegraphics[width=0.33\linewidth]{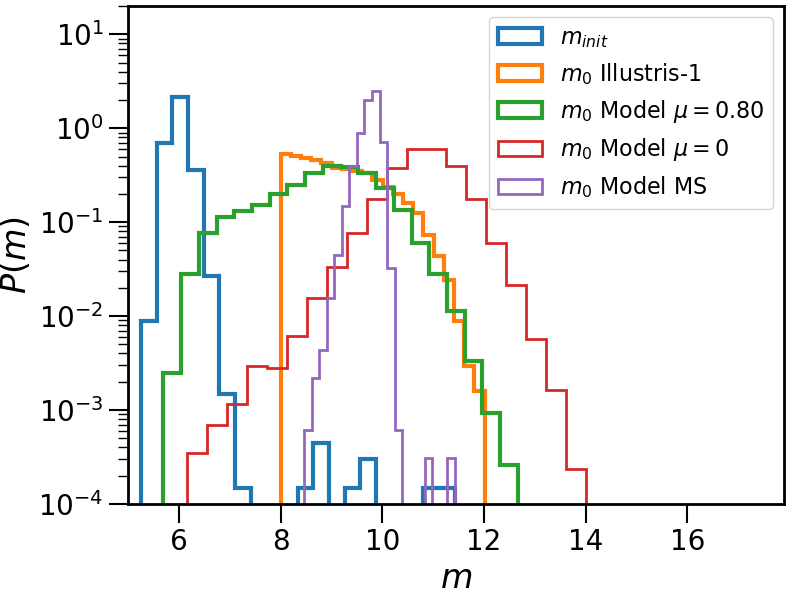}
	\includegraphics[width=0.33\linewidth]{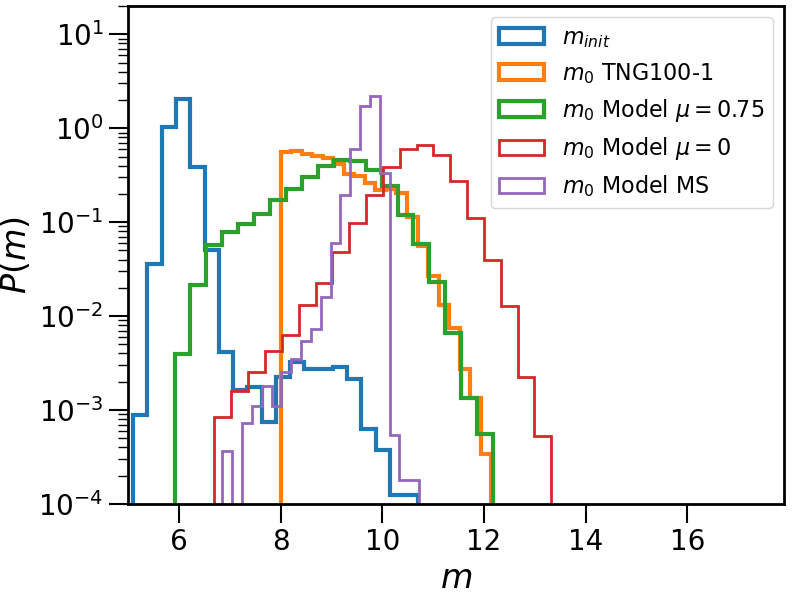}
	\caption{
		The  initial stellar mass function of SFHs (blue line), stellar mass functions at $z=0$ in simulations (orange line), and stellar mass functions of galaxies at $z=0$ from models(green, red, purple lines).
		The green lines represent the  model with a mass-loss rate that can best match the distribution of stellar mass observed in simulations.
		The red lines represent the model with a mass-loss rate of $0$.
		The purple lines represent the  model of SFHs with no variations and mass-loss rate of $0$ (\Eqn{EquMt}).
		The models and simulations in one subplot share the same initial stellar mass function (blue line).
		From left to right, figures show the results from the simulation \ttt, \ill\ and \tng\ respectively.
	}
	\label{FigMF}
\end{figure*}

To calibrate the performance of $\mu$ and variation part of modelled SFH, we compare the final stellar mass functions from simulations and different models.
\Fig{FigMF} shows the results of our test.
The blue lines show the distribution of $m_{init}$, the SFHs' initial stellar mass.
Be aware that, for one simulation, its corresponding models have the same distribution of $m_{init}$.
The $m_{init}$ distribution in \ttt\ clearly distinguishes from those in \ill\ and \tng.
The distribution of $m_{init}$ is not only affected by the physics and initial mass function.
It is also affected by the resolution and algorithm used to detect progenitors.

After growth following \Eqn{equ:model3}, the modelled SFHs generate stellar mass distributions quite comparable to that of simulations.
With appropriate mass-loss fractions, the distributions of $m_0$ from models (green line) can overlap with those from simulations (orange line).
The mass-loss rate has a vital role in shaping the stellar mass function, as can be seen.
Without it, i.e., when $\mu=0$, the galaxies in our models will be about $1$ magnitude oversized than simulations.
The effect of mass-loss rate in tuning over-sizing of galaxies is more important in \ill\ and \tng\ than in \ttt.
On the other hand, mass-loss rate mainly affects the amplitudes of stellar mass function.
The slopes are not changed when mass-loss rate is different.
Therefore, whether using a constant or time dependent mass-loss rate affects little on the final slopes of stellar mass function.

The variations of SFHs is also crucial for shaping the slopes of stellar mass functions, as seen in \Fig{FigMF}.
We present the $m_0$ distributions obtained by the model exclusively with mass growth along the MS (\Eqn{EquMt}, referred as ``MS model'' here after) for comparison(purple lines).
With only the MS part, the final distributions of $m_0$ are
likes to keep the shape of distribution of initial stellar mass.
This is effect is significant in \ill\ and \tng.

In \Fig{FigMF}, the low mass ends of stellar mass functions are
not recovered by our model.
There are too many small galaxies in our models.
This implies that the small galaxies may have growth path different from our model.

\subsubsection{Average PSDs of SFHs}

\begin{figure*}
	\includegraphics[width=0.33\linewidth]{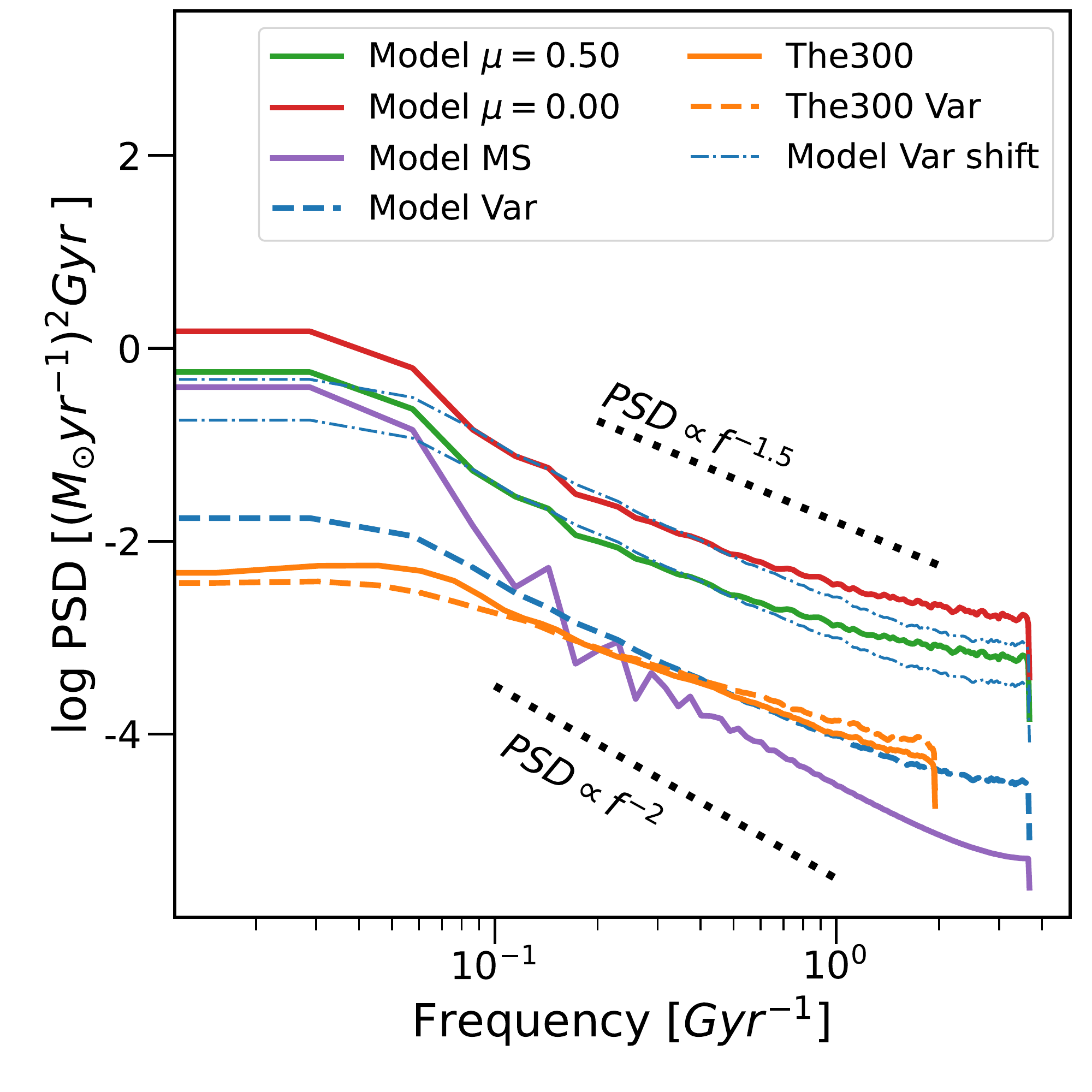}
	\includegraphics[width=0.33\linewidth]{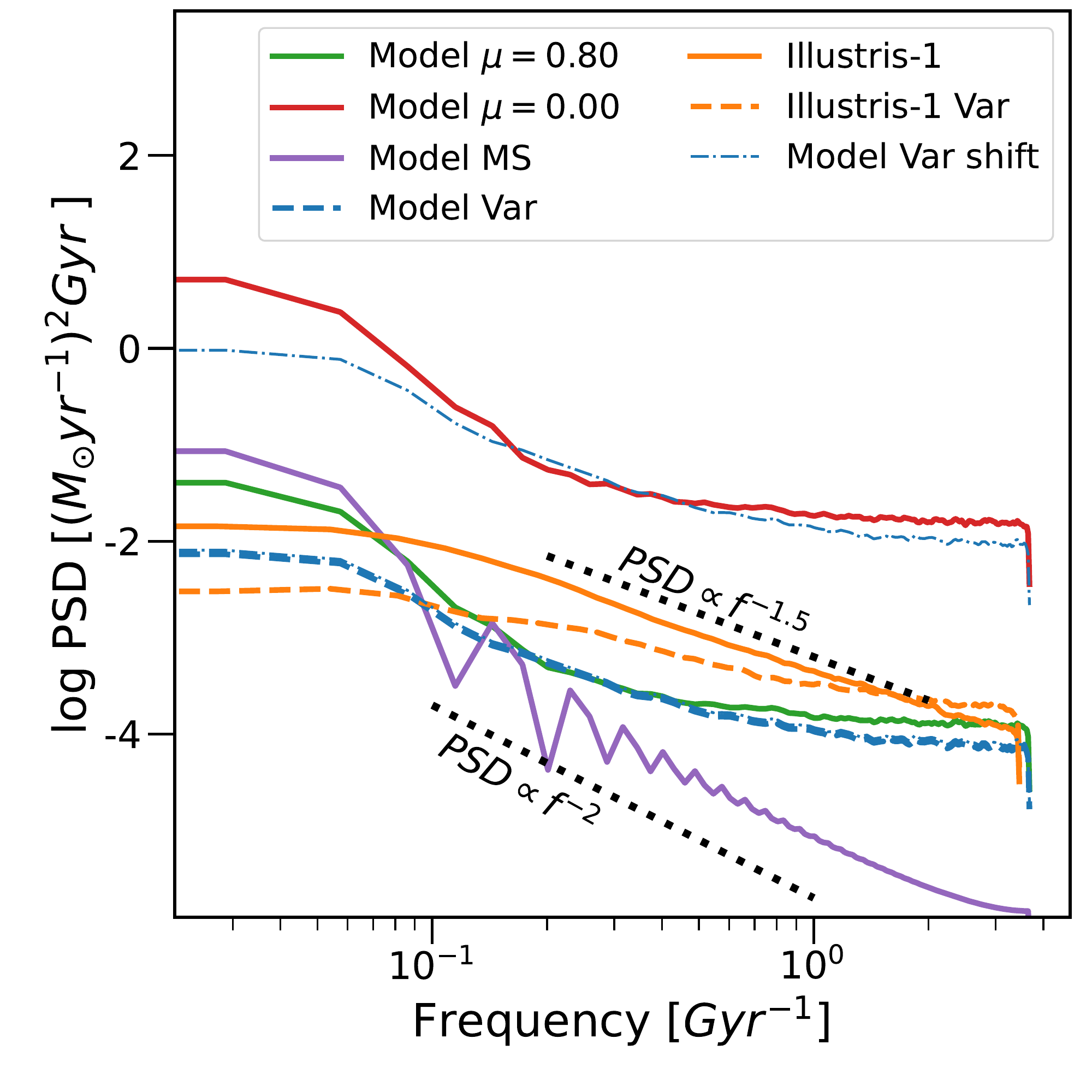}
	\includegraphics[width=0.33\linewidth]{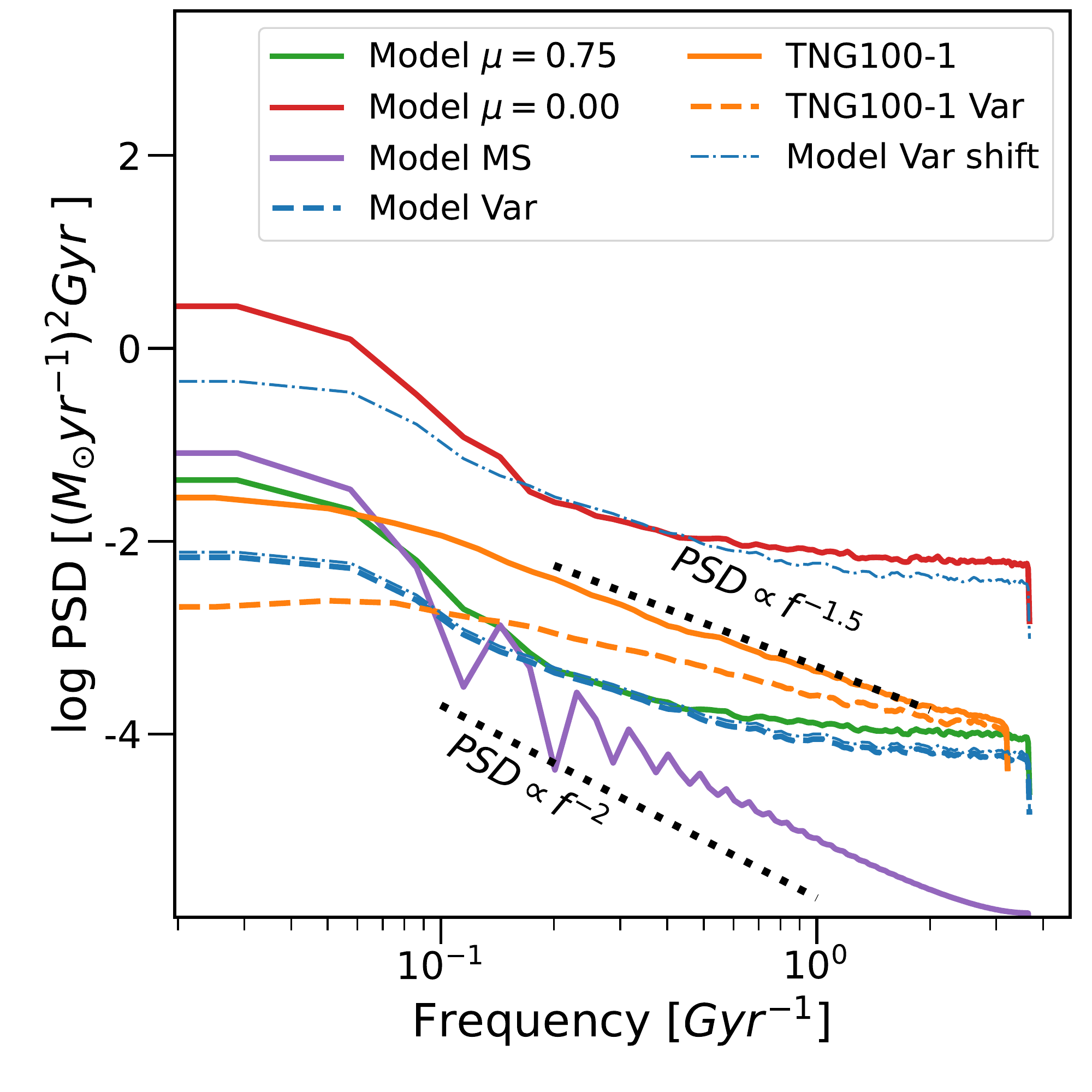}
	\caption{
		The average PSDs of SFHs from simulations and models.
		The average PSDs of SFHs from simulations are plotted with orange solid lines.
		The average PSDs of variation histories from simulations are plotted with orange dashed lines, which are labeled with ``Var'' in legend.
		The average PSDs of modelled SFHs with the best fitted mass-loss rate are plotted with green solid lines.
		The average PSDs of modelled SFHs without mass-loss are plotted with red solid lines.
		The average PSDs of modelled SFHs generated only by tracks along the MS are plotted with purple solid lines.
		The average PSDs of modelled variation histories are plotted with blue dashed lines.
		The blue dash-dotted lines are shitted copies of average PSDs of variation histories.
		They are shown in the purpose to compare the PSDs between variations and the complete SFHs.
		From left to right, figures show the results from the simulation \ttt, \ill\ and \tng\ respectively.
	}
	\label{FigPSDmean}
\end{figure*}

\begin{figure}
	\includegraphics[width=1\linewidth]{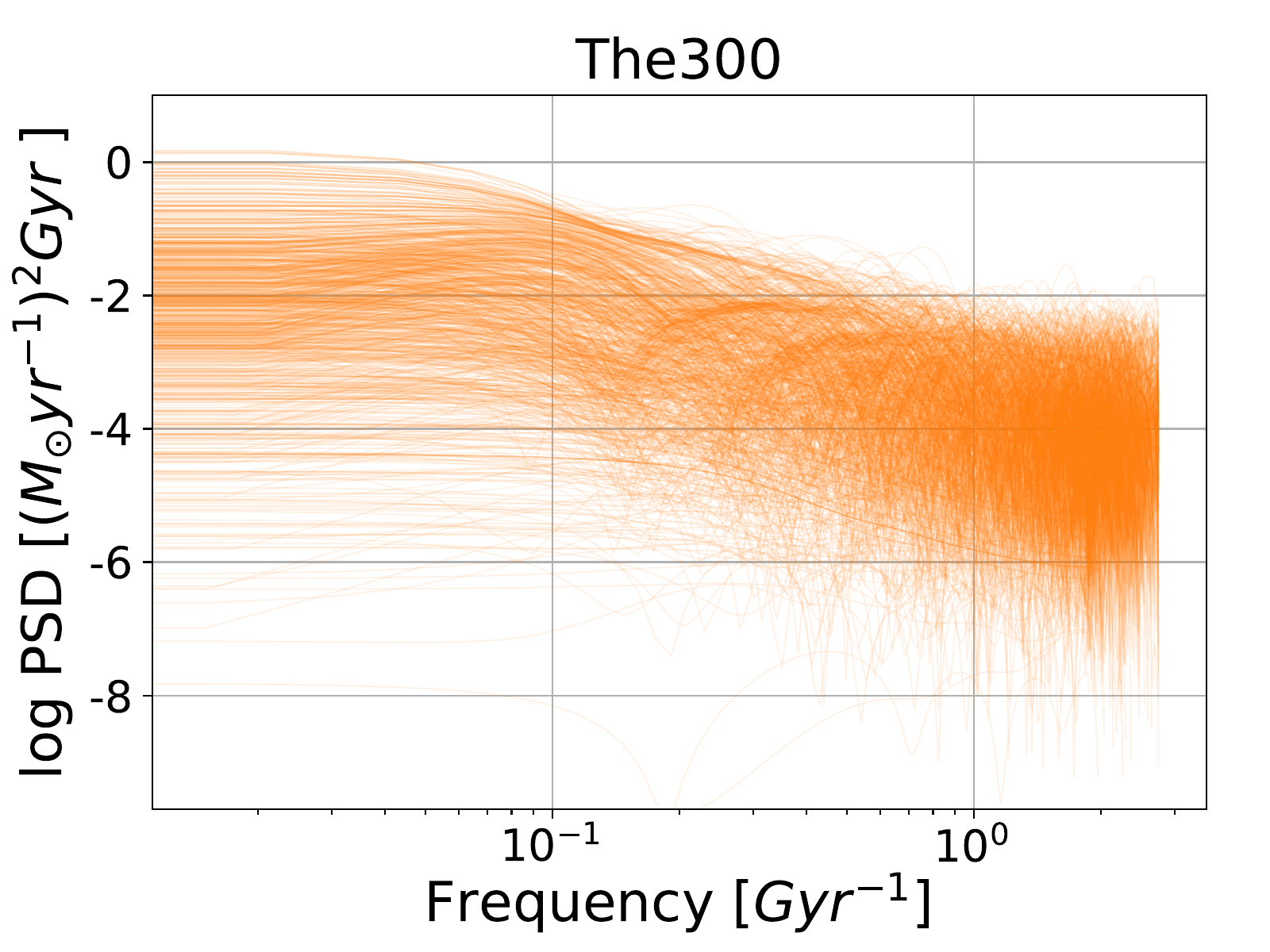}
	\includegraphics[width=1\linewidth]{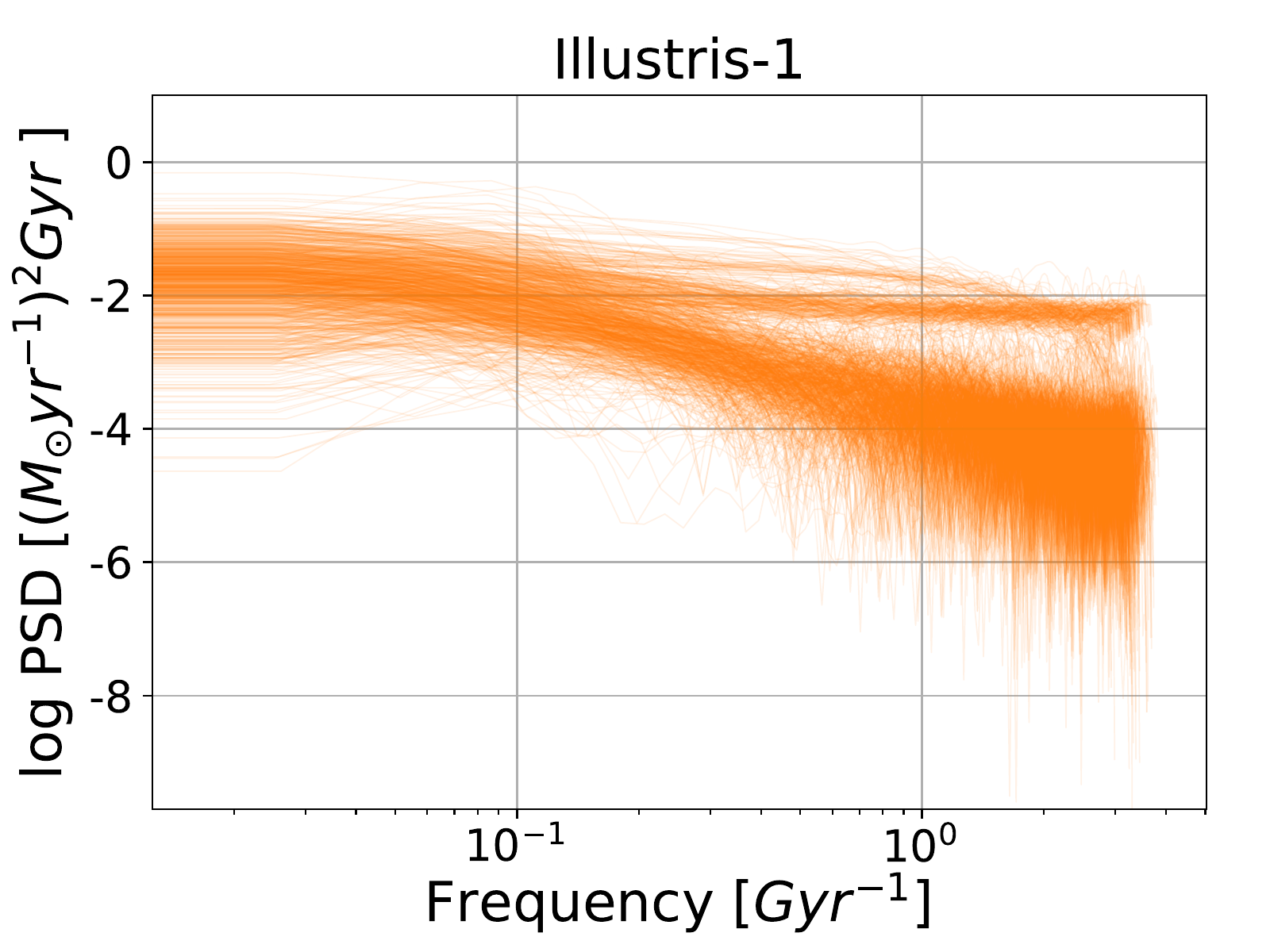}
	\includegraphics[width=1\linewidth]{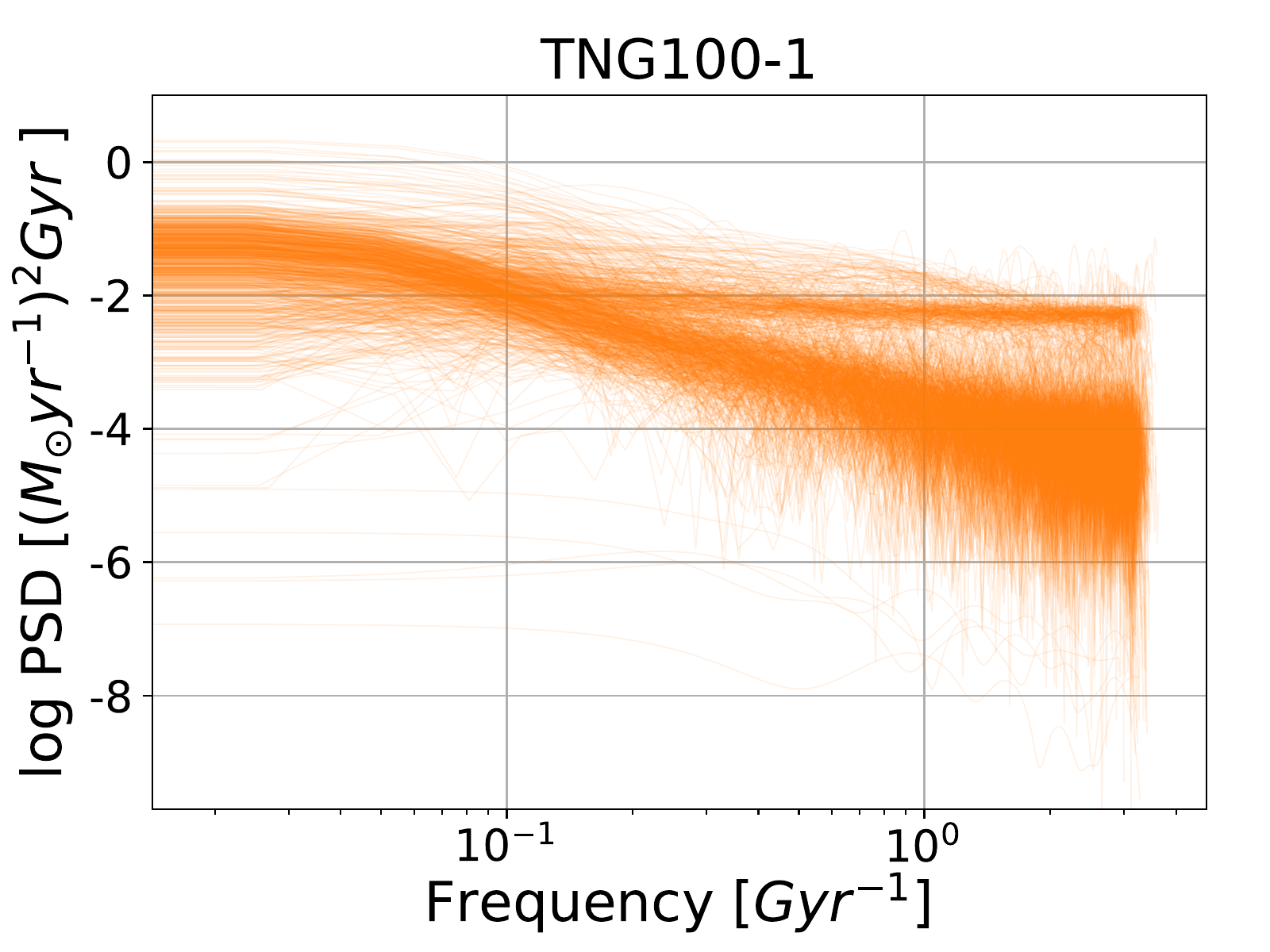}
	\caption{
		The PSDs of 1000 randomly selected SFHs from simulation \ttt\ (top), \ill\ (middle) and \tng\ (bottom).
	}
	\label{FigPSDsim}
\end{figure}

\Fig{FigPSDmean} shows the PSDs of SFHs from simulations and our models.
Three simulations' average PSDs for SFHs all have a slope of $-1.5$.
The simulation \ttt\ is different from \ill\ and \tng\ in the PSDs of variations histories.
With a slope of $-1.4$,
the average PSD of variation histories in \ttt is very close to the average PSD of SFHs.
In \ill\ and \ttt\, PSDs of SFHs and variation histories are clearly distinguished.
Their PSDs of variation histories have less power at lower frequency (larger time scale) region, with a slope of $-1$.
This suggest that the contributions of MS part and variation part in SFHs are different between \ttt\ and \ill/\tng.
Be aware that, although the average PSDs of SFHs from three simulations look close, the PSDs of individual galaxies differ significantly.
As \Fig{FigPSDsim} shows, the PSDs of SFHs in \ttt\ are more scattered but stay within one band.
In \ill\ and \ttt\ PSDs are less scattered, but show at least three distinguished bands.

\cite{Iyer2020} showed the average PSDs of SFHs from \ill\ and \tng\ in different stellar mass bin.
They find that these PSDs have a slope around $-2$.
There are two breaks in \ill\ around $\sim 0.6 - 1 Gyr$ and $\sim 2.6-4.2 Gyr$, and have one break in \tng\ around $\sim 1.1 - 2.6 Gyr$ for average PSDs in their work.
Our analysis suggest a slope of $-1.5$ for the PSDs of SFHs from simulations.
Be aware that two works use distinct definitions of star formation rate.
We use the instantaneous SFR given by the gas particles in each snapshots of simulations.
\cite{Iyer2020} tracks the SFH using the ages of star particles.
The discrepancy between this work and \cite{Iyer2020} shows the impact of different SFR indicator.
The breaks of PSDs we found are weak but still can be identified around frequency of $1 Gyr^{-1}$, as shown in \Fig{FigPSDsim}.
PSDs in this work are for all galaxies, whereas the data in \cite{Iyer2020} were divided into different stellar mass bins.
According to \cite{Iyer2020}, the breaks alter depending on the mass bin.
As a result, the breaks of average PSDs will be obscured for the entire sample.

Although the PSDs of variation histories are well matched between simulations and models, large discrepancy emerges when the MS part is taken into account in our model.
When the MS part is added to variation part, the PSDs rise at both low  and high frequency ends.
It makes our modelled SFHs have shallower slopes at high frequency and steeper slopes at low frequency.
Additionally, the mass-loss rate affect the total amplitudes of PSD, which does nothing with the slope of PSD.
The PSDs of modelled SFHs are not improved when we try to use the median evolutionary mass-loss rate shown in \Fig{FigMassloss}.
We need SFHs with higher temporal resolutions to solve this problem.

\section{Discussion \& Conclusion}
\label{sec:con}

In this work, we have investigated the evolution of SFR in three simulations \ttt, \ill\ and \tng.
We have proposed a mathematical model to match the SFR history of individual galaxy in the three simulations. The model, based on Brownian random motions on the SFR-mass diagram, turned out to reproduce the major features of galaxy SFHs.
Specifically, our model suggests that the SFR of a galaxy evolves according to this general law:
\begin{equation}
	\begin{aligned}
		\Psi(t_L)   & =\Psi_{\rm MS}(t_L,M_*(t_L))+\Delta(t_L)                          \\
		            & =[(at_L+b)m+ct_L+d]+[\alpha t_L+\beta+\mathscr{A}\times B_H(t_L)] \\
		            & =(at_L+b)m+(c+\alpha)t_L+(d+\beta)+\mathscr{A}\times B_H(t_L)     \\
		\Psi        & \equiv logSFR                                                     \\
		m           & \equiv logM_*                                                     \\
		\mathscr{A} & \sim \mathscr{N}(\mu_A, \sigma_A)
	\end{aligned}
	\label{EquFinal}
\end{equation}
where we
separate the SFH of a galaxy
into two parts: the trajectory following main sequence ($\Psi_{\rm MS}$) and variation component ($\Delta$). For each simulation, we use this model to fit their individual SFHs and get the best fit parameters.

In \Sec{sec:ms}, we
have discussed the evolution of the main sequence, based on the function $\Psi_{\rm MS}(t_L,M_*(t_L))$.
We have noticed that the main sequences differ between simulations, in both the slopes and intercepts.
Moreover, none of the main sequences in three simulations match the observational main sequence,
despite the fact that they can all statistically reproduce some observational quantities, such as the star formation rate density and stellar mass function.
The differences of MS can provide quantitative information to understand the effect of the physics behind the simulations.

Another important component of SFR  evolution is the variation part of the MS, which is linked to the internal or external galaxy processes that regulate the SFR and producing sudden variations. We are motivated by previous works \cite[e.g., ][]{Kelson2014,Caplar2019,Tacchella2020,Iyer2020} to assume that the variations of the SFR are stochastic processes that can be represented by
inclined  fractional Brownian motions. In \Sec{sec:var} we have introduced our method to reproduce the variation in SFH.
We have used the mean variation, mean squared variation, star burst time, quenching time, star burst duration and quenched duration to constrain the parameters of our variation history model.
The resulting models can predict the majority features of the time series of the variational histories, including their ACF and PSD.
Although the model do not fully recover all these  quantities from simulations, we have shown that
the fractional Brownian motion can reproduce majority of variation histories.
On the other hand, the divergence between models and simulations suggests that some processes like prolonged quenching and noisy-like fluctuation are not negligible.
According to our results, the SFHs in \ttt\ contain more quench stages besides Brownian motions, compared with other two simulations. On the other hand, \ill\ and \tng\ SFHs need multiple kinds of stochastic process, like white noise, in addition to the fractional Brownian motion to reproduce their features (see \Fig{FigPSD}, \Fig{FigPSDsim}).

We try to combine two parts together in \Sec{sec:fullsfh}.
The complete model can recover the stellar mass function.
But the PSDs of modelled SFHs is quite different from simulations.
We suggest that it is caused by the inaccuracy of mass-loss rate and low temporal resolution of SFH.

\cite{Caplar2019} proposed method quite similar to our model to construct the stochastic process of $\Delta_{\rm MS}$ history.
The variation of SFH in their model was defined through a power spectrum density with a functional form of a broken power-law, where the key characteristics are the timescale of uncorrelation $\tau_{\rm break}$ and slope of power-law $\alpha$.
The fractional Brownian motion is basically a subset of the stochastic process produced by the broken power-law.
Since the broken power-law method can reproduce stochastic series with all ranges of PSDs.
The fractional Brownian motion fixes its power-law slope to $-2H-1$ when $H<0.5$ and to $-2$ when $H>0.5$.
The Hurst parameter in fractional Brownian motion is quite close to the burstiness parameter in \cite{Caplar2019}.
It can be related to $\tau_{\rm break}$ and slope of PSD.
Figure A1 in \cite{Caplar2019} shows the relations between Hurst parameter, $\tau_{\rm break}$ and the slope.
The power slope $\alpha\simeq-2$ is favored in most theoretical and observational works \citep[e.g., ][]{Caplar2019}.
But our model suggests that the variations in simulations are more likely to have a slope shallower than that, e.g., $\sim -1.4$ for \ttt\ and $\sim -1.1$ for \ill\ and \ttt.
Moreover, we find that the PSDs of variations and SFHs do not have the same slope, which is observed in the simulation \ill\ and \ttt.

In future works, we will improve our model's flaws.
Firstly, we will experiment with various stochastic process outside fractional Brownian motion.
We might try assigning different stochastic process to galaxies of various types or masses.
Secondly, quenching process need to be considered beside a normal stochastic process.
Third, we need to established a compatible function for the mass-loss rate in simulations.
Moreover a simulation with higher output frequency can help us improve the model.
As we find in this work, the large time step between snapshots brings uncertainties in constructing the SFH of an individual galaxy.

On the other hand,  we will explore the link between parameters of our mathematical models and the sub-grid physics in simulations, which will help calibrating the influence of sub-grid physics in simulations.

% \normalem
\section*{Acknowledgements}
\addcontentsline{toc}{section}{Acknowledgements}

XL is supported by the NSFC grant (No. 11803094), and the Science and Technology Program of Guangzhou, China (No. 202002030360).

YW is supported by NSFC grant No.11733010, NSFC grant No.11803095 and the Fundamental Research Funds for the Central Universities.

WC is supported by the European Research Council under grant number 670193 and by the STFC AGP Grant ST/V000594/1. He further acknowledges the science research grants from the China Manned Space Project with NO. CMS-CSST-2021-A01 and CMS-CSST-2021-B01.

NRN acknowledges financial support from the “One hundred top talent program of Sun Yat-sen University” grant N. 71000-18841229.

This work has received financial support from the European Union's Horizon
2020 Research and Innovation program under the Marie Sklodowskaw-Curie
grant agreement number 734374, i.e., the LACEGAL project.
The authors would like to thank The Red Espa\~nola de Supercomputaci\'on for
granting us computing time at the MareNostrum Supercomputer of the BSC-CNS
where most of the cluster simulations have been performed. Part of the
computations with {\textsc{Gadget-X}} have also been performed at the
'Leibniz-Rechenzentrum' with CPU time assigned to the Project 'pr83li'. The
authors would like the acknowledge the Centre for High Performance Computing
in Rosebank, Cape Town for financial support and for hosting the ``Comparison
Cape Town" workshop in July 2016. The authors would further like to
acknowledge the support of the International Centre for Radio Astronomy
Research (ICRAR) node at the University of Western Australia (UWA) in the
hosting the precursor workshop ``Perth Simulated Cluster Comparison" workshop
in March 2015; the financial support of the UWA Research Collaboration Award
2014 and 2015 schemes; the financial support of the ARC Centre of Excellence
for All Sky Astrophysics (CAASTRO) CE110001020; and ARC Discovery Projects
DP130100117 and DP140100198. We would also like to thank the Instituto de
Fisica Teorica (IFT-UAM/CSIC in Madrid) for its support, via the Centro de
Excelencia Severo Ochoa Program under Grant No. SEV-2012-0249, during the
three week workshop ``nIFTy Cosmology" in 2014, where the foundation for this
this project was established.

Most analysis of this work is done on the Kunlun HPC in SPA, SYSU.

The authors thanks the Illustris projects for providing the data.

The authors contributed to this paper in the following ways: GY, FRP, AK, CP
and WC formed the core team that provided and organized the simulations and
general analysis of data.
AA \& Giuseppe Murante managed the {\sc Gadget-X} simulation and
provided the data. The specific data analysis for this paper was led by YW, NRN, XL, XK.

YW wrote the text.
All authors had the opportunity to
proof read and provide comments on the paper.

\section*{Data Availability}
\addcontentsline{toc}{section}{Data Availability}
The data underlying this article will be shared on reasonable request to the corresponding author.

\bibliographystyle{mnras}
% \bibliography{C:/Users/wango/OneDrive/Work/Bib/Lib}
\bibliography{Lib}

\appendix

\section{Main sequence at different redshifts}

\begin{figure*}
	\includegraphics[width=0.33\linewidth]{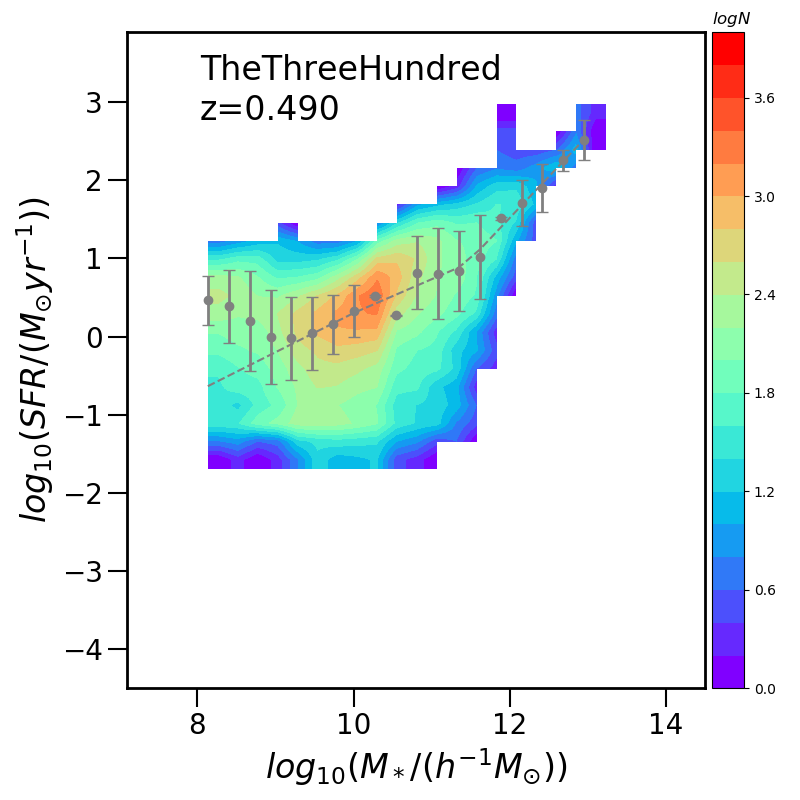}
	\includegraphics[width=0.33\linewidth]{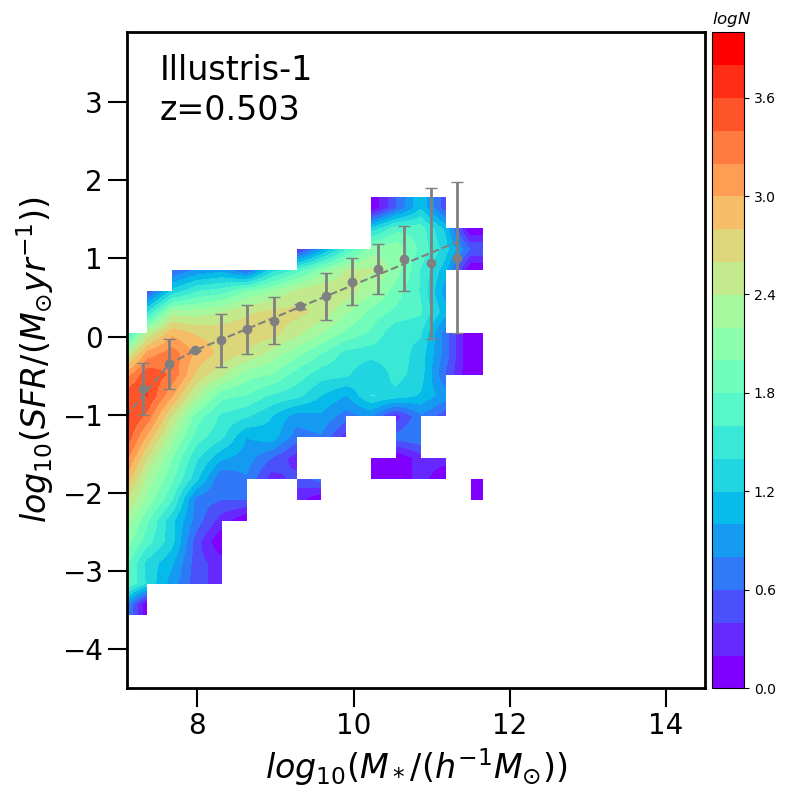}
	\includegraphics[width=0.33\linewidth]{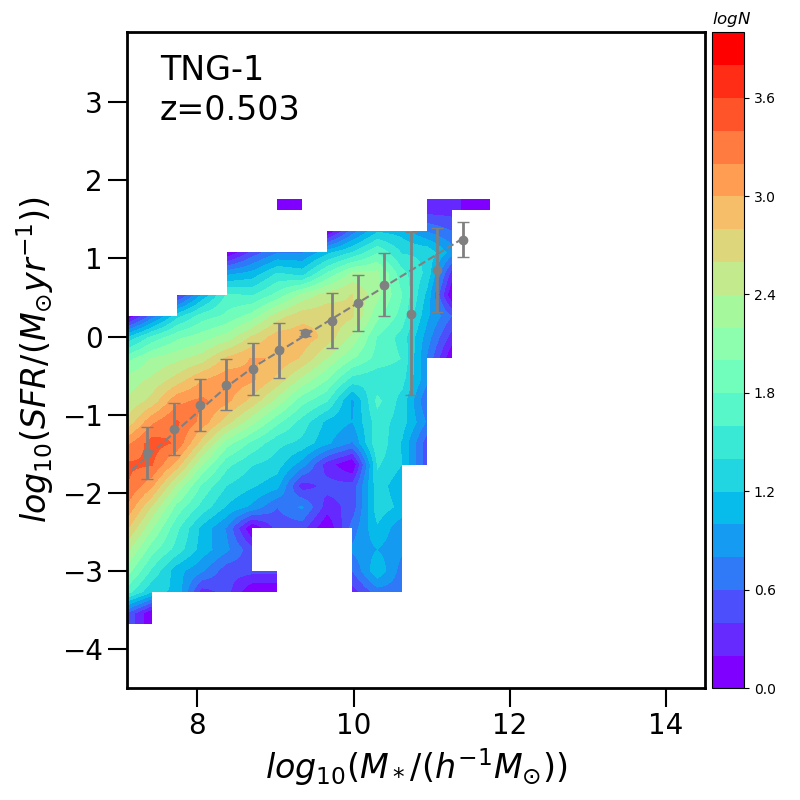}
	\includegraphics[width=0.33\linewidth]{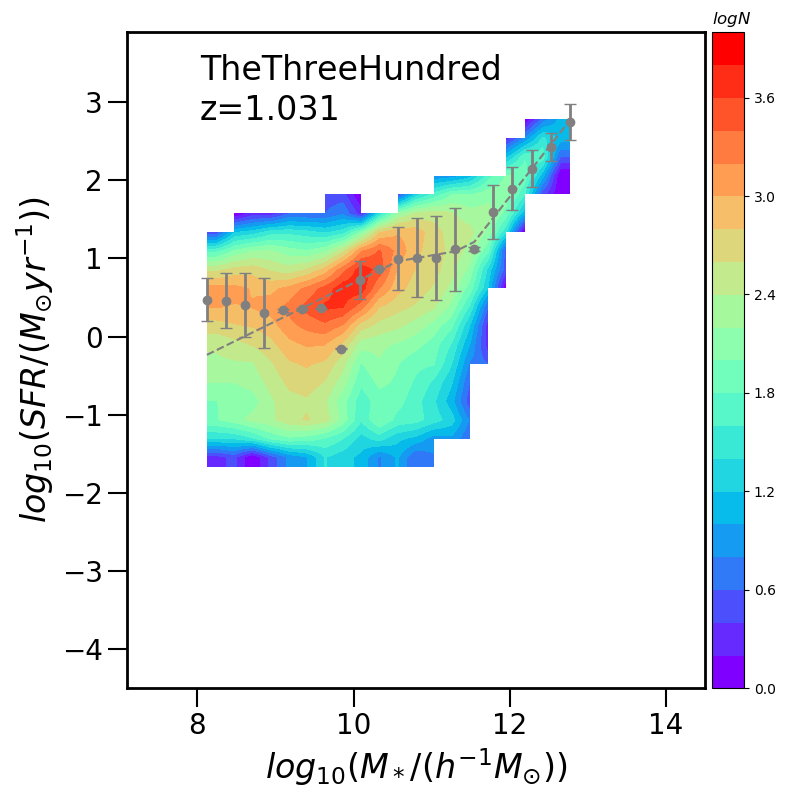}
	\includegraphics[width=0.33\linewidth]{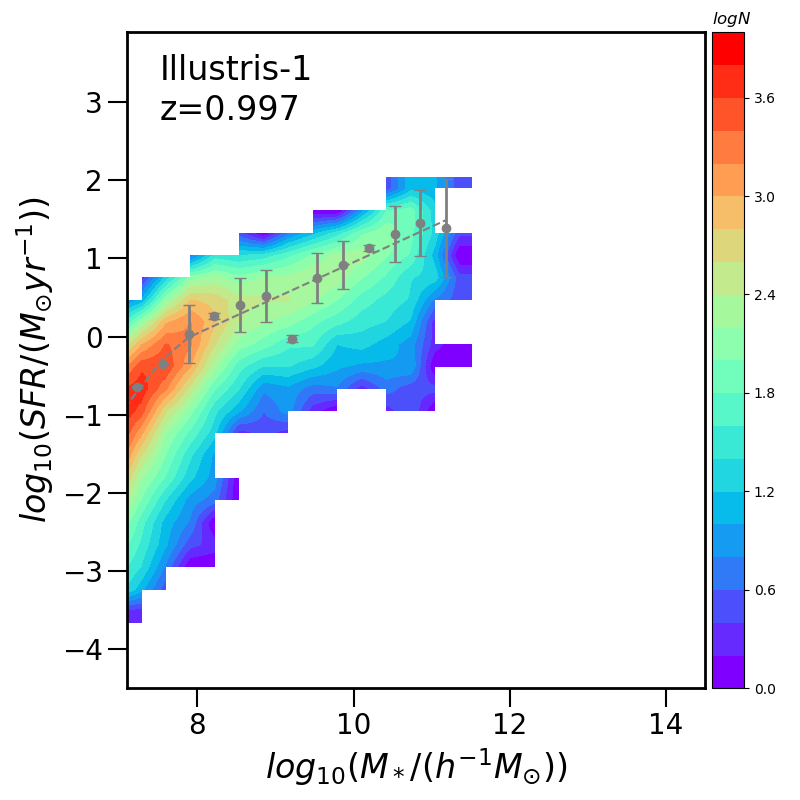}
	\includegraphics[width=0.33\linewidth]{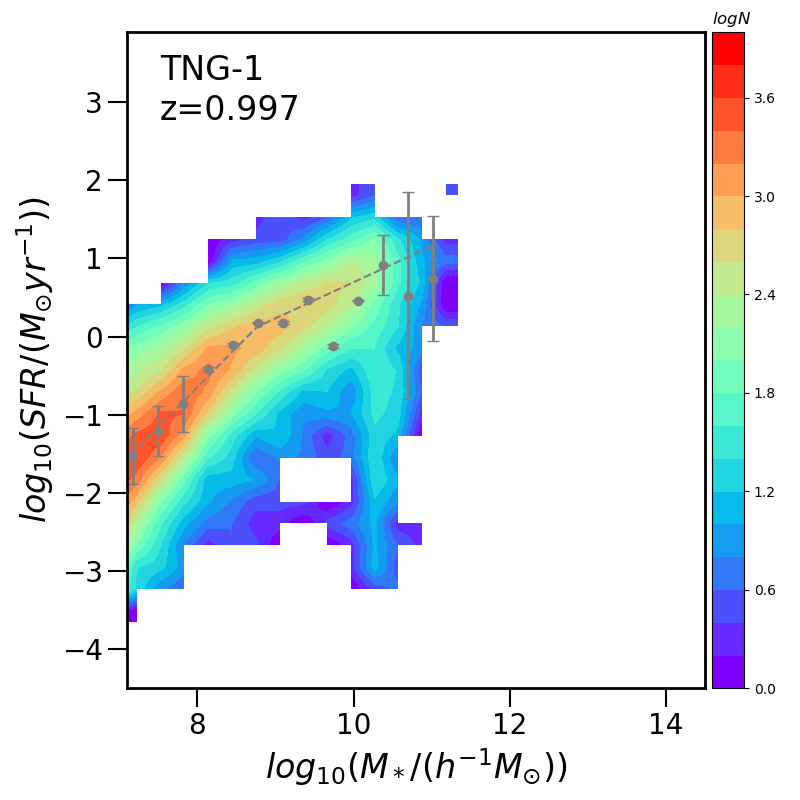}
	\includegraphics[width=0.33\linewidth]{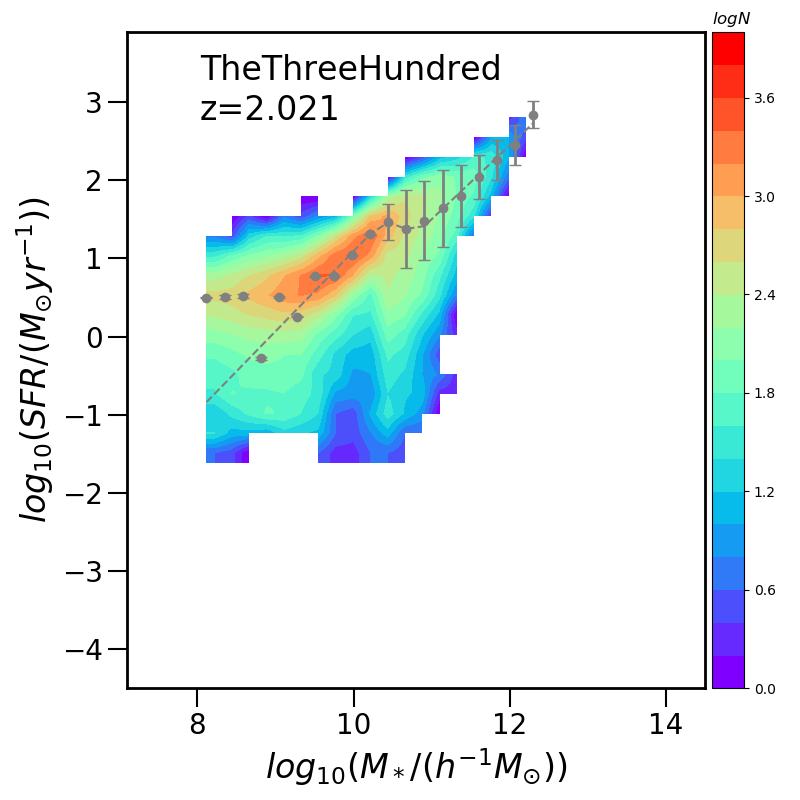}
	\includegraphics[width=0.33\linewidth]{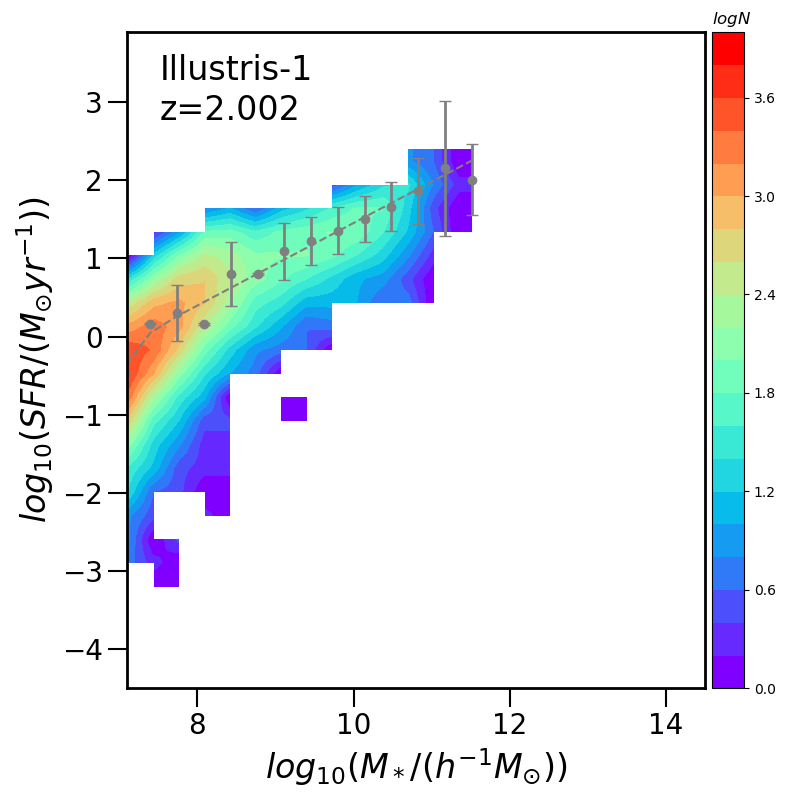}
	\includegraphics[width=0.33\linewidth]{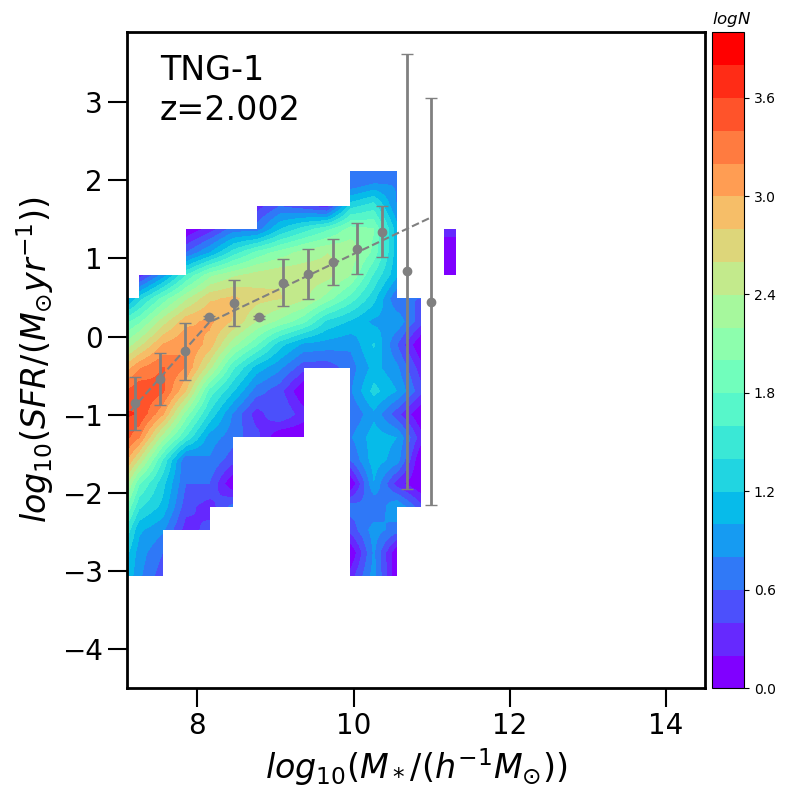}
	\caption{The contour of $SFR-M_*$ distribution of galaxies in  simulations \ttt\ (left), \ill\ (middle) and \tng\ (right) at higher redshifts. The pattern is identical to that in Fig.\ref{FigMSz0}.
	}
	\label{FigMSzH}
\end{figure*}

\Fig{FigMSzH} shows the distribution of SFR versus stellar mass of galaxies at high redshift in three simulations, as well as the shape of the main sequence.

\section{Comparison of Main Sequence between simulations and observations}
\label{sec:mscompare}

For comparison with observations, we additionally include the MS parameters from
\cite{Speagle2014} and \cite{Iyer2018} in \Tbl{TabMSPara}.
\cite{Speagle2014} derived the MS from a compilation of 25  papers.
\cite{Iyer2018} found an evolving MS in the CANDELS GOODS-S survey.
We convert their MS formula to be in the same units as ours.
For example, in \cite{Iyer2018}'s text , the original description of MS is as follows:
\begin{equation}
	\begin{aligned}
		\log SFR= & (0.80\pm0.029-0.017\pm0.010t_{\rm univ})\log M_* \\
		-         & (6.487\pm0.282-0.039\pm0.008t_{\rm univ})
	\end{aligned}
	\label{EquIyer1}
\end{equation}
By applying $t_L\simeq13.82 Gyr-t_{\rm univ}$, $\Psi=\log SFR$ and $m=\log (M_*h)$ to equation above, we get:
\begin{equation}
	\begin{aligned}
		\Psi_{\rm MS,obs}= & (0.017\pm0.010t_{\rm L}+0.57\pm0.031)m \\
		-                  & 0.042\pm0.008t_{\rm L}-6.04\pm0.302.
	\end{aligned}
	\label{EquIyer2}
\end{equation}

Both the \cite{Speagle2014} and \cite{Iyer2018} studies imply that the MS has a higher slope and somewhat lower intercept at early stages.
In \cite{Speagle2014}, the stellar mass range of galaxies is around $10^{9.7} \sim 10^{11.1} M_{\rm \odot}$.
The samples in \cite{Iyer2018} range from  $10^7$ to $10^{11}$$M_{\rm \odot}$.

	Lower mass galaxies have MS that are closer to the MS in \cite{Iyer2018} because they are less massive than $\sim 7$ in \ill  or $sim 8$ in \tng .
	For MS of galaxies with a mass greater than $10^8 M_{\rm \odot}$, the discrepancy between simulations and observations can not be ignored.

	It's important to note that the observations do not all agree on how MS is evolving.
	For example, \cite{Whitaker2012} found that the slope of MS is $0.70-0.13z$, which is in direct contrast to \cite{Speagle2014} ($0.84-0.026t_{\rm univ}$) and \cite{Iyer2018}($0.80-0.017t_{\rm univ}$).
	These uncertainties are most likely caused by SFR measurements and sample selection.
	\cite{Whitaker2014} found that the evolution of the MS slope may differ for different SFR indicators
	at $M_*<10^{10.2}M_{\rm \odot}$.

	Generally, there are distinctions between simulations and simulations, as well as between simulations and observations.
	To reconcile all this contradictory evidence, it is important to match the evolution of the SFR main sequence between simulations and observations.
	Indeed, in order to fully understand the physical processes underlying the evolution of the SFR main sequence and possibly improve the hydro-dynamical recipes in simulations, it seems crucial to 1) homogenize the observational results and 2) match the definition of the quantities derived from simulations and observations.

	\section{Analytical term of mass growth following MS}
	\label{ap1}

	From \Eqn{EquMt}, we can expand the $\Psi_0$ and $k$:
	\begin{equation}
		\begin{aligned}
			\frac{dM_*(t_L)}{dt_L}= & -10^{\Psi_0}M_*^k        \\
			=                       & -10^{ct_L+d}M_*^{at_L+b}
		\end{aligned}
	\end{equation}

	The expression $at+b$ represents the evolution of slope $k$, while the $ct+d$ represents the evolution of intercept $\Psi_0$.
	The value of $a,b,c,d$ can be found in \Tbl{TabMSPara}.
	The solution to this equation is complicated.
	However, because the slope of MS $k$ changes little over time, resulting in a very small $a$ in \Eqn{EquMt}, we can simplify the solution by taking $a\simeq0$ ( i.e., $k=b$).
	Then the solution to \Eqn{EquMt} is:

	\begin{equation}
		M_*(t) =\begin{cases}
			[-\frac{1-b}{c ln10}10^{ct_L+d}+C_0(1-b)]^{\frac{1}{1-b}} & b\neq1 \& c \neq 0 \\
			[-(1-b)10^dt_L + C_0(1-b)]^{\frac{1}{1-b}}                & b\neq1 \& c=0      \\
			C_0 e^{\frac{-10^{ct_L+d}}{c ln10}}                       & b=1 \& c\neq 0     \\
			C_0 e^{-10^dt_L}                                          & b=1 \& c=0         \\
		\end{cases}
		\label{EquMtFull}
	\end{equation}
$C_0$ is an arbitrary constant.

	\label{sec:mu}
	\begin{figure*}
		\includegraphics[width=0.3\linewidth]{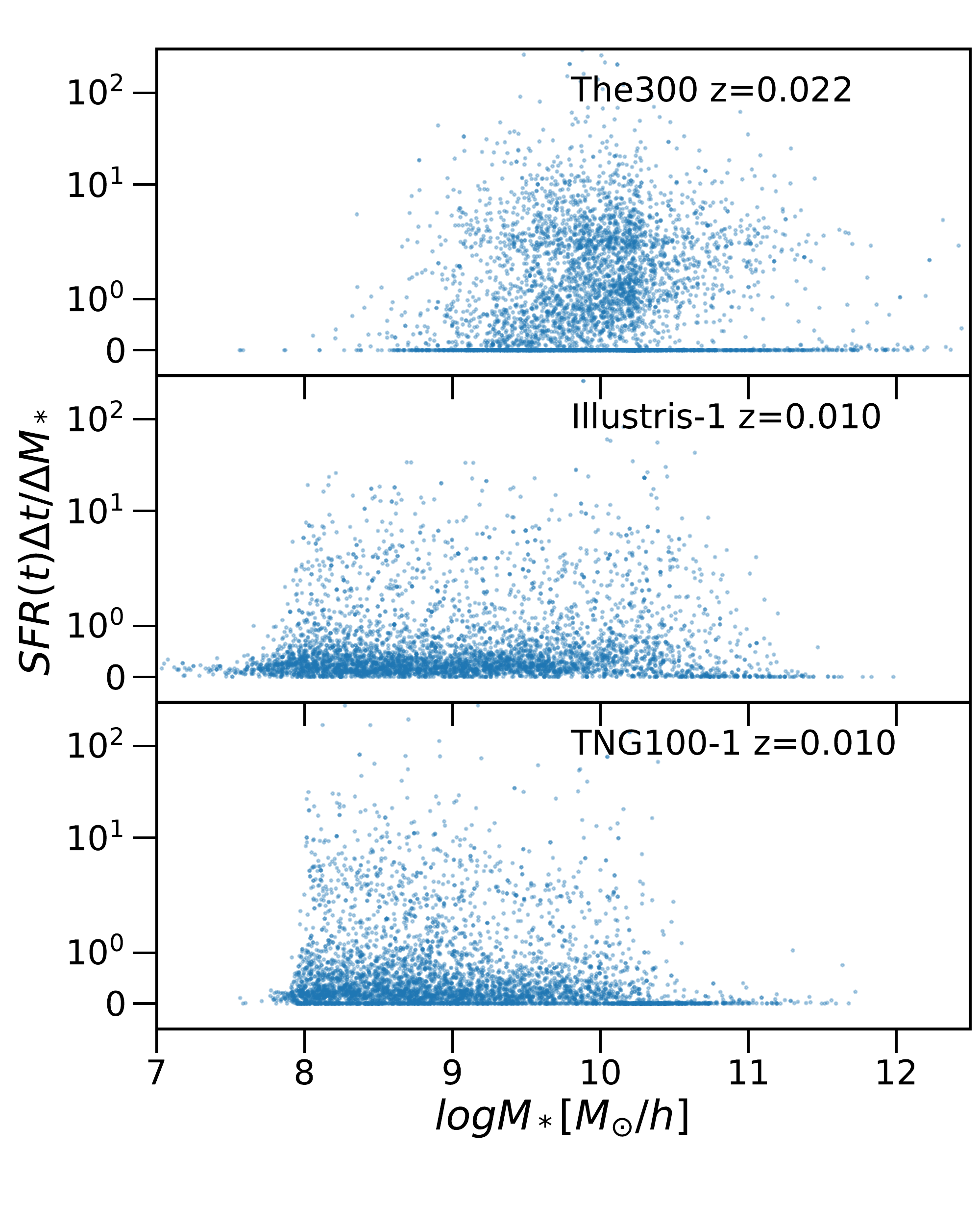}
		\includegraphics[width=0.3\linewidth]{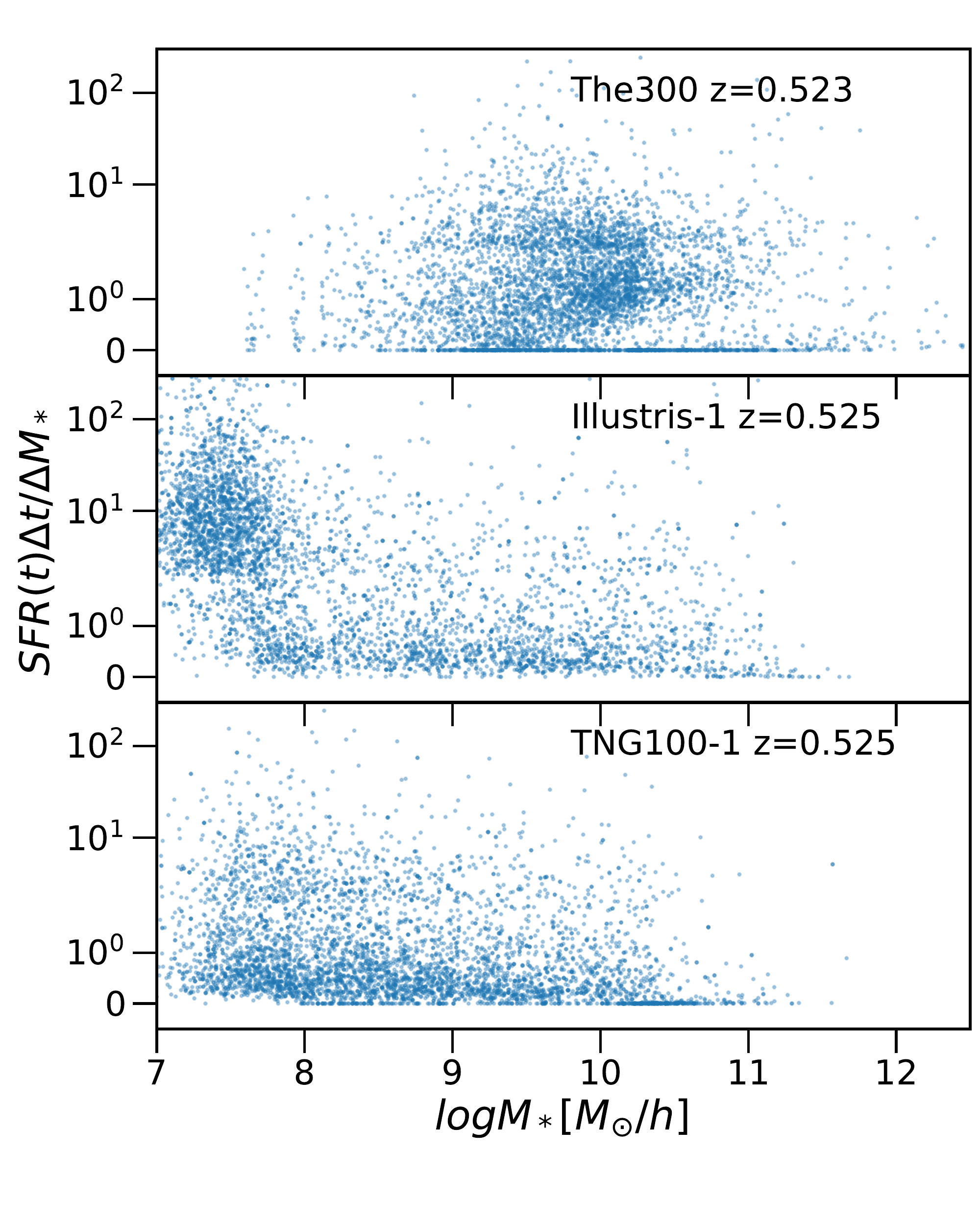}
		\includegraphics[width=0.3\linewidth]{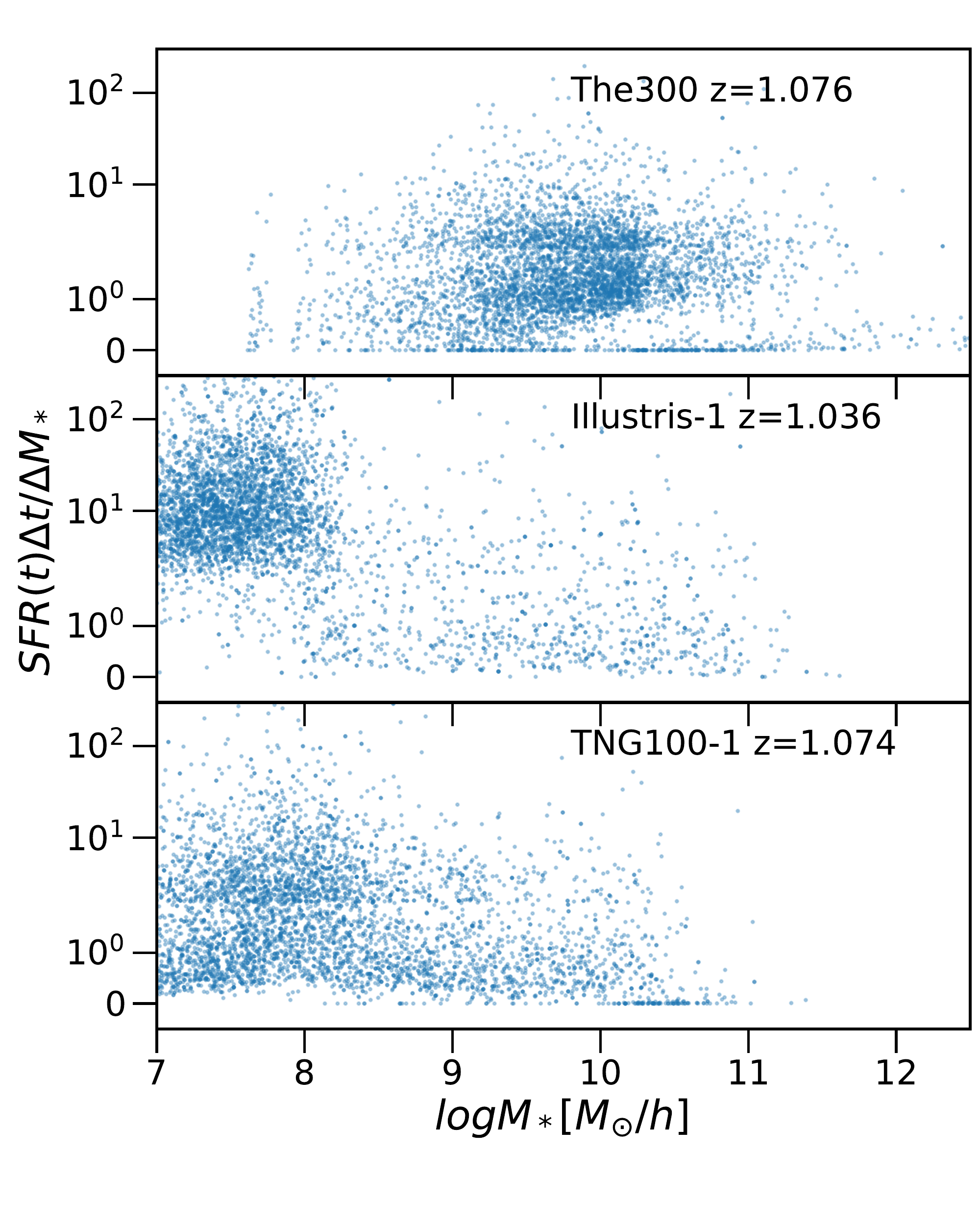}
		\caption{
			The mass dependence of the mass increment ratio predicted by SFR ($SFR\times \Delta t$) to the actual mass increment in the same time interval for three snapshots from each simulation.
			In each snapshot, 2000 galaxies are randomly selected.
			We leave out the $\Delta M_*\le0$ points.
			Subplots from top to bottom show scatter plots from  simulations \ttt, \ill\ and \tng, respectively.
			Each column shows results from snapshots with close redshifts.
		}
		\label{FigMasslossM}
	\end{figure*}

	\begin{figure}
		\includegraphics[width=0.8\linewidth]{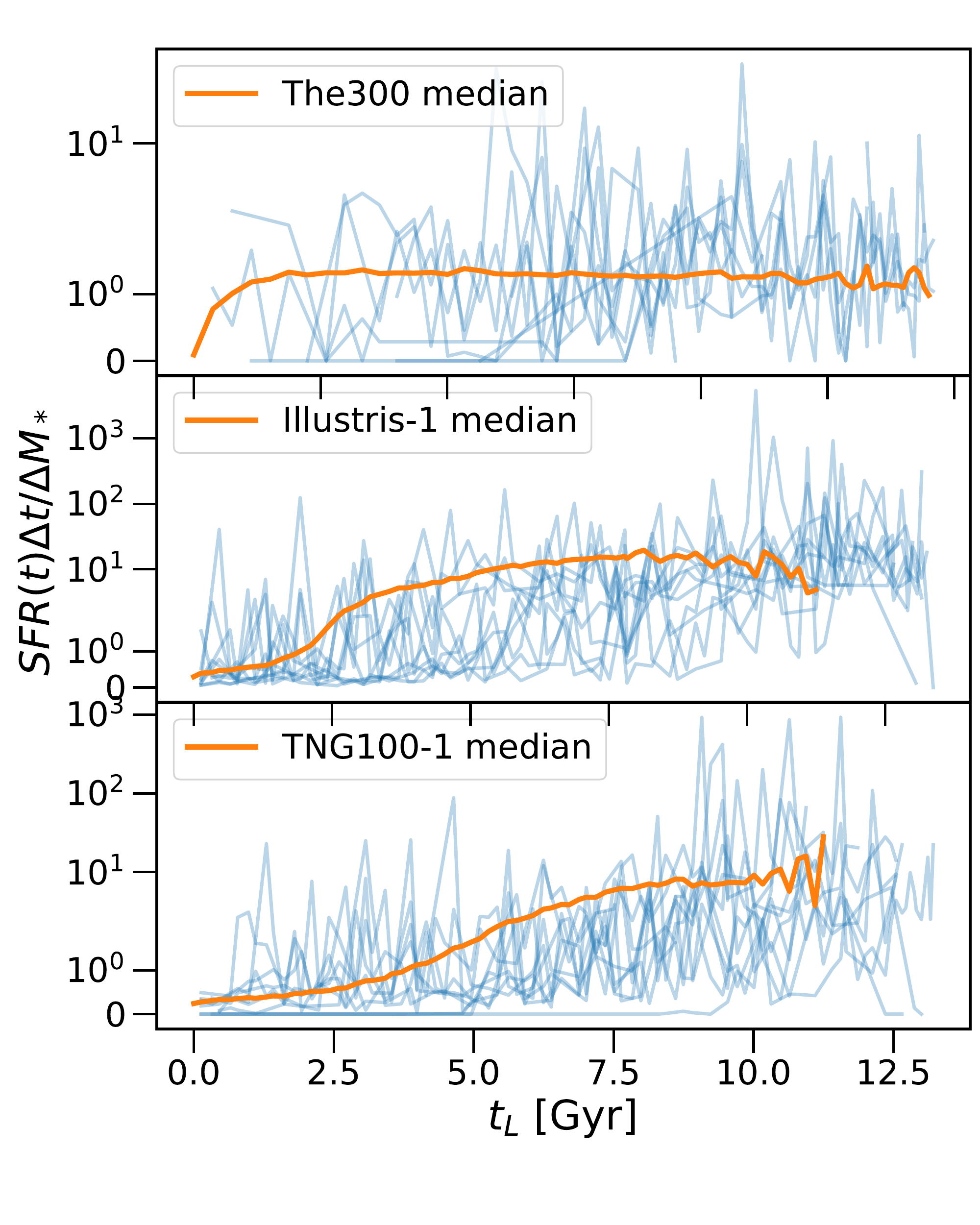}
		\caption{
			The time dependence of the mass increment ratio predicted by SFR ($SFR\times \Delta t$) to the actual mass increment within the same time interval for all snapshots from simulations.
			The relation between mass increment ration ratio and mass-loss rate is $SFR(t)\Delta t/\Delta M_* = 1/(1-\mu(t))$.
			We do not directly plot the curves of $\mu(t)$ because there are points with $SFR(t)=0$ that result in infinity values for $\mu(t)$.
			We leaf out the $\Delta M_*\le0$ points.
			The blue lines show the ratio of $10$ randomly chosen SFHs.
			The orange lines show the median ratio of $2000$ randomly chosen SFHs.
			Each subplot shows data from one simulation, \ttt, \ill\ and \tng\ from top to bottom, respectively.
		}
		\label{FigMassloss}
	\end{figure}

	In \Eqn{EquMtFull}, a galaxy's stellar mass grows exponentially or in a power law (when $k\neq 1$ and the intercept does not change with time).
	According to \Tbl{TabMSPara}, the MSs in all three simulations and other works follow the case that $b\neq1$ and $c\neq0$

	\section{Time and mass dependence of mass-loss rate}

	The time dependence of the mass-increment rate is shown in \Fig{FigMassloss}.
	The mass dependence of the mass-increment rate is shown in \Fig{FigMasslossM}.
	To visualize the data better,
	we use mass-increment ratio $SFR(t)\Delta t/ \Delta M$, i.e., $1/(1-\mu(t))$, instead of the mass-loss rate $\mu$.
	Because those points with $SFR(t)=0$ will lead to infinity values for $\mu(t)$.

	\Fig{FigMassloss} depicts the change of mass-increment rates of
	individual SFHs and their median trends in three simulations.
	As can be seen, the mass-increment ratio of individual SFH is volatile.
	The main uncertainties come from the large fluctuations of star formation, mergers, and gas flows hidden between two snapshots.
	Because the time interval is substantially larger than the time scale for instantaneous SFR of particles in simulations,
	the integration of instantaneous SFR across two snapshots has a significant bias.
	In this case, strictly following the mass-loss rate in simulations is almost impossible.
	Alternatively, using the average mass-loss rate in \Eqn{equ:model3} will modify the pace of mass growth, and subsequently change the  characteristics such as PSD and ACF of the SFH.

	Three simulations have different median mass-loss-rates-time relations.
	The median $SFR(t)\Delta t/ \Delta M$ in \ttt is slightly larger than $1$ and nearly constant when $t_L>\sim 1 Gyr$.
	It indicates that the mass-loss rate $\mu$ in \ttt is small and stable.
	In the very recent time, star formation has stopped contributing to mass growth.
	Mergers are the primary source of galaxy growth during this period.
	In \ill\ and \tng, the median $SFR(t)\Delta t/ \Delta M$ is around $10$ at early time and gradually decreases.
	It means that there is significant mass loss at early time and the contribution of star formation to mass growth decreases gradually.

	In \Fig{FigMasslossM}
	the mass-increment ratio does not show obvious mass dependency.
	There is a clear division line at a mass-increment ratio of $\sim 2$(i.e., $\mu\sim0.5$) to separate samples into two groups.
	In \ttt, the mass-loss ratio is independent of $M_*$ for low $\mu$ galaxies but slightly decreases with stellar mass for high $\mu$ samples.
	The mass-loss rate appears to be independent of $M_*$ in \ill and \tng\.
	The number of low $\mu$ samples increases at higher redshifts.

In summary, the mass-loss rate in \ttt is more likely mass dependent, while in \ill and \tng it is more time dependent.

\label{lastpage}
\end{document}